\documentclass[twocolumn]{aastex62}

\pdfoutput=1
\received{ , 2020}
\revised{ , 2020}
\accepted{ , 2020}
\submitjournal{ApJS}
\shorttitle{Pulse Properties of Prompt Gamma-rays of Swift/BAT Short GRBs}
\shortauthors{X.-J., Li et al.}
\usepackage{amsmath}
\usepackage{graphicx}
\usepackage{multirow}
\usepackage{subfigure}
\begin{document}
\title{Temporal Properties of Precursors, Main peaks and Extended Emissions of Short GRBs in the Third Swift/BAT GRB Catalog}
\author{X. J. Li}
\author{Z. B. Zhang*}
\author{X. L. Zhang}
\author{H. Y. Zhen}
\affil{College of Physics and Engineering, Qufu Normal University, Qufu 273165, People's Republic of China;*Email: astrophy0817@163.com }
\begin{abstract}
A comprehensive study is given to short gamma-ray bursts (sGRBs) in the third Swift/BAT GRB Catalog from
December 2004 to July 2019. We examine in details the temporal properties of the three components in the prompt gamma-ray emission phase, including precursors, main peaks and extended emissions (EE). We investigate the similarity of the main peaks between one-component and two-component sGRBs. It is found that there is no substantial difference among their main peaks. Importantly, comparisons are made between in the single-peaked sGRBs and the double-peaked sGRBs. It is found that our results of main peaks in Swift/BAT sGRBs are essentially consistent with those in CGRO/BATSE ones recently found in our paper I. Interestingly, we suspect, besides the newly-found MODE I/II evolution forms of pulses in BATSE sGRBs in paper I, that there would have more evolution modes of pulses across differently adjacent energy channels in view of the Swift/BAT observations. We further inspect the correlation of the main peaks with either the precursors or the EEs. We find that the main peaks tend to last longer than the precursors but shorter than the EEs. In particular, we verify the power-law correlations related with peak fluxes of the three components, strongly suggesting that they are produced from the similar central engine activities. Especially, we compare the temporal properties of GRB 170817A with other sGRBs with EE and find no obvious differences between them.

\end{abstract}

\section{Introduction} \label{sec:intro}
Gamma-ray bursts (GRBs) are sudden releases of gamma-rays, lasting from milliseconds to thousands of seconds. The typical energy range
of GRBs is from tens of keV to several MeV. The soft gamma-ray/hard X-ray emission is usually called ``prompt emission"
of GRBs. According to the classification criterion based on the duration distribution \citep{1993ApJ...413...L101}, short gamma-ray
bursts (sGRBs), for which T$_9$$_0$ $<$2 s, likely originate from the mergers of compact binaries involving Double Neutron Star
(DNS) and Neutron Star Black-Hole (NSBH) systems \citep{1986ApJ...308...L43, 1989nature...340...126, 1992ApJ...395L..83N}. The prompt
emission properties of sGRBs are distinguished from long GRBs (lGRBs), for example, sGRBs are found to have negligible spectral lag
\citep {2000APJ...534...248, 2006MNRAS...373...729} and harder spectra than lGRBs \citep {1993ApJ...413...L101}.
Swift/BAT (The Burst Alert Telescope) has detected about 120 sGRBs in the past $\sim$14 years since it was successfully launched in November 2004 and achieved several important breakthroughs during the study of the prompt emission and the early afterglow \citep{2004ApJ...611...1005, 2016ApJ...829...7}.

In the internal shock model, as the jet propagates, a faster shell meets a slower one and interacts in the form of an
internal shock, which is thought to cause the prompt emission of GRBs \citep{1994APJ...430...L93, 1997APJ...490...92, 1999PR...308...L43}. Although the light curves of the prompt emissions of GRBs have very irregular and complex structures, some of these can be divided into some individual or overlapping pulses which contain the key information of internal
energy dissipation and radiative mechanisms \citep{1996APJ...459...393, 2005APJ...627...324}. In general, the prompt gamma-ray emission
of GRBs may consist of various emission episodes, including precursors, main peaks and
extended emissions (EEs) or parts of them. For some GRBs, the three components are usually bright enough to be detected easily. For instance, the extraordinarily bright GRB 160625B was found to have three
emission episodes separated by long quiescent intervals \citep{2018Nature...2...69}.

\cite {1974ApJ...194...L19} reported a probable precursor prior to the three main impulses with the deviation 3.1 $\sigma$ from
background. They pointed out that the precursor was not initiated by its most
explosive phase. The precursor might come from the photospheric emission and have black
body-like spectrum \citep [e.g.,][]{1991Nature...350...592,2000ApJ...543...L129, 2002MNRAS...336...1271,2007MNRAS...380...621}.
For the lGRBs, the jet launched by the central engine makes its way out of the stellar envelope of the progenitor star and releases
thermal emission as the shock breakout precursors \citep{2002APJ..331...197}. When the progenitor system is a NS-NS system and
or magnetar, the similar shock breakout precursor could occur in sGRBs if a dense wind is released
from the center engine \citep{2014ApJ...788...L8}. \cite{2007APJ...670...1247} discussed that
the central engine might undergo a second collapse and the precursor may be due to the initial weak jet. They reported that some of the initial
jet materials fall back onto the central engine when it manages to penetrate the stellar mantle and will be accreted by
the central engine. The core collapse or
binary merger could produce a temporarily stable intermediate object as a ``spinar", which supports a large range of precursor energies \citep{2008MNRAS...383...1397, 2009MNRAS...397...1695}. Different precursor models can be used to explain the diversity of the quiescent period timescales, the spectral and temporal properties and so on. Observationally, significant differences were not found not only
between precursor and main peak but also for the GRBs with/without precursor
\citep [e.g.,][]{2010ApJ...723...1711, 2014ApJ...789...145}. Moreover, there was no obvious or mild correlations between
 the precursor and the main peak \citep [e.g.,][]{1995ApJ...452...145, 2005MNRAS...357...722, 2008APJ...685...L19, 2009AA...505...569,
 2015MNRAS...448...2624}. By reanalysing the observed pulse light curves, one can check the previous findings  or exclude some untenable theoretical models for the origin of the precursors.

The EE component as soft $\gamma$-ray emission or
hard X-ray afterglow occurring after the main prompt emission is another important messager. Postburst emission component might be a
feature of BATSE bursts, which was interpreted as the hard X-ray afterglow occurring \citep [e.g.,][]{2001AA...379...L39,2002APJ...567...1028,2005Science...309...1833}.
\cite{2002Mazets} discussed a special kind of the ``short" burst, the initial emission of which was spike-like and accompanied
by low intensity EE for tens of seconds. \cite{2006Nature...444...1044} found that the temporal lag and peak luminosity of long
GRB 060614, were similar to those of the sGRBs.
By studying a large BATSE sample, \cite{2006APJ...643...266} found that a handful of GRBs were somewhat similar to GRB 060614.
They showed that the extended components were always softer than
the initial spikes. Note that GRB 060614 is a sGRB-like long burst with the EE component \citep{2006Nature...444...1053}, which is very similar to GRB 050709, a short burst with the EE lasting about 130s. Interestingly, both of them are found to associate with a macro-nova and share with the same origin \citep{2015ApJ...811...L22,2016NC...7...12898}.
Although the EE tails within some GRBs have been confirmed,
their physical natures are still in debate. For example, the sGRBs with EE tail may be produced by the formation and early evolution of
a highly magnetized, rapidly rotating neutron star \citep[e.g.,][]{2008MNRAS...385..1455,2012MNRAS...419...1537B}. \cite{2011MNRAS...417...2161} proposed
a short-duration jet powered by heating due to $\nu$$\tilde{\nu}$ annihilation and a long-lived Blandford-Znajek jet to describe
the initial pulses (IP) and the EE segment. The lifetime of the accretion process, which is divided into multiple emission episodes
by the magnetic barrier, may be prolonged by radial angular momentum transfer and the accretion disk mass is critical for producing
the observed soft EE \citep{2012Apj...760...63}. \cite{2017MNRAS...470...4925} explained the EE tail with a process of fallback accretion on to a new born magnetar. The temporal and spectral characteristics of the IP and the EE of the sGRBs, or
the sGRBs with/without EE, were extracted and compared \citep[e.g.,][]{2010APJ...717...411, 2011APJ...735...23, 2013MNRAS...428...1623,
2015MNRAS...452...824, 2015ApJ...811...4, 2016ApJ...829...7, 2018MNRAS...481...4332}. \cite{2017ApJ...846...142} argued that the EE components reflect the central engine activities instead of the external environments associated with afterglows in general. Thus, it would be very intriguing to
analyse whether observational temporal properties of EE tail were similar or dissimilar to those of precursor
or main peak.

For the first time, we systematically investigate the temporal properties of the fitted sGRB pulses in the third Swift/BAT catalog and present a joint temporal analysis of the three prompt emission components across four energy channels in one-component and two-component sGRBs.
The data preparation and sample selection are given in Section 2. The results are presented in Section 3, in which, we pay special attention to a direct comparison with
our recent results of BATSE sGRBs (Li et al. 2020, hereafter paper I).
Finally, we discuss and summarize the results in Section 4 and 5, respectively.
\section{SAMPLE PREPARATION} \label{sec:DATE PREPARATION }
\subsection{Data and Method}
We construct our initial sGRBs sample using the parameter T$_9$$_0$ from the third Swift GRBs Catalog from December 2004 to July 2019, which corresponds to about all the sGRBs detected by the Swift/BAT. The sample comprises 124 sGRBs including additional sGRB 090510 \citep [e.g.,][]{2010ApJ...720...1008, 2010ApJ...723...1711,
2010ApJ...716...1178, 2013ApJ...772...62}, sGRB 050724 \cite[e.g.,][]{2007APJ...655...989,2007PTRSLS...365...1281}
and GRB 060614 \citep [e.g.,][]{2006Nature...444...1053,2006Nature...444...1044, 2007APJ...655...L25, 2013MNRAS...428...1623,
2014ApJ...789...145, 2015ApJ...811...4, 2016ApJ...829...7}.

The mask-weighted light curve data of sGRBs are taken from
the Swift website \citep{2016ApJ...829...7}\footnote{\url{https://swift.gsfc.nasa.gov/results/batgrbcat/}} for four energy channels, labeled with Ch1 (15-25 keV), Ch2 (25-50 keV), Ch3 (50-100 keV), and Ch4 (100-350 keV). We calculate the background noise (1$\sigma$) and define the effective sGRBs signal at a level of S/N $>$3. Although the increase in bin size can reduce the
level of background noise fluctuations, it might change the potential pulse structure.
Because the total BAT energy band with good localization is narrower
and the signals of sGRBs are relatively weaker than those of the lGRBs, we fit the sGRB pulses of different energy bands with a small bin size of 8ms till the potential GRBs signals can be identified significantly, except some sGRBs with precursor or EE, i.e. GRB 071112B and GRBs 060614, 150101B, 170817A. For these four GRBs, we fit the pulse light curves with other bin sizes
of 2ms, 16ms, 64ms or 1s instead. Several points need to be cautioned for these GRBs. First, the precursor or EE weak pulse structure can all be identified in a corresponding energy band. Second, the detection points of the effective GRB signal should be relatively more enough to ensure a successful fit. In addition,
for the possible weak signals, we combine the adjacent energy channels into one channel in order to increase the statistical reliability.
Considering the features of duration
and noise level of sGRBs, the mask-weighted light curves data are intercepted from
1$-$2T$_9$$_0$ prior to the BAT trigger time and 2$-$3T$_9$$_0$ posterior to the trigger time to
further enhance the fitting accuracy. The detailed methods to identify the pulse numbers of a burst have been described in paper I. The pulse shapes depend on the final choice of fit. In this study, we have
used the least chi-square criterion together with a residual analysis to evaluate the goodness of our fits, as done in paper I and other previous works \citep[e.g.,][]{1996APJ...459...393,2005APJ...627...324,2003APJ...596...389,Peng2006,2011ApJ...740...104,2019ApJ...876...89}. In spite of this, this high-dimensional nonlinear regression fit should instead be performed using unbinned maximum likelihood estimation using the photon arrival times \citep{2002Fraley,2000McLachlan,2008McLachlan,2014Tartakovsky}. Subsequently, we will apply the powerful EM method to study how many pulses within a burst in our subsequent paper.

Several authors have proposed some relatively simple functions to describe the pulse light curves of GRBs \citep{2002APJ...566...210, 2003APJ...596...389,
1996APJ...459...393, 2005APJ...627...324, 2007APJ...662...1093, 2007ApJ...670...565, 2016ApJS..224...20Y}. Among these functions, the ``KRL'' function proves the most flexible profiles of individual GRB pulses \citep{2003APJ...596...389} and can be written as \begin{equation}\label{equation:1}
f(t)=f_m(\frac{t+t_0}{t_m+t_0})^r[\frac{d}{d+r}+\frac{r}{d+r}(\frac{t+t_0}{t_m+t_0})^{(r+1)}]^{-(\frac{r+d}{r+1})},
\end{equation}
where \emph{r} and \emph{d} determine the rise and the decay shapes of an individual pulse, $f_m$ represents the peak flux, $t_m$ is the peak time, $t_0$ is the offset from the pulse start time to the trigger time. Simultaneously, the five parameters in the ``KRL'' model have been interpreted in theory by \cite{2005MNRAS...363...1290}. As applied in our paper I, this empirical function will be utilized again in this study.

To investigate the prompt emission mechanisms or classify GRBs, many temporal properties of GRBs pulse light curves have been studied \citep[e.g.,][]{1996APJ...459...393, 2005APJ...627...324, 2001AA...380...L31, 2002AA...385...377,2003APJ...596...389,2007APSS...310...19,2011ApJ...740...104, 2014ApJ...783...88,2015ApJ...815...134,2018ApJ...855...101,2019ApJ...883...70,2019-190510440}, but bimodal distributions are still preferred \citep[e.g.,][] {1993ApJ...413...L101,2008AA...484...293,2016APSS...361...257,2017APSS...362...70,
2018ApSS...363...223,2016MNRAS...462...3243,2018PASP...130...054202,2015AA...581... 29,2019ApJ...870... 105,2019ApJ...887... 97}.
In this study, the pulse properties including peak amplitude (f$_m$), peak time (t$_m$), full width
at half maximum (FWHM), rise time (t$_r$) and decay time (t$_d$) as well as asymmetry (t$_r$/t$_d$) will be investigated in details for different kinds of Swift sGRBs. The systematic errors of pulse measurements are estimated with error propagation using the same methods described in paper I according to \cite{2006ChJAA...6...312}. In particular, if the diverse parameters of the pulse shapes have strong covariance the relationships among these pulse properties could not reflect underlying behaviors. Fortunately,  we calculate the covariance matrix and the correlations matrix and find that there are no significant correlations between different pulse parameters derived from our fits.

\subsection{Selection Criteria of Precursor and EE Candidates }
 \cite {1995ApJ...452...145} had a conclusion that there are only 3\% GRBs with precursor in their 1000 BATSE
 GRBs. \cite {2010ApJ...723...1711} found that 8\%-10\% of Swift/BAT sGRBs display precursor. Further studies showed that roughly 5\% of Fermi/GBM sGRBs and 18\% of Fermi/GBM lGRBs have precursor \citep{2015PHD...???...???}. \cite{2017ApJ...43...1} find the precursors existing in less than 0.4\% of
the SPI-ACS/INTEGRAL sGRBs. There is no obviously objective criterion to define the ``precursor''. In general, the peak flux of precursor is smaller than those of main events while the flux falls below the background level before the start
of the main event \cite[e.g.,][]{2008APJ...685...L19, 2010ApJ...723...1711}. \cite{2014ApJ...789...145} pointed out that precursors can be triggered and non-triggered events. Considering all above aspects, we identify a significant precursor
when it fulfills the following three conditions:

(1) The precursor is effective as the detection points are at least 3$\sigma$ above the background in the whole energy range of 15-350 keV.

(2) The precursor at least includes three detection points and its peak flux is smaller than that of the main peak (see also \citealt{2008APJ...685...L19,2010ApJ...723...1711}).

(3) The precursor can be detected prior or posterior to the BAT trigger time. Simultaneously, quiescent period duration between the precursor and the main peak is well-defined (or, the precursor flux has fallen into the background level when the main peak starts) (see also \citealt{2008APJ...685...L19,2010ApJ...723...1711}).

\begin{figure*}
\centering
\gridline{
\fig{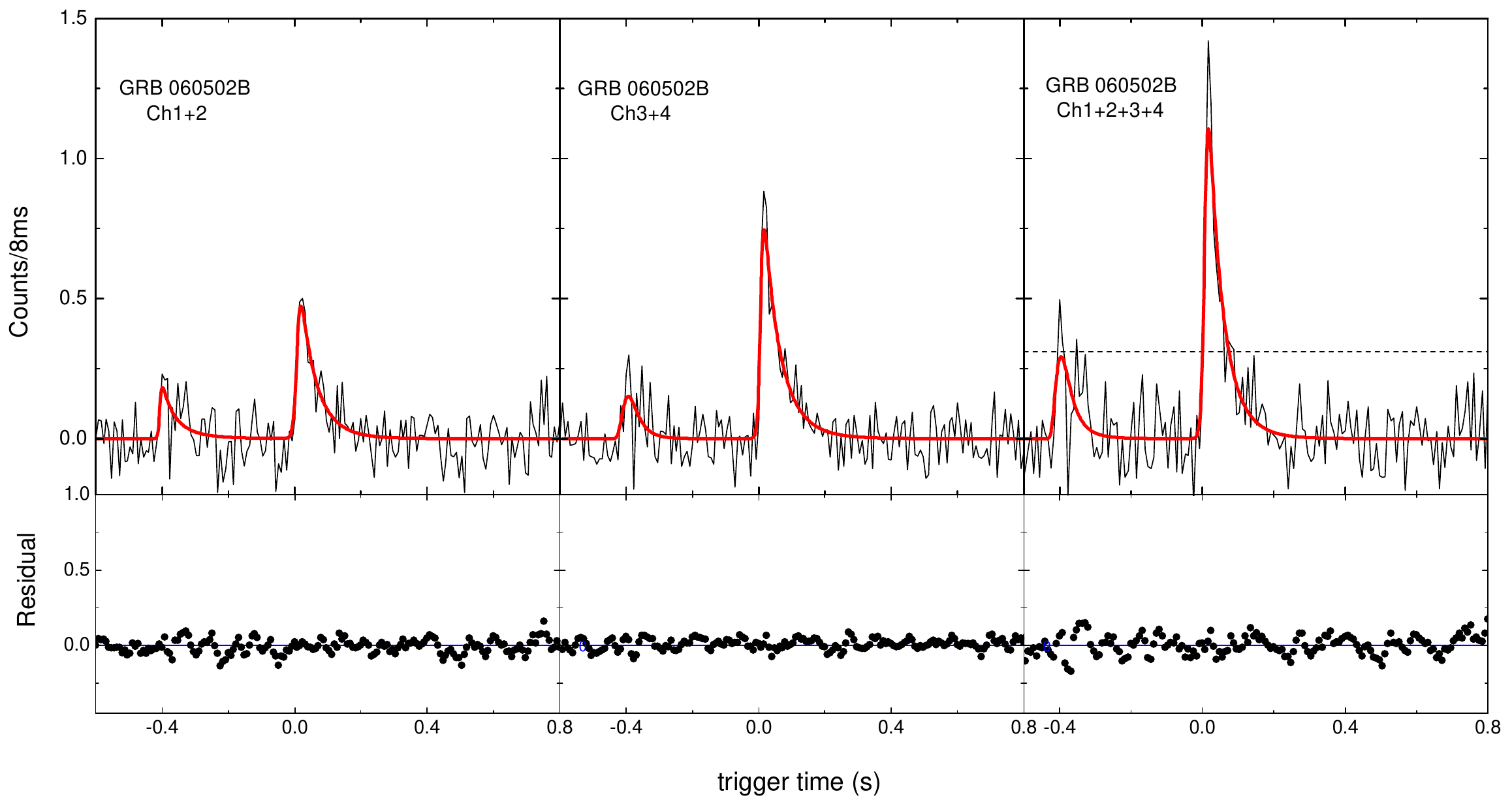}{0.5\textwidth}{(a)}
\fig{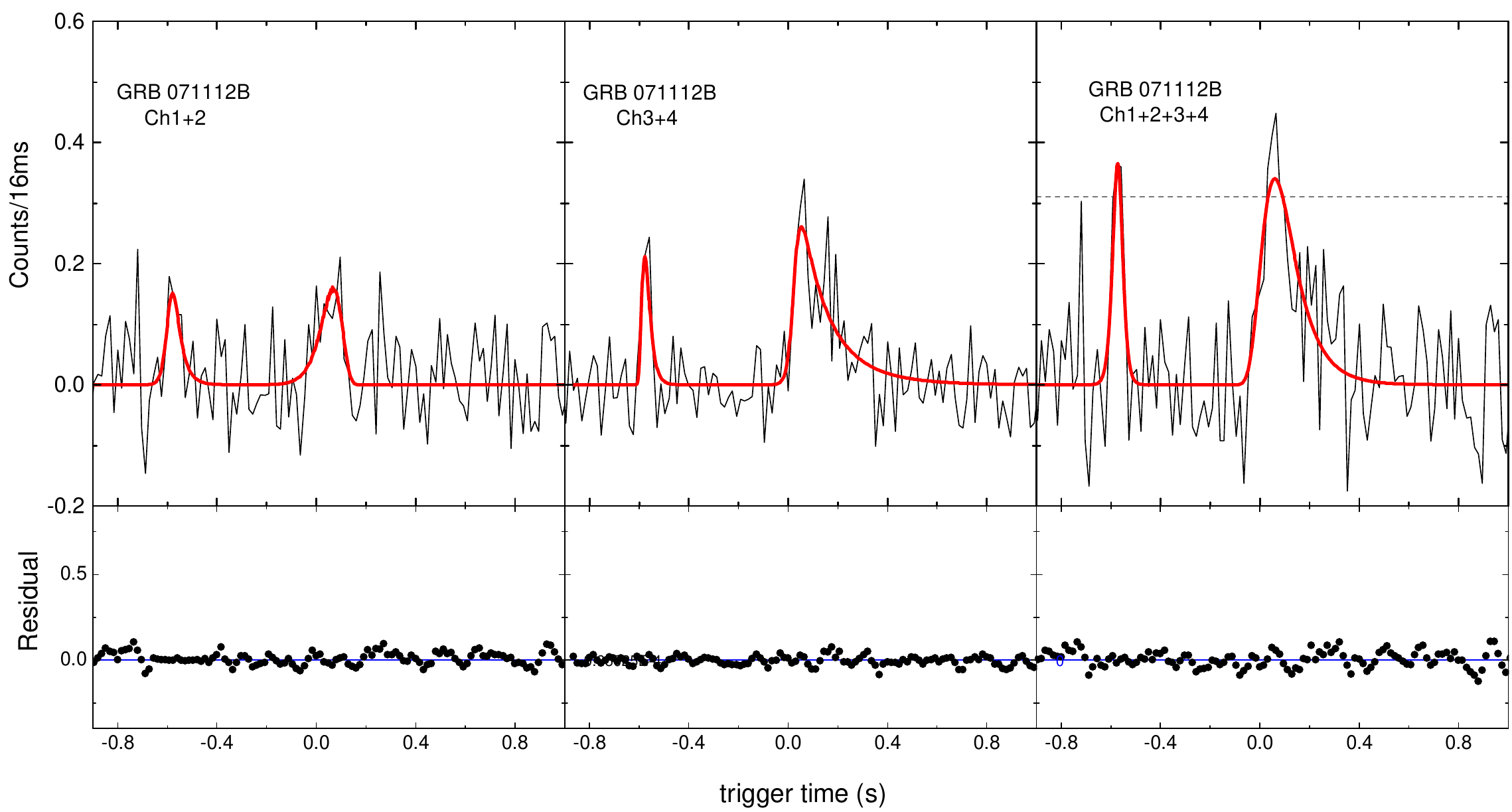}{0.5\textwidth}{(b)}
          }
\gridline{
\fig{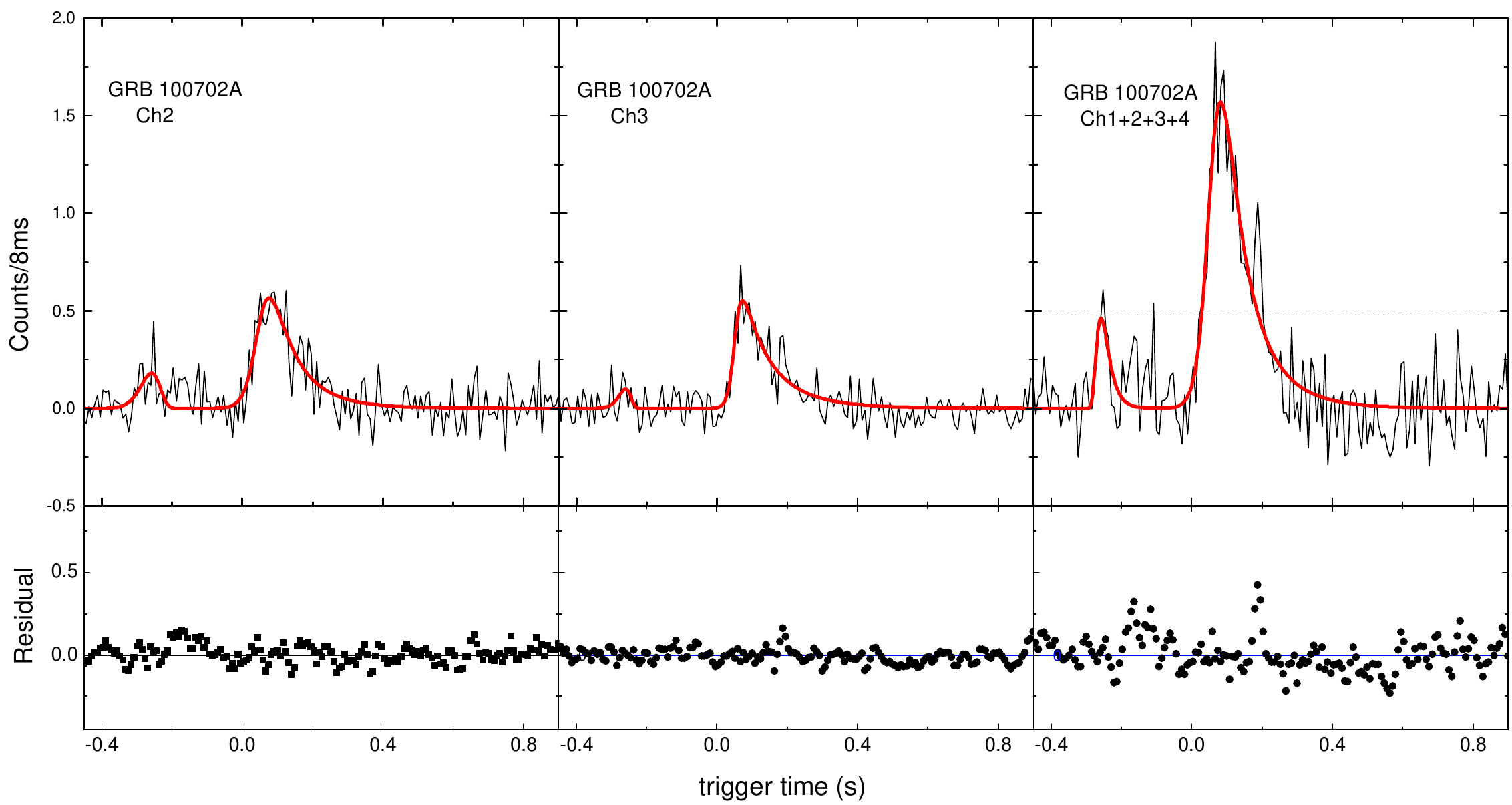}{0.495\textwidth}{(c)}
\fig{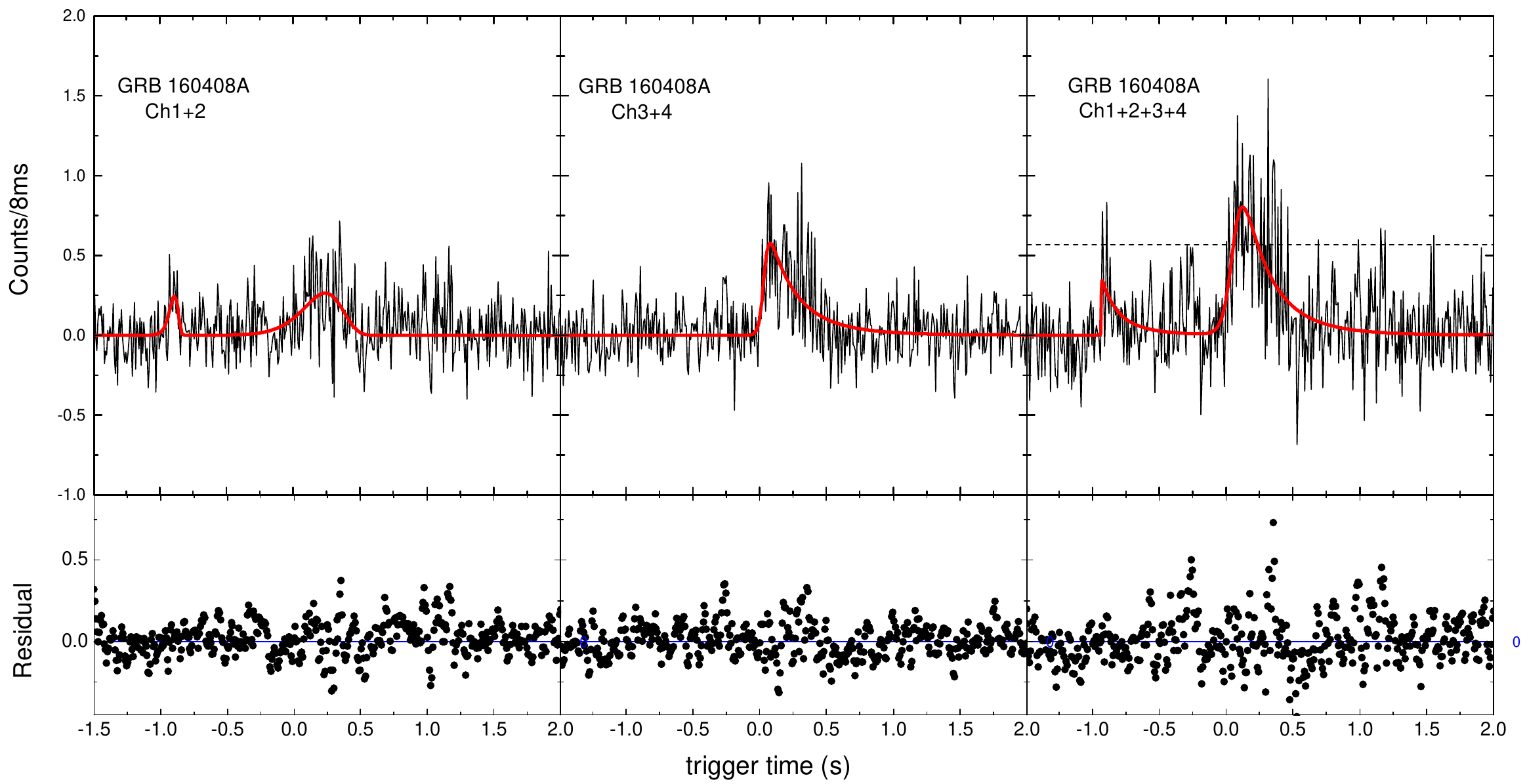}{0.5\textwidth}{(d)}
          }
\gridline{
\fig{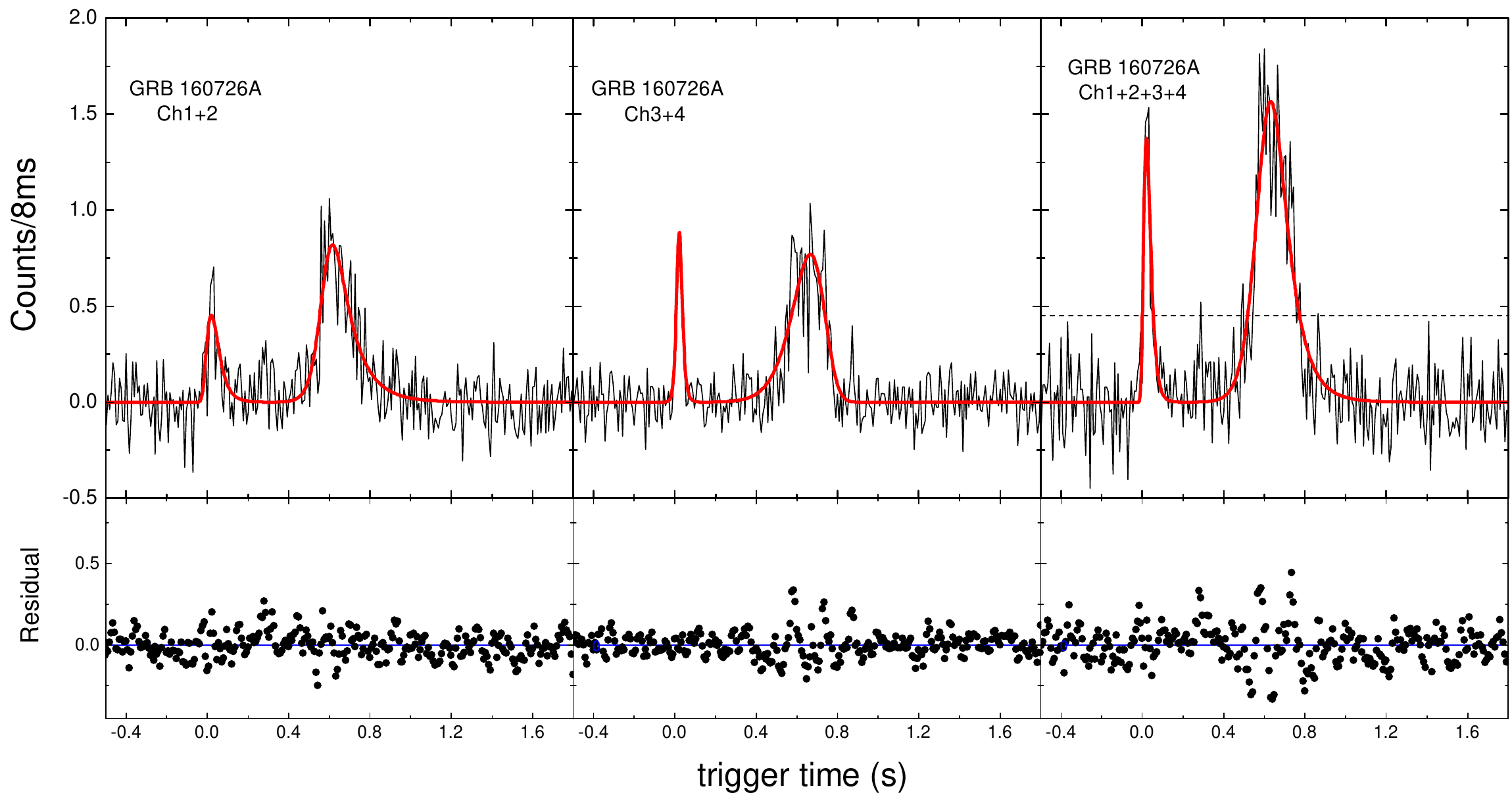}{0.5\textwidth}{(e)}
\fig{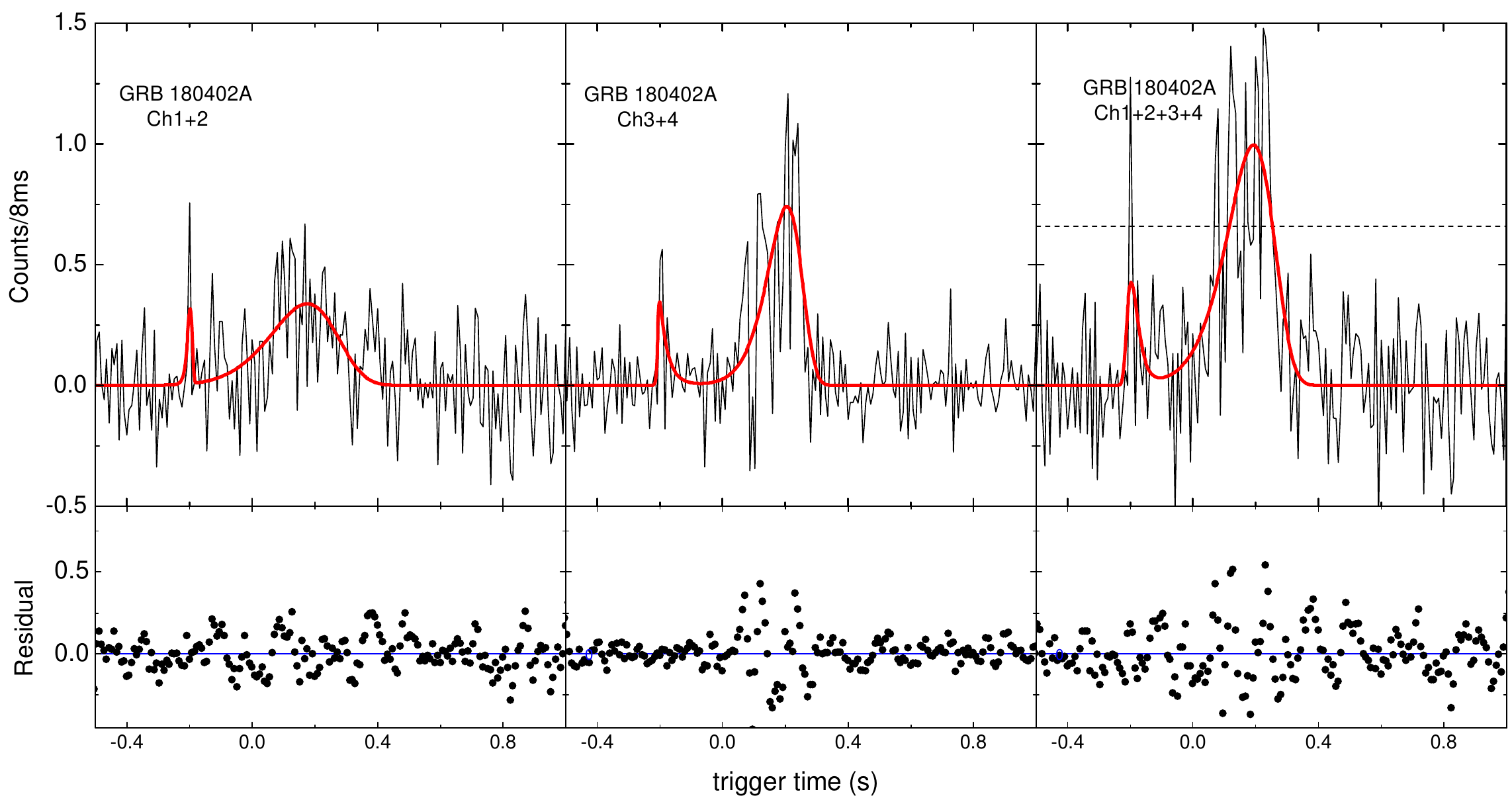}{0.5\textwidth}{(f)}
          }
\caption{Examples of pulses in the Pre+SPs. For comparison, the pulses of the two individual channels and the combined energy channel for each sGRBs are analyzed. The horizontal dotted black lines mark a 3$\sigma$ confidence level. \label{fig:precursor1}}
\end{figure*}

\begin{figure*}
\centering
\gridline{
          \fig{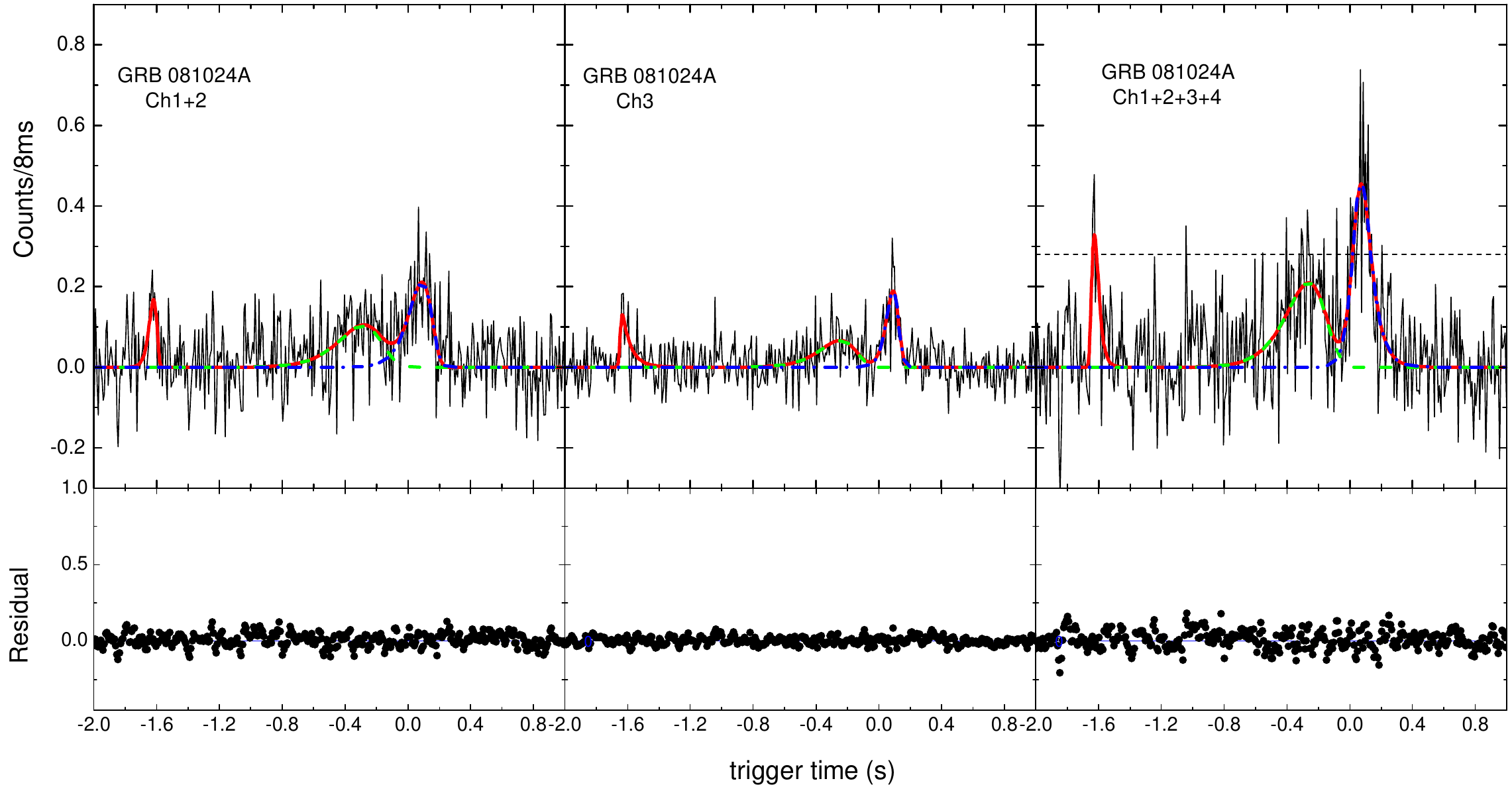}{0.5\textwidth}{(a)}
          \fig{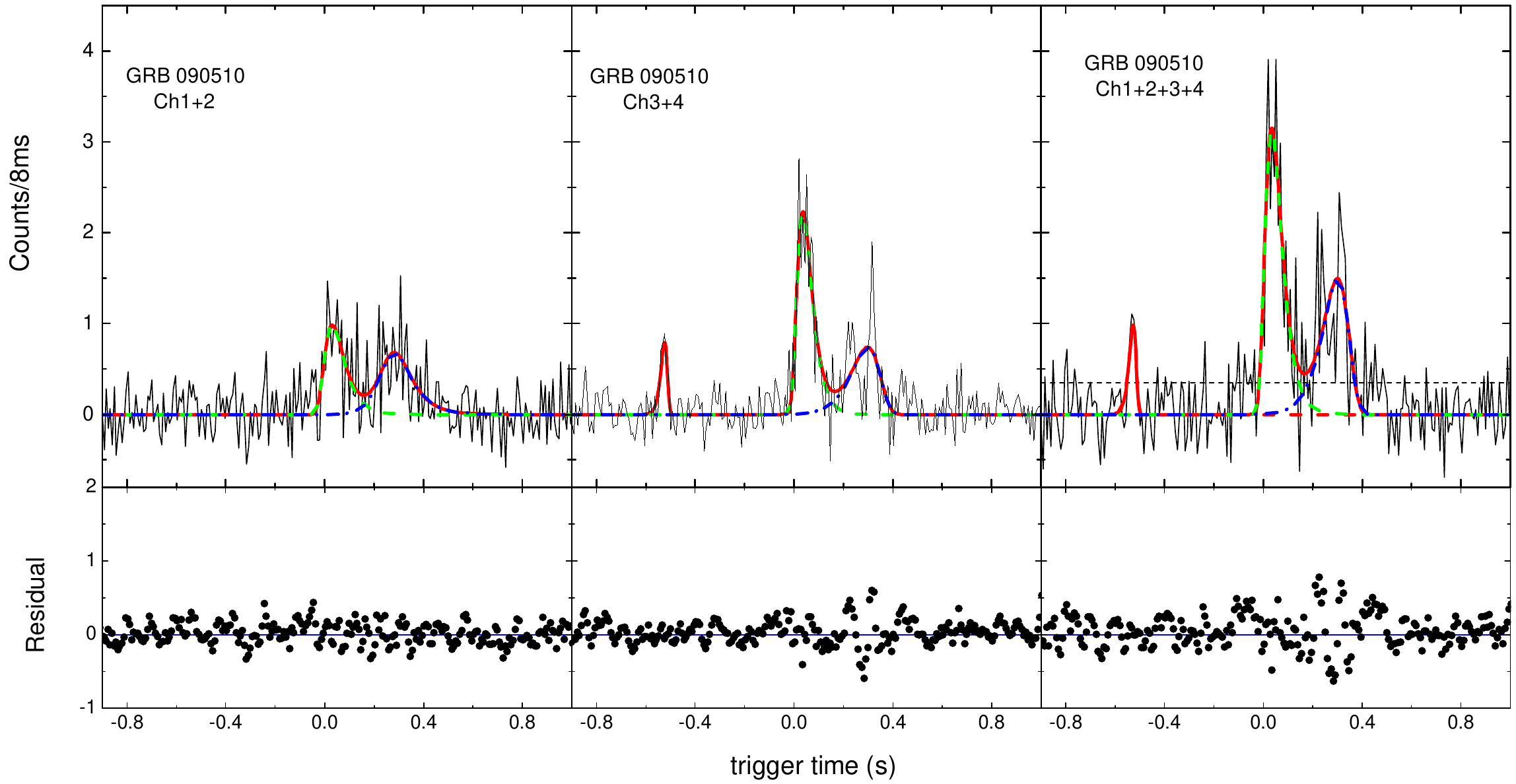}{0.5\textwidth}{(b)}
          }
 \gridline{
          \fig{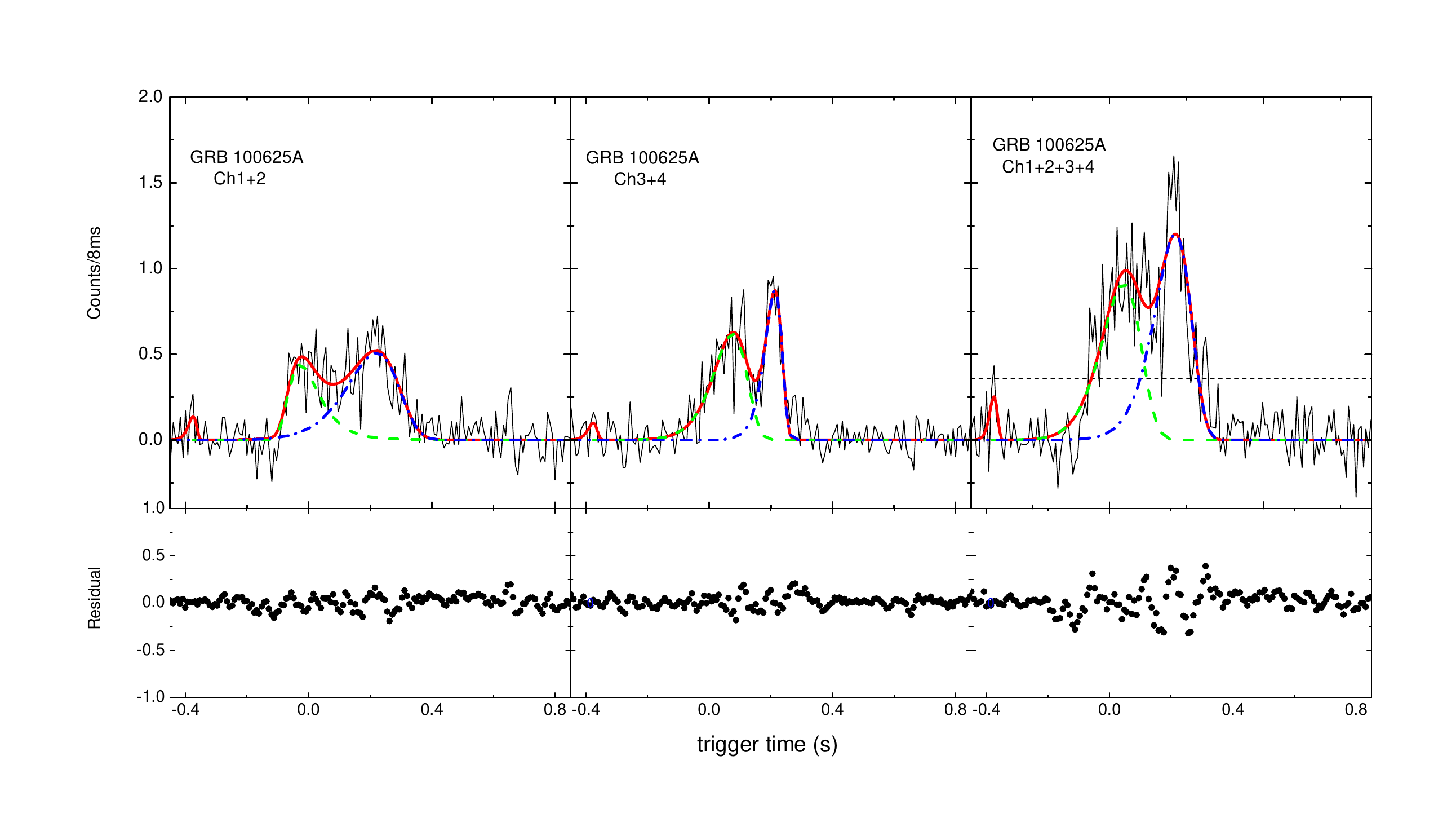}{0.5\textwidth}{(c)}
}
\caption{Examples of pulses in the Pre+DPs. For comparison, the pulses of the two individual channels and the combined energy channel for each sGRBs are analyzed. The horizontal dotted black lines mark a 3$\sigma$ confidence level.\label{fig:precursor2}}
\end{figure*}

 Figure \ref{fig:precursor1} and \ref{fig:precursor2} show six single-peaked sGRBs (SPs) and three double-peaked sGRBs (DPs) with precursors. Totally, 25 precursor pulses across different energy bands have been successfully identified.
 It is worthy to point out that two precursors are found in GRB 090510 \citep{2010ApJ...723...1711}. In this work, however, only one precursor occurring at t $\sim$ 0.5 s prior to the main peaks is confidently discriminated as a real precursor detected by the Fermi/GBM in the higher energy bands \citep{2009Nature...462...331,2010ApJ...723...1711}, will be carefully reanalyzed. We stress that it is a very challenging result since most precursors are very fainter and generally detected before the main outbursts in the lower energy channels.

\begin{figure*}
\centering
\gridline{
\fig{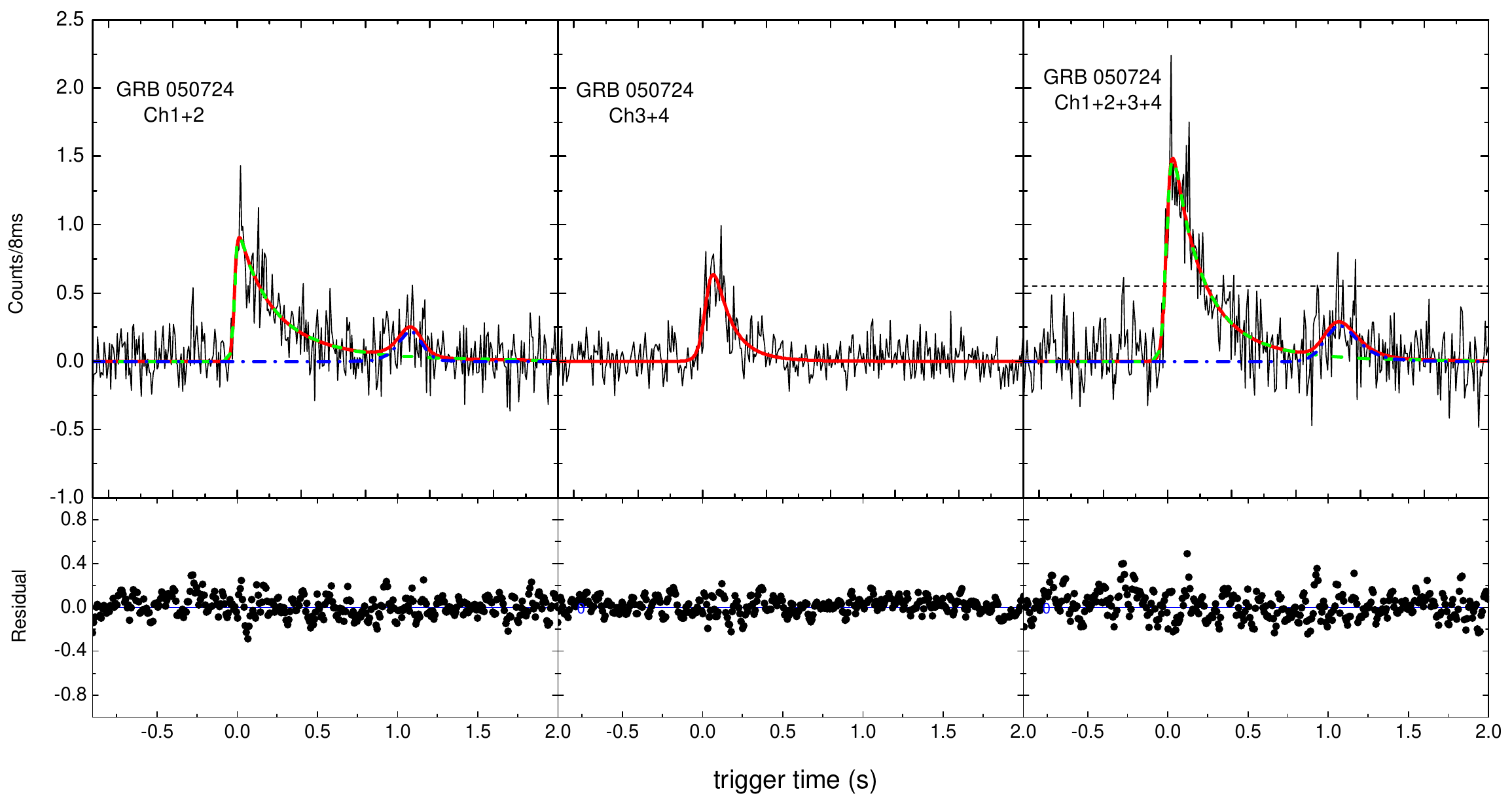}{0.5\linewidth}{(a)}
\fig{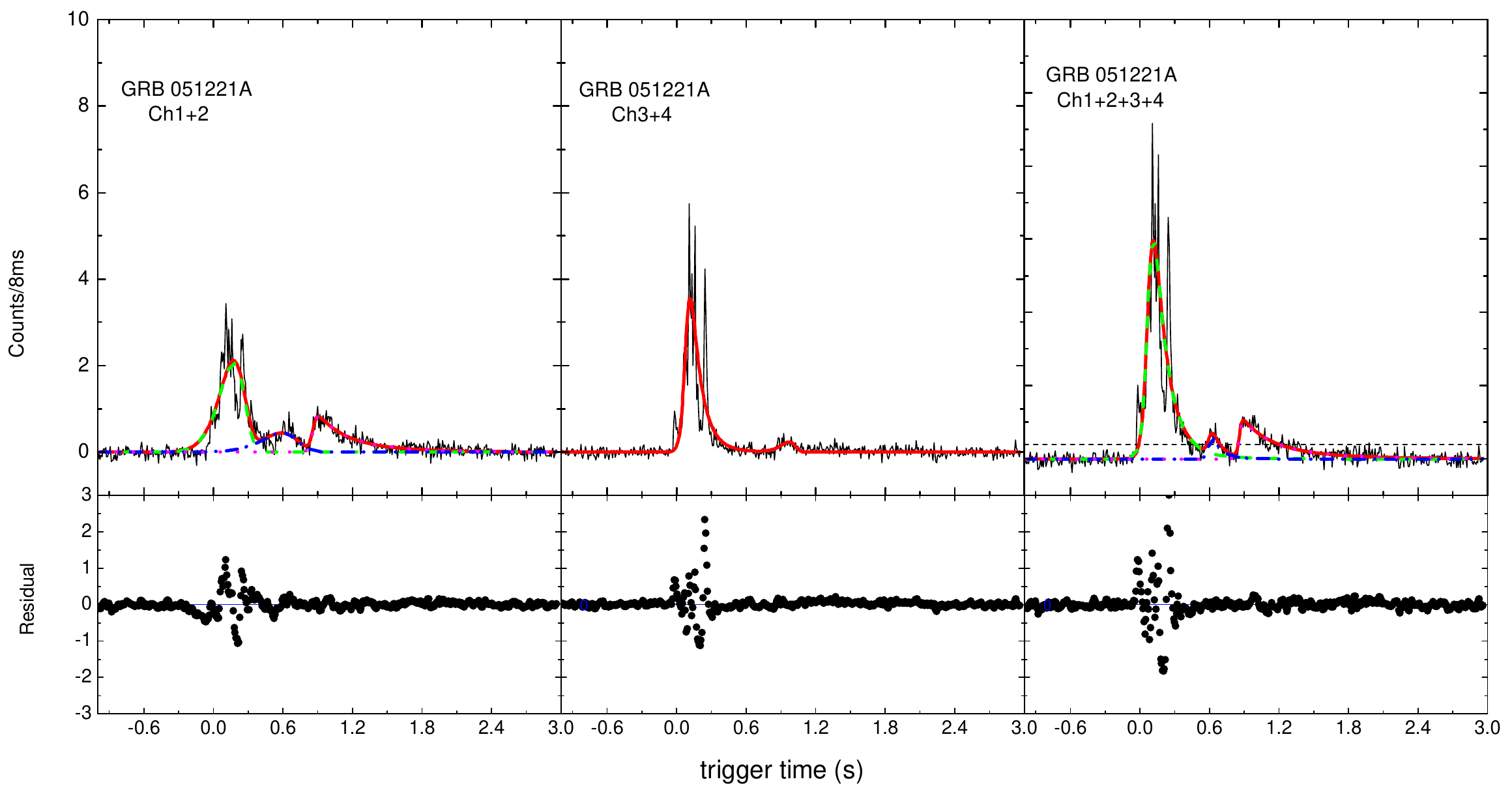}{0.5\textwidth}{(b)}
          }
 \gridline{
 \fig{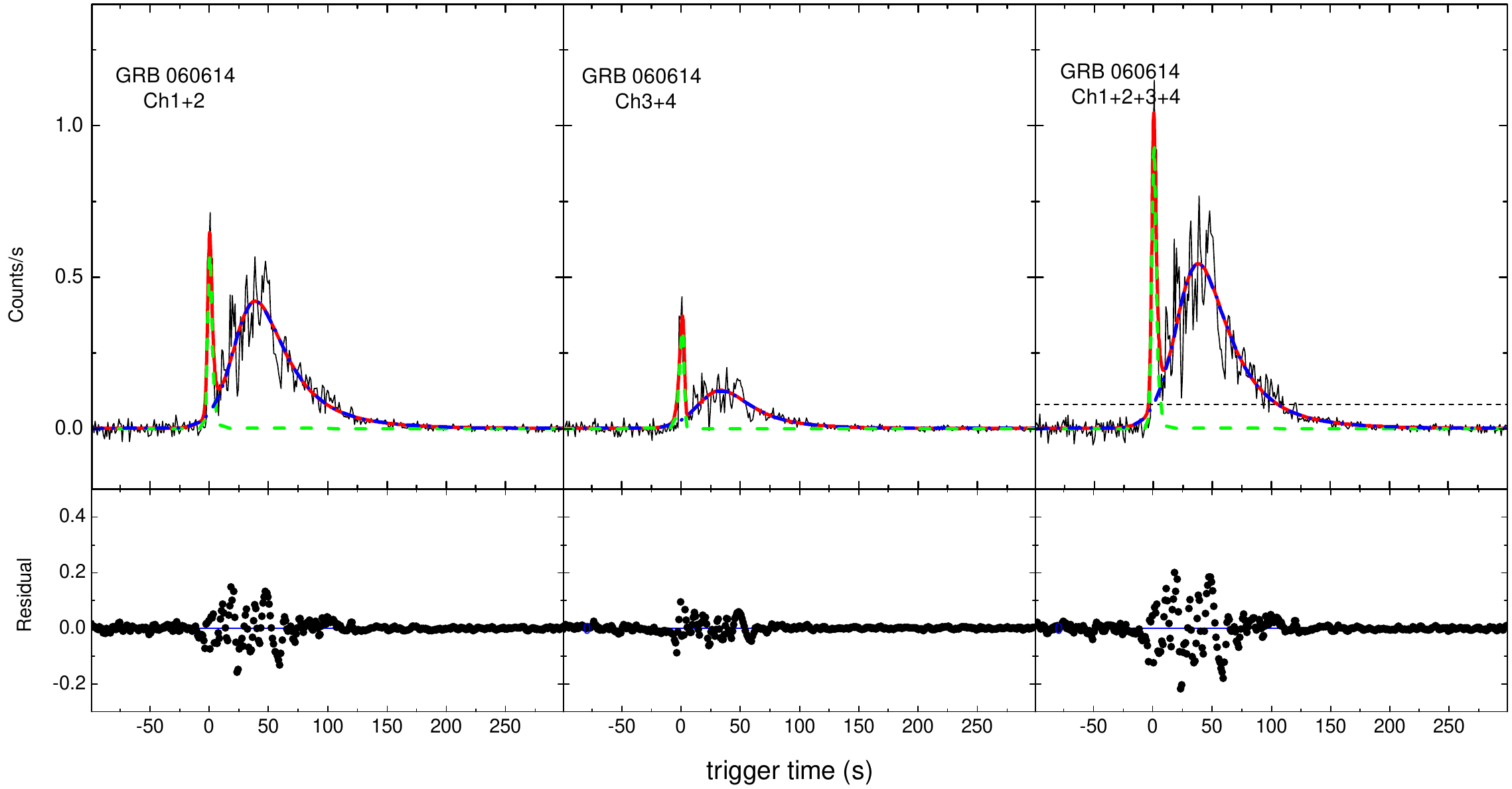}{0.5\textwidth}{(c)}
 \fig{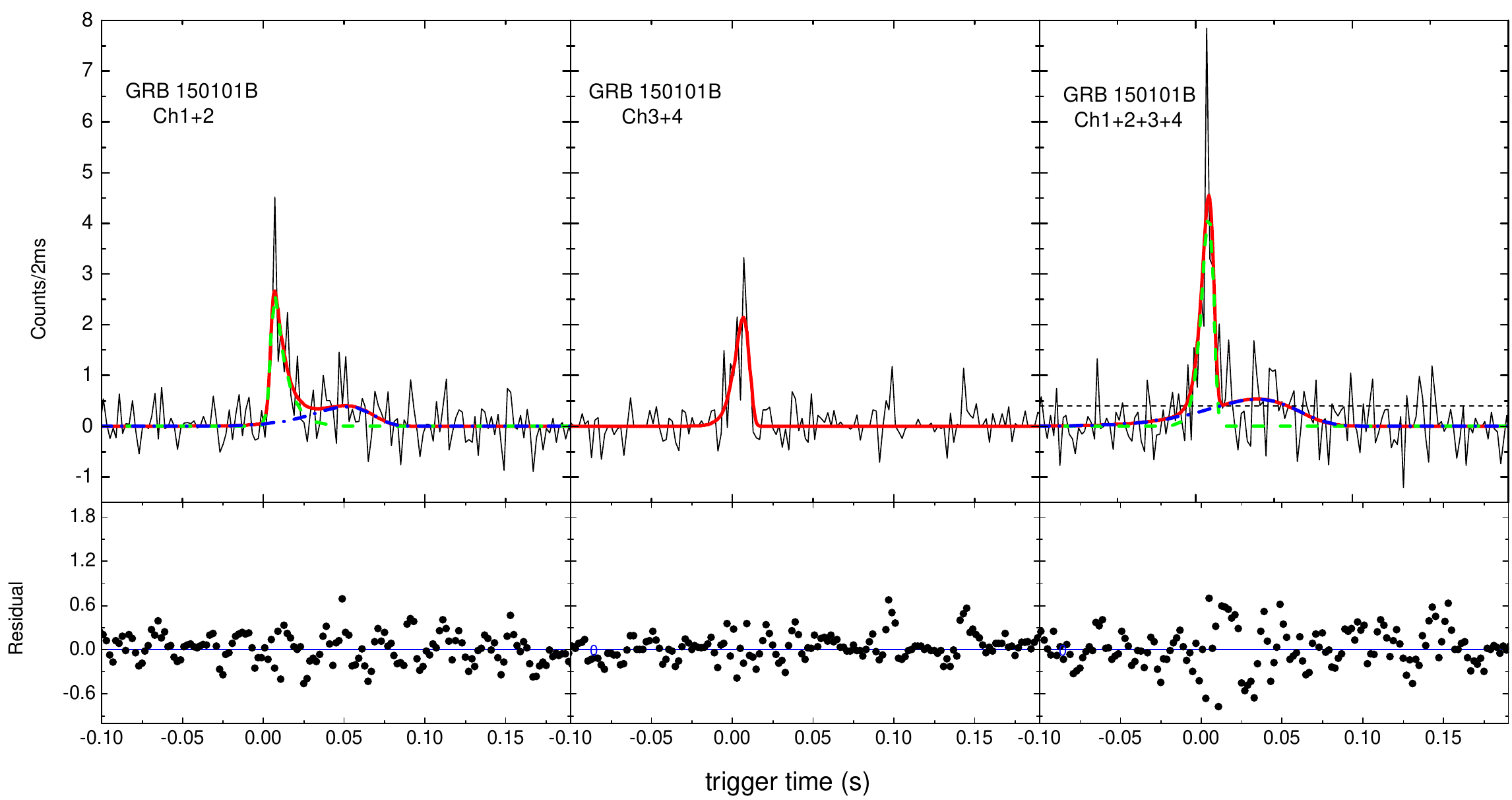}{0.5\textwidth}{(d)}
          }
           \gridline{
 \fig{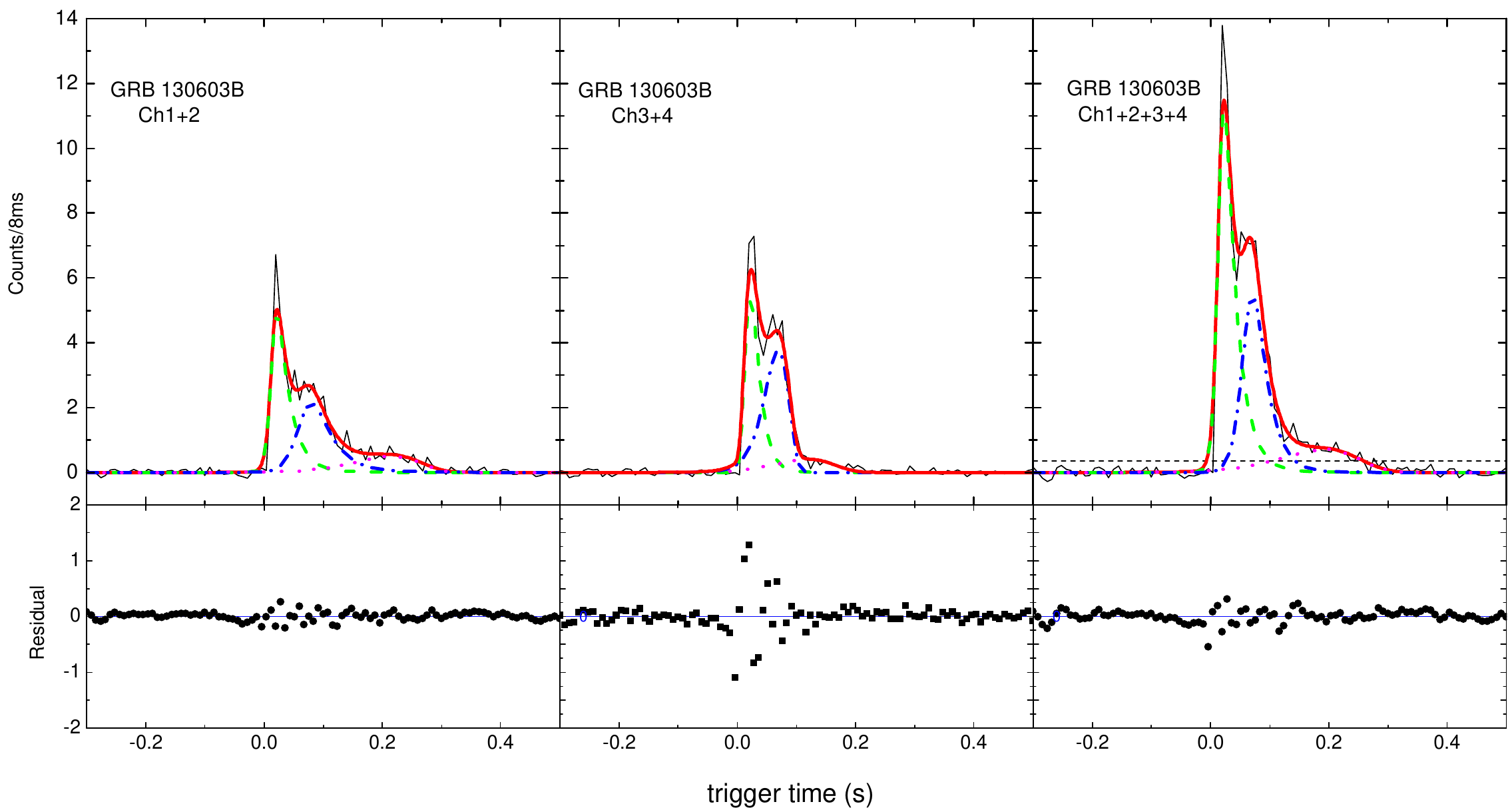}{0.5\textwidth}{(e)}
 \fig{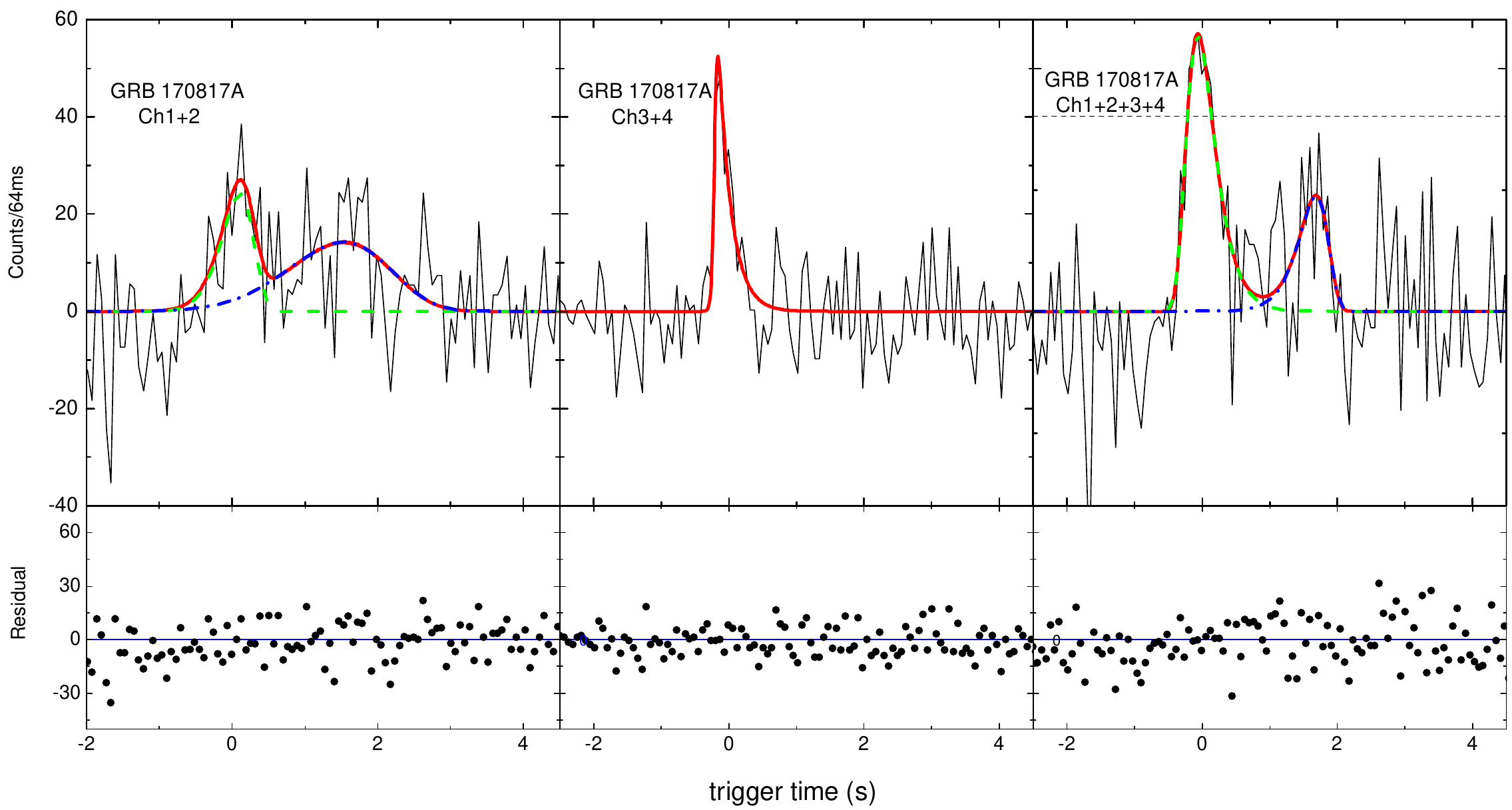}{0.5\textwidth}{(f)}
          }
\caption{Examples of pulses in the SPs+EE and DPs+EE (GRB 130603B). For comparison, the pulses of the two individual channels and the combined energy channel for each sGRBs are analyzed. The horizontal dotted black lines mark a 3$\sigma$ confidence level. \label{fig:EE}}
\end{figure*}

\cite{2013MNRAS...428...1623} searched 11 from 256 BATSE GRBs with EE to unveil the BATSE population of a new hybrid class of GRBs similar to
GRB 060614. Using Bayesian Block (BB) methods, \cite{2010APJ...717...411,2011APJ...735...23} found $\sim$25\% of 51 Swift/BAT sGRBs with EE component. \cite{2016ApJ...829...7} reported the fraction
of sGRBs with EE to be 1.19\% in the third Swift/BAT catalog. Here, we extend our search to include the GRBs with EE tails
reported in literatures \cite[e.g.,][]{2006APJ...643...266,2017ApJ...846...142,Yu2020,Zhangxiaolu2020}. According to \cite{2019ApJ...876...89}, we adopt the following three criteria:

(1) The peak flux of the EE is well lower than that of the main peak.

(2) The signal-to-noise ratios of the main peak and the EE tails should be at least S/N$>3$ above background.

(3) The EE parts must be brightest in lower energy channel than 50 keV and are  too weak to be distinguished effectively in higher energy channels, say 50-350 keV.

We have to stress that we would like to include as many sGRBs with EE as possible in our EE sample. Although the EE tails reflect long-lasting activities with a typical timescale of
$\lesssim$ 10$^3$ s \cite[e.g.,][]{2001AA...379...L39, 2002APJ...567...1028, 2006APJ...643...266, 2008MNRAS...385..1455,2014ApJ...789...145, 2017ApJ...846...142},
most of them cannot be well fitted by a pulse function because of their low signal-to-noise ratios as illustrated as in \cite{2018ApJ...855...101} and our paper I. So we just choose GRB 050724 \citep{2005Nature...438...988,2010ApJ...723...1711,2016ApJ...829...7,2017ApJ...846...142,2018MNRAS...481...4332}, GRB 051221A
\citep{2006MNRAS...372...L19, 2011ApJ...734...35, 2017ApJ...846...142}, GRB 060614
\citep{2006Nature...444...1044,2006APJ...643...266}, GRB 130603B \citep{2015ApJ...802..119K}, and GRB 150101B \citep{2018NC...9...4089,2018APJL...863...L34, 2019ApJ...876...89, 2019MNRAS} as the subsample with EE. For comparison, the EE sample also includes the first gravitational-wave associated GRB 170817A detected by the Fermi Gamma-ray Burst Monitor (GBM), whose main
peak is dominant over an energy
range of 50-300 keV and its soft tail is stronger below 50 keV \citep{2017ApJL...848...L14,ZhangBB2018,2019ApJ...876...89,2020MNRAS...492...3622}. Finally, twelve typical EE tail pulses across different energy bands are fitted (see Figure \ref{fig:EE}). Totally, Table \ref{tableEE} lists 67 typical sGRBs with three components, including 10 precursor events ($\sim$15\%) and 17 EE events ($\sim$25\%). It is necessary to point out that there are no sGRBs with both the precursor and the EE components in our sample.
\section{RESULTS}
In this section, the pulse features of the main peaks not only in one-component and two-component sGRBs but also in the SPs and the DPs are compared. On the other hand, the correlations of the main peaks with either the precursors or the EEs in the two-component sGRBs are inspected. As done in paper I, the M-loose double-peaked sGRBs (Ml-DPs) and the M-tight double-peaked sGRBs (Mt-DPs) will be investigated separately. For convenience, we name the sGRBs with precursor or EE as Pre+sGRBs or sGRBs+EE hereafter.
\subsection{Main Peaks}
\subsubsection{One-component versus Two-component sGRBs}
In order to investigate whether the temporal properties of the main peaks in one-component sGRBs differ from
those of two-component sGRBs, comparisons are displayed
in Figures \ref{fig:trtd}$-$\ref{fig:fmFWHM}.
In Figure \ref{fig:trtd}$-$\ref{fig:fmFWHM} (a), we find that the main peaks of the one-component SPs tend to be similar to those of the two-component SPs. In Figures \ref{fig:trtd}$-$\ref{fig:fmFWHM} (b), the main peaks of both the first pulses (1st) and the second pulses (2nd) of the two-component Mt-DPs tend to be similar to those of the one-component Mt-DPs. There is no two-component Ml-DPs in our sample. Note that, we can see from Figures \ref{fig:trtd}$-$\ref{fig:fmFWHM} that there is no significant evolutions on energy in both the SPs and the Mt-DPs. Thus we combine adjacent two channels into one channel for the two-component sGRBs and compare them with ones of the one-component sGRBs in two individual channels. These comparisons suggest that the main peaks in either one-component or two-component sGRBs tend to have no significant difference and are
likely to share the similar physical mechanism.

\subsubsection{SPs versus DPs}
\begin{figure*}
\centering
\gridline{
\fig{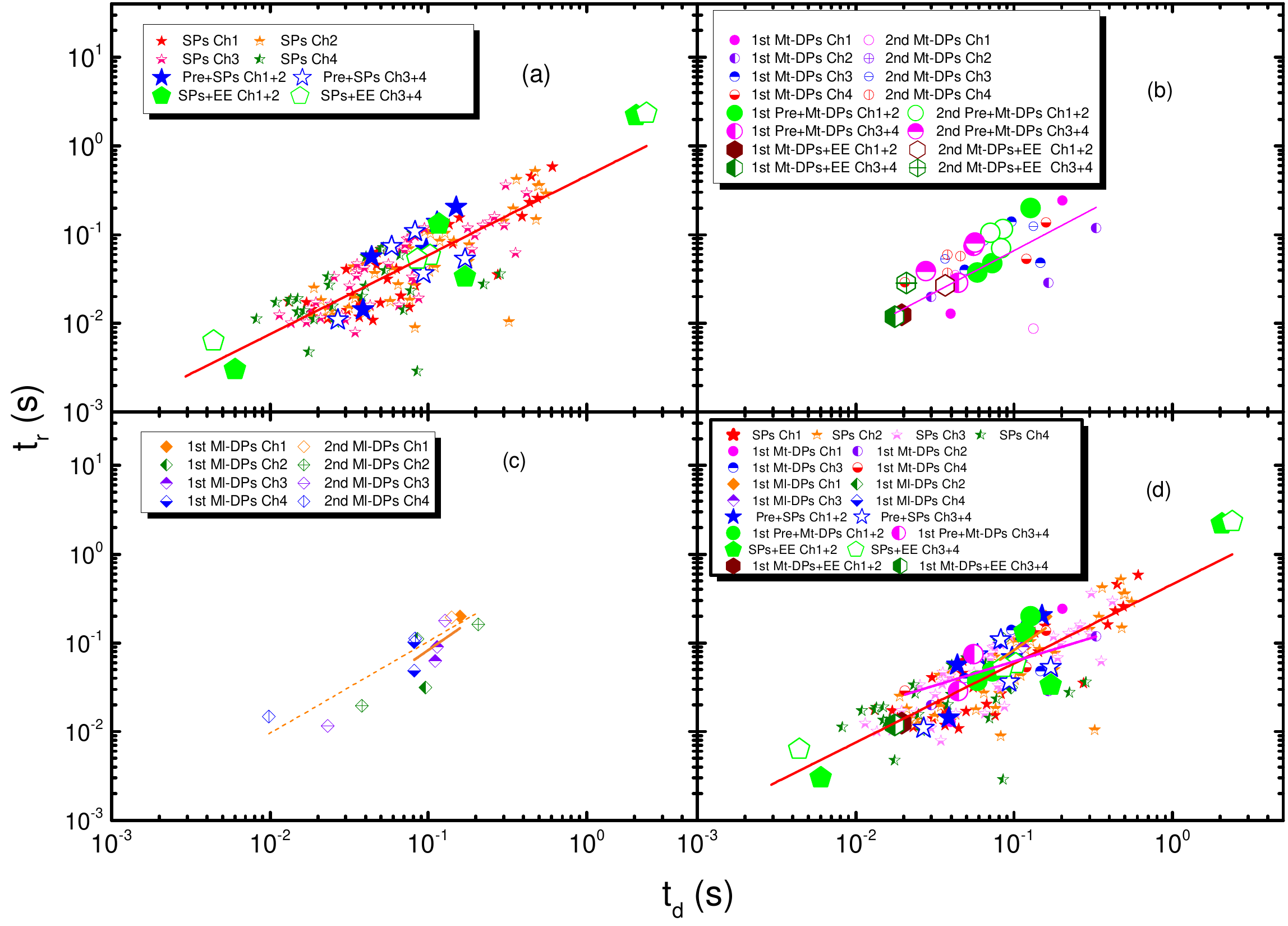}{1\textwidth}{}
          }
\caption{The t$_r$ vs. $t_d$ of the sGRB pulses in the main peaks. In each panel, (a) SPs, (b) Mt-DPs, (c) Ml-DPs and
(d) Comparisons of the SPs and the 1st pulses in two kinds of the DPs.
The lines are the best fits, solid lines for the 1st main peaks and dotted lines for the 2nd ones, respectively.
\label{fig:trtd}}
\end{figure*}

\begin{figure*}
\centering
\gridline{
\fig{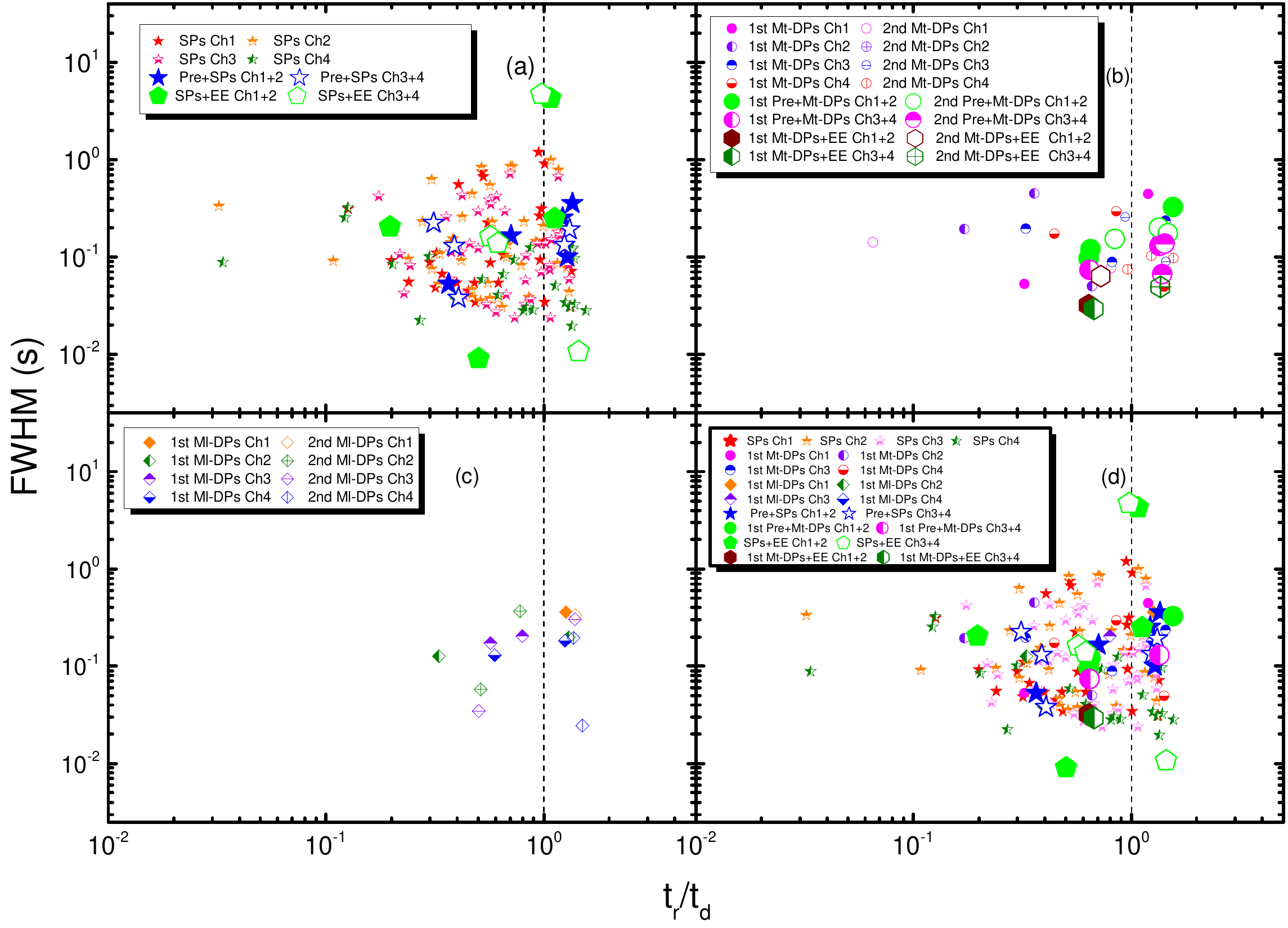}{1\textwidth}{}
          }
\caption{ The FWHM vs. $t_r/t_d$ of the sGRB pulses in the main peaks. The symbols are as same as Figure \ref{fig:trtd}.
\label{fig:FWHM-asy}}
\end{figure*}

\begin{figure*}
\centering
\gridline{
\fig{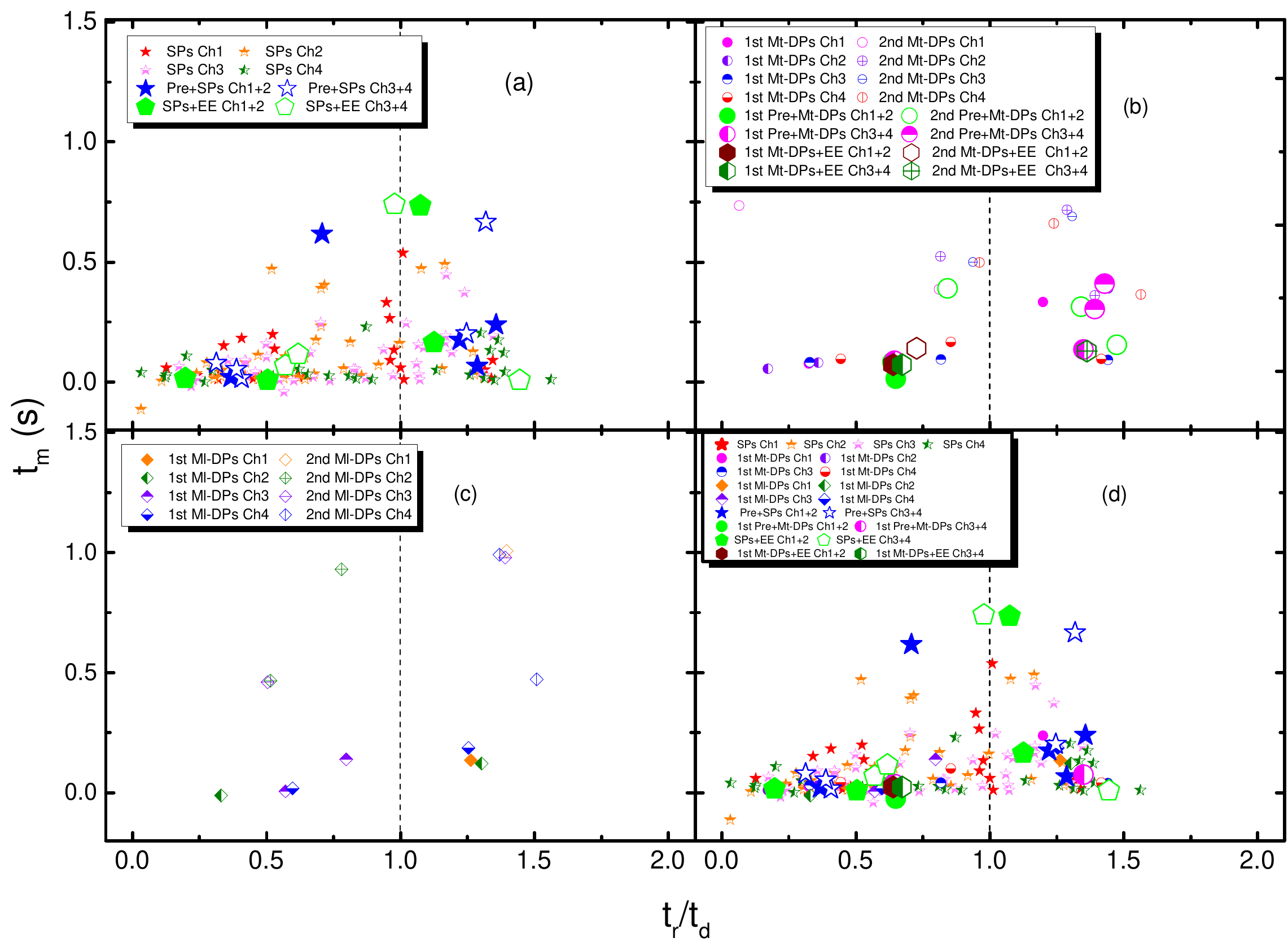}{1\textwidth}{}
}
\caption{ The $t_m$ vs. $t_r/t_d$ of the sGRB pulses in the main peaks. The symbols are as same as Figure \ref{fig:trtd}.
\label{fig:tm-asy}}
\end{figure*}

\begin{figure*}
\centering
\gridline{
\fig{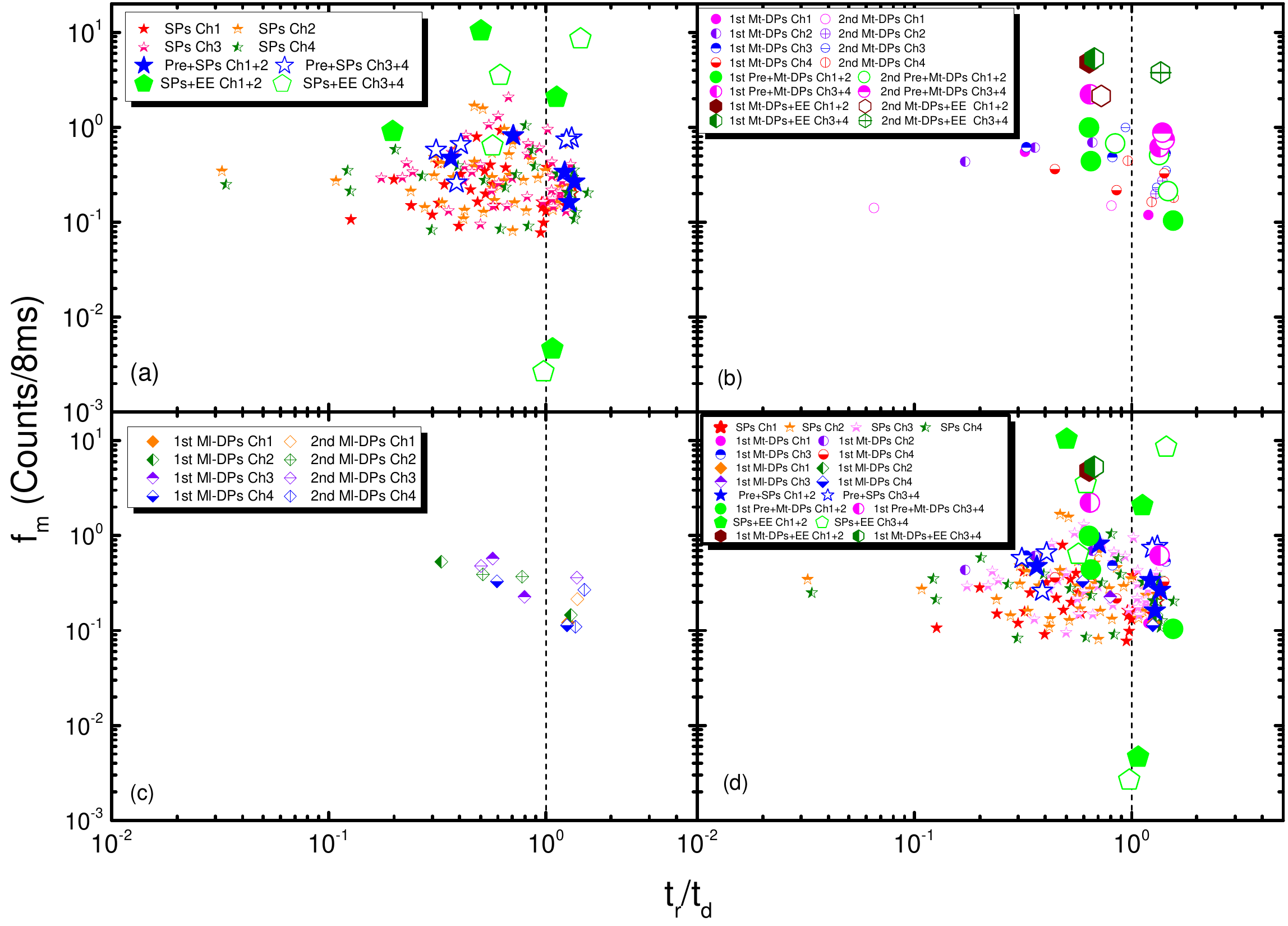}{1\textwidth}{}
          }
\caption{ The $f_m$ vs. $t_r/t_d$ of the sGRB pulses in the main peaks. The symbols are as same as Figure \ref{fig:trtd}.
\label{fig:fm-asy}}
\end{figure*}

\begin{figure*}
\centering

\gridline{
\fig{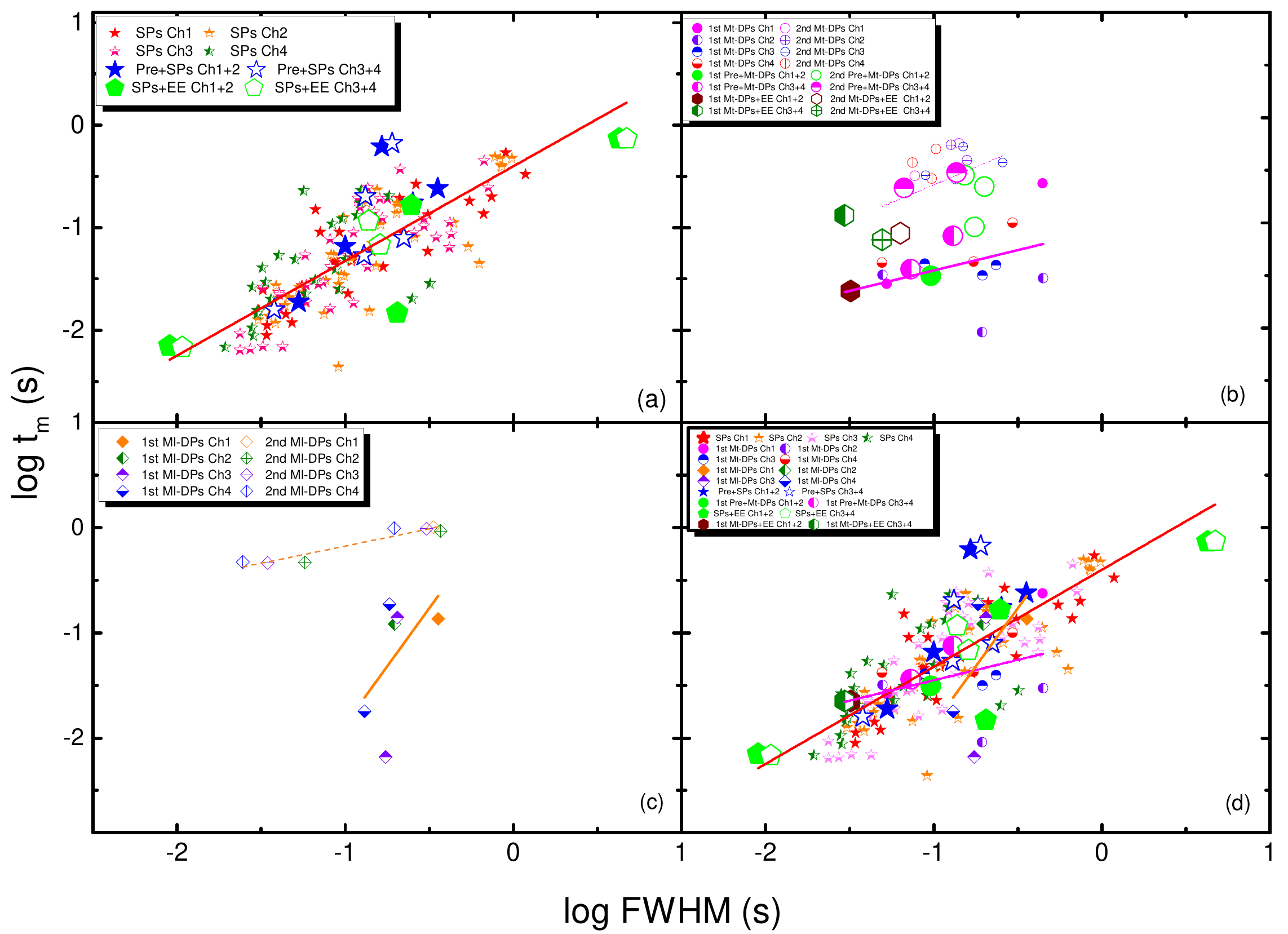}{1\textwidth}{}
          }
\caption{The log$t_m$ vs. logFWHM of the sGRB pulses in the main peaks. The values of $t_m$ in this figure are positive. The symbols are as same as Figure \ref{fig:trtd}.
\label{fig:tmFWHM}}
\end{figure*}

\begin{figure*}
\centering

\gridline{
\fig{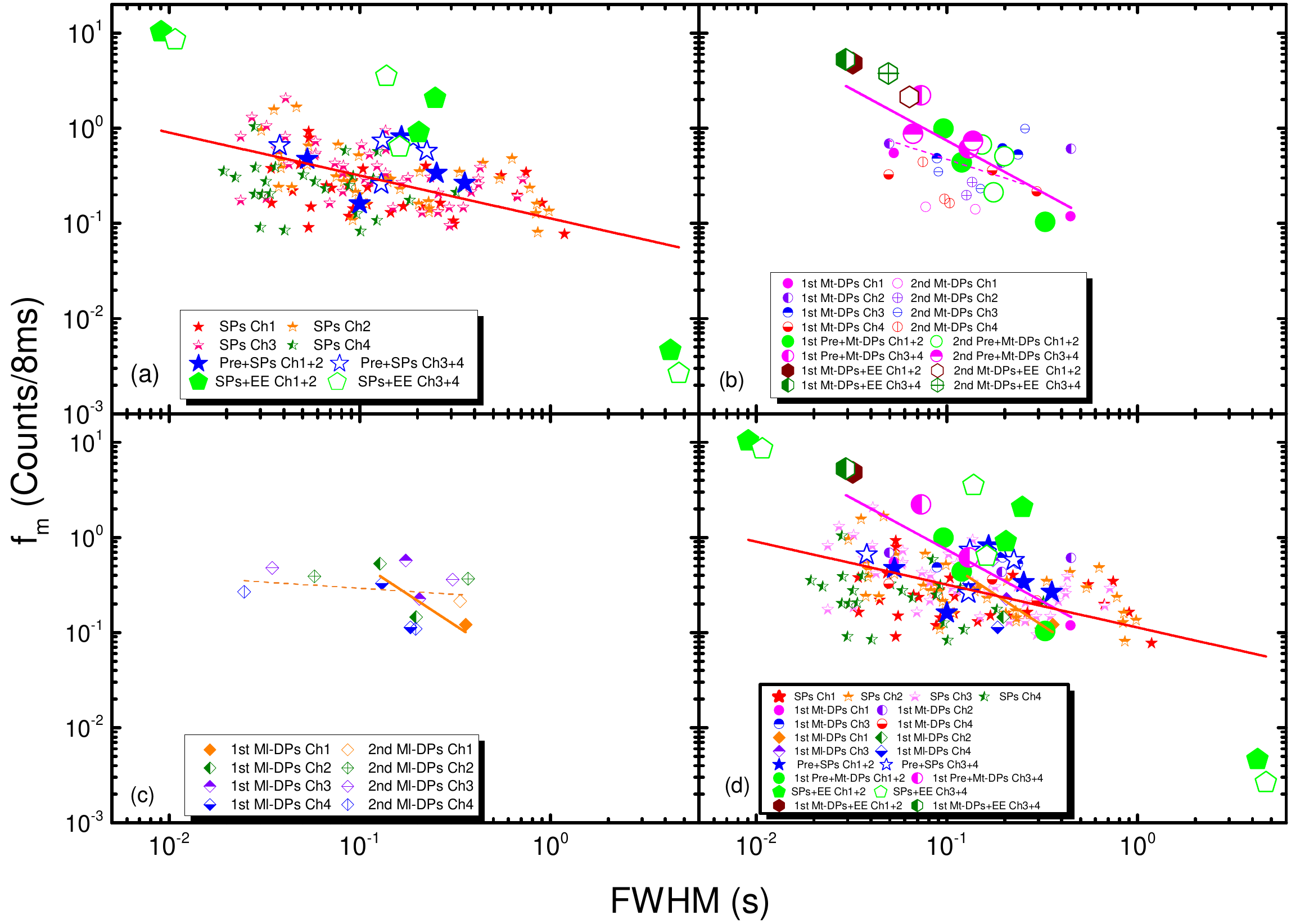}{1\textwidth}{}
          }
\caption{The $f_m$ vs. FWHM of the sGRB pulses in the main peaks. The symbols are as same as Figure \ref{fig:trtd}.
\label{fig:fmFWHM}}
\end{figure*}

In order to reveal the individual or
collective temporal characteristics of the main peaks
between the SPs and the DPs, we compare the analysis results in Figures \ref{fig:trtd} (a)  $-$ \ref{fig:fmFWHM} (a) with the reuslts of Figures \ref{fig:trtd} (b) $-$ \ref{fig:fmFWHM} (b) and \ref{fig:trtd} (c) $-$ \ref{fig:fmFWHM} (c).

It is found that the t$_r$ and the t$_d$ of the main peaks
of all kinds of the sGRBs are self-similar with a
power-law form as t$_r \sim t_d$$^\beta$ in Figure \ref{fig:trtd},
The detailed fitting results
are summarized in Table \ref{tab:tabletrtd}. Interestingly, these results
for all kinds of the sGRBs are consistent with
the BATSE sGRBs, especially for the SPs. For
two kinds of DPs, the power-law correlations are
tighter or looser than those of the BATSE DPs.

The relations of the FWHM, the t$_m$ and the f$_m$ with
the asymmetry appear to be not evident in Figures
\ref{fig:FWHM-asy}$-$\ref{fig:fm-asy}. It needs to note that the
weak dependence of the t$_r$/t$_d$ on the FWHM or the
t$_m$ was found in the BATSE Ml-DPs, instead of the Swift/BAT sGRBs. It may implies that these temporal parameters could evolve with the energy bands of detectors.

Figure \ref{fig:tmFWHM} illustrates that the t$_m$ increases with the FWHM, following a power-law behavior t$_m \sim FWHM^\mu$, which are similar to the results found by \cite{2005APJ...627...324} for the long-lag, wide-pulse BATSE
GRBs and our previous results for BATSE sGRBs. The
detailed fitting results are summarized in Table \ref{tab:tabletmFWHM}. The
f$_m$ and the FWHM are an anti-correlated power-law relation with the
BATSE lGRBs \citep{2002AA...385...377}, similar to our previous conclusion
for the BATSE sGRBs. Figure \ref{fig:fmFWHM} shows that the
f$_m$ is generally anti-correlated with the FWHM with a
power-law form of f$_m \sim FWHM^\nu$. The detailed fitting
results are summarized in Table \ref{tab:tablefmFWHM}. It is worthy to point
out that the power-law index of the SPs is consistent with that of the BATSE SPs.
For two kinds of the DPs, the power-law relations of the first pulses are tighter than those of the second pulses.

In Table \ref{tab:tableasy}, we find that the asymmetries of the main peaks for all kinds
of sGRBs are limited from 0.03 to 1.56 and the
mean asymmetry is 0.79 which is almost equal
to the value of the BATSE sGRBs and lager than the value
of 0.65 for a sample of 100 bright BATSE sGRBs found
by \cite{2001AA...380...L31}. The result is well agreement
with the value of 0.81 obtained by \cite{2007APSS...310...19}.
The really interesting result is that the mean asymmetry of the SPs is still very similar to the values
of the 1st pulses of the two subclasses of the
DPs, which we found in the BATSE sGRBs. A K-S test
to the cumulative distributions of the t$_r$/t$_d$ between
the SPs and the 1st pulses of the Mt-DPs
gives a p-value of 0.50 showing they are not significantly different. Similarly, a K-S test to the distributions
of the t$_r$/t$_d$ between the SPs and the 1st
pulses of the Ml-DPs gives a a p-value of 0.50 showing they
also are not significantly different.

\subsubsection{Dependence of Pulse Width on Energy}

\begin{figure*}[ht]
\centering
\gridline{
\fig{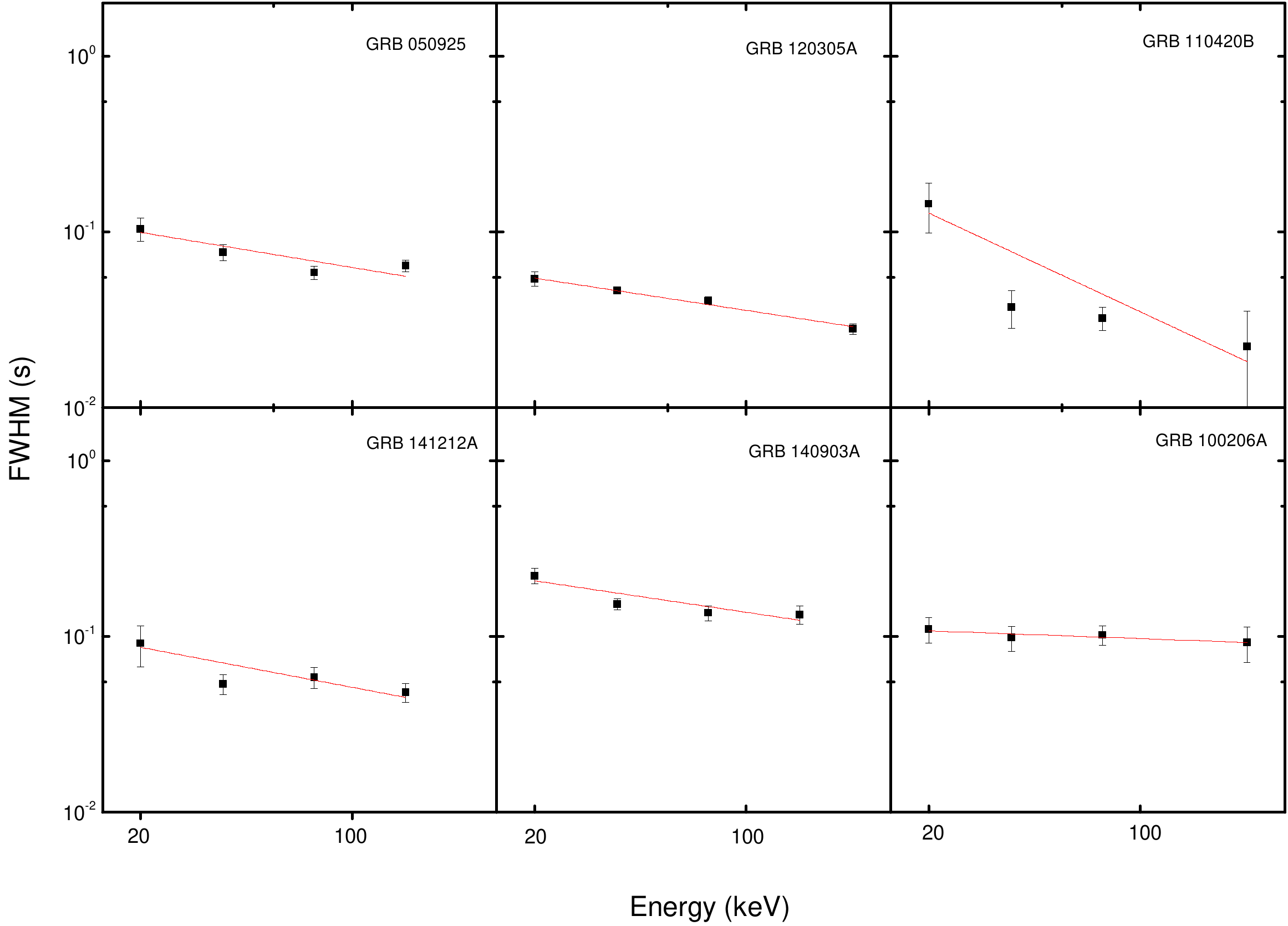}{0.45\linewidth}{(a)}
\fig{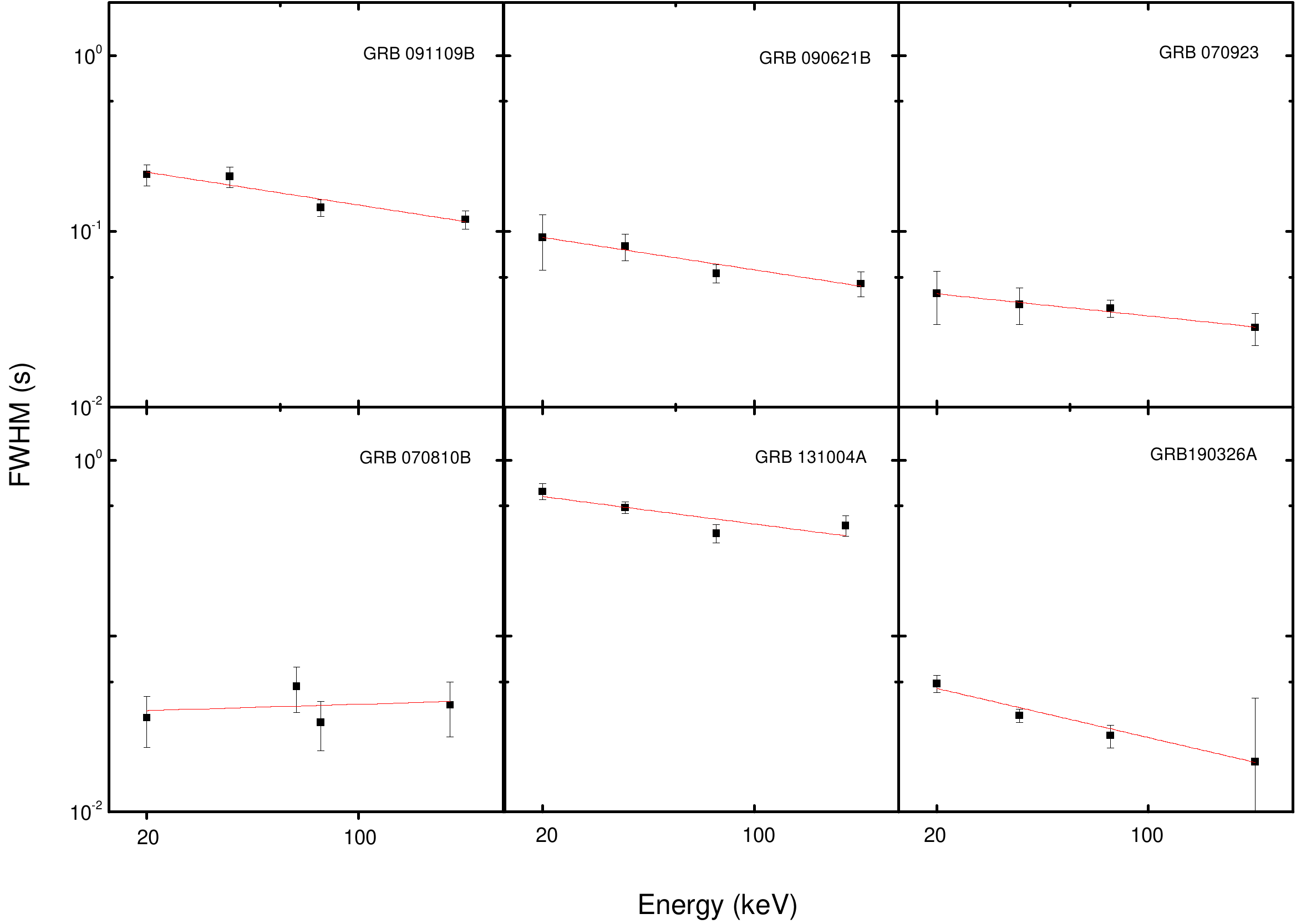}{0.45\linewidth}{(b)}}
\gridline{
\fig{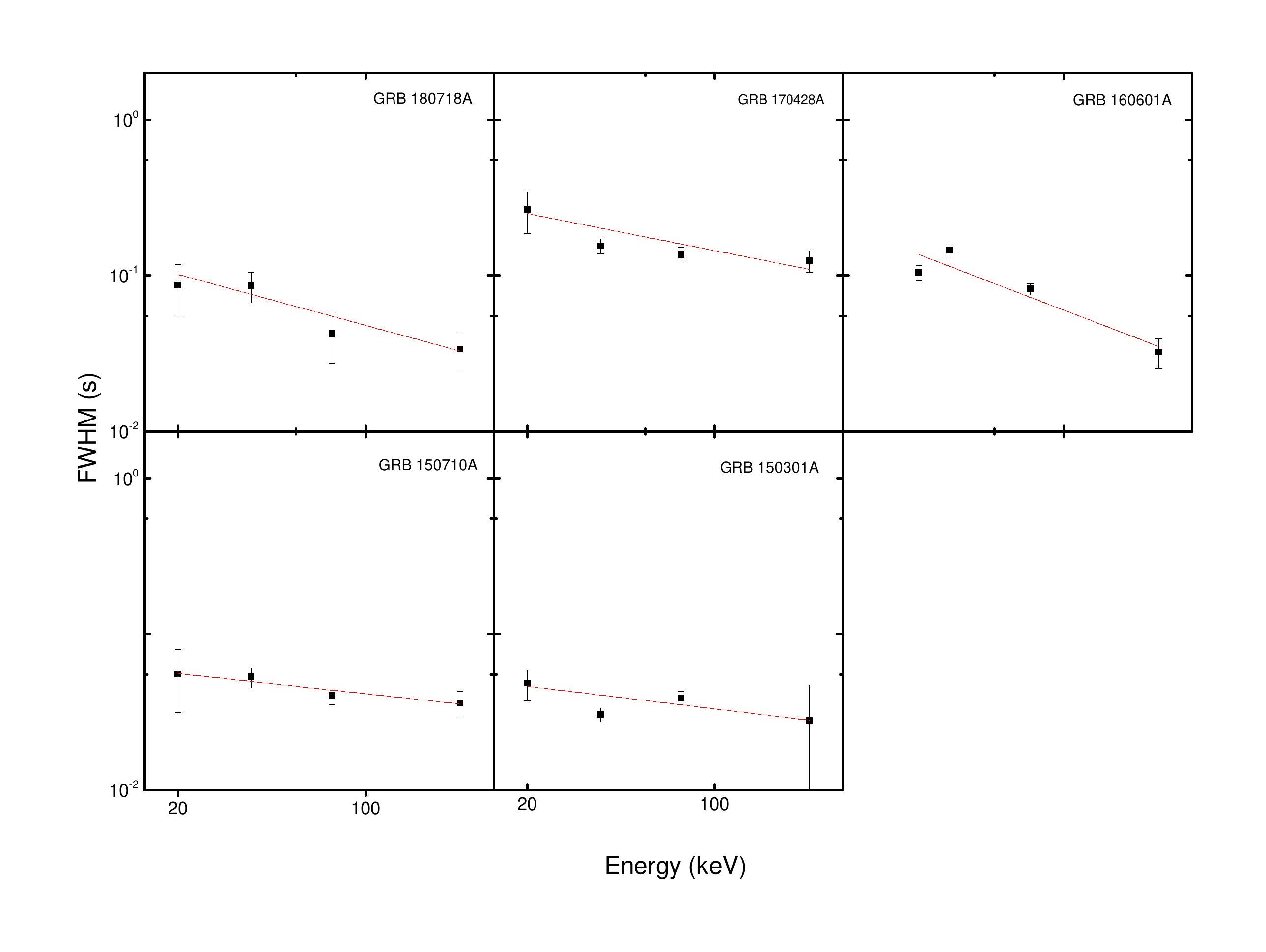}{0.5\linewidth}{(c)}}
\caption{The FWHM vs. average photon energy of
the 17 SPs. The solid line stands for the best power-law fit to the observations. \label{fig:singleFandE}}
\end{figure*}

\begin{figure*}[ht]
\centering
\gridline{
\fig{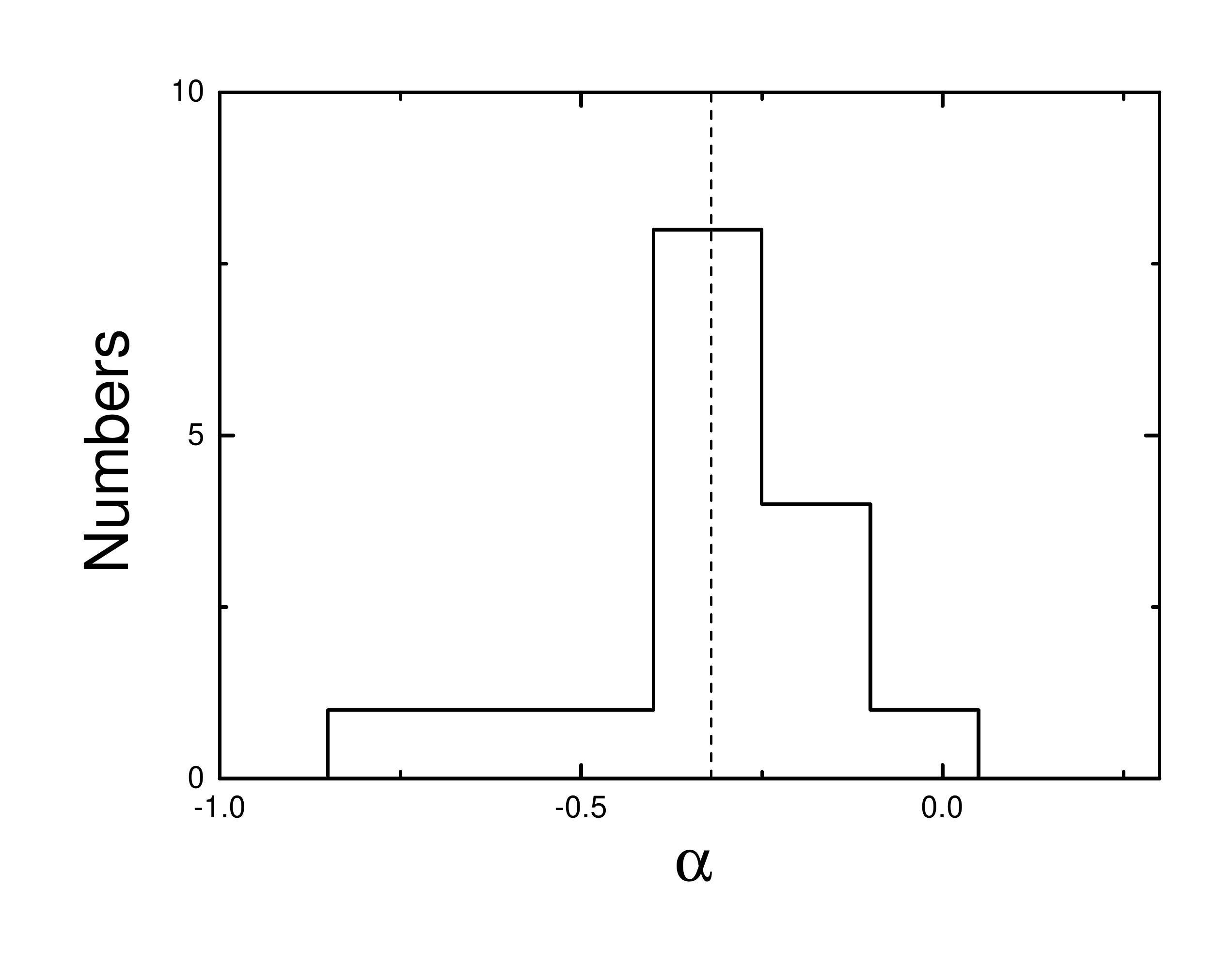}{0.5\linewidth}{}}
\caption{Distribution of the negative power-law indexes in $FWHM \sim E^\alpha$ for the 16 SPs. The vertical red dashed line shows the mean value of $\alpha \sim$ $-$0.32. \label{fig:single-index}}
\end{figure*}

\begin{figure*}[ht]
\gridline{
	\centering
\fig{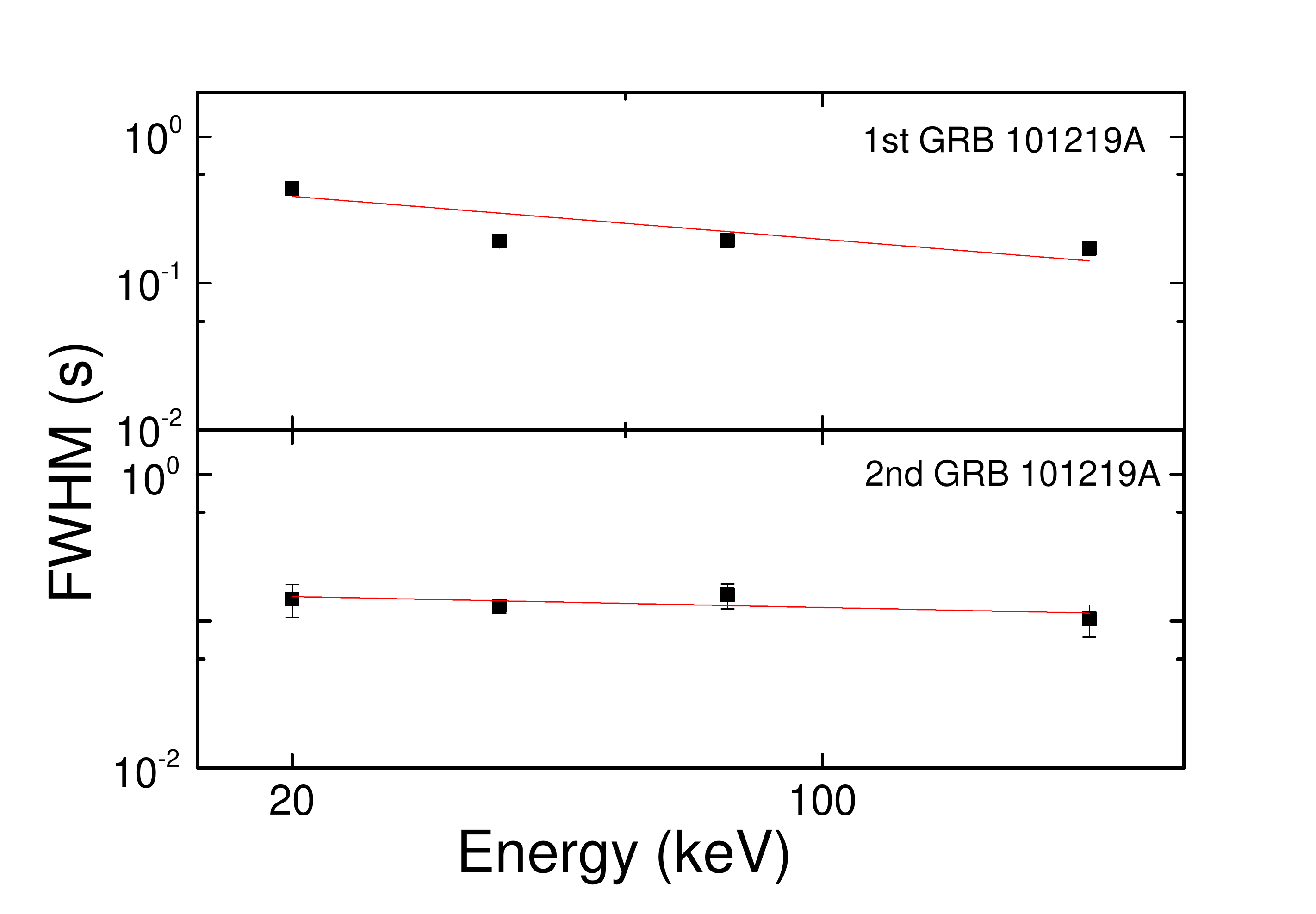}{0.3333\linewidth}{(a)}
\fig{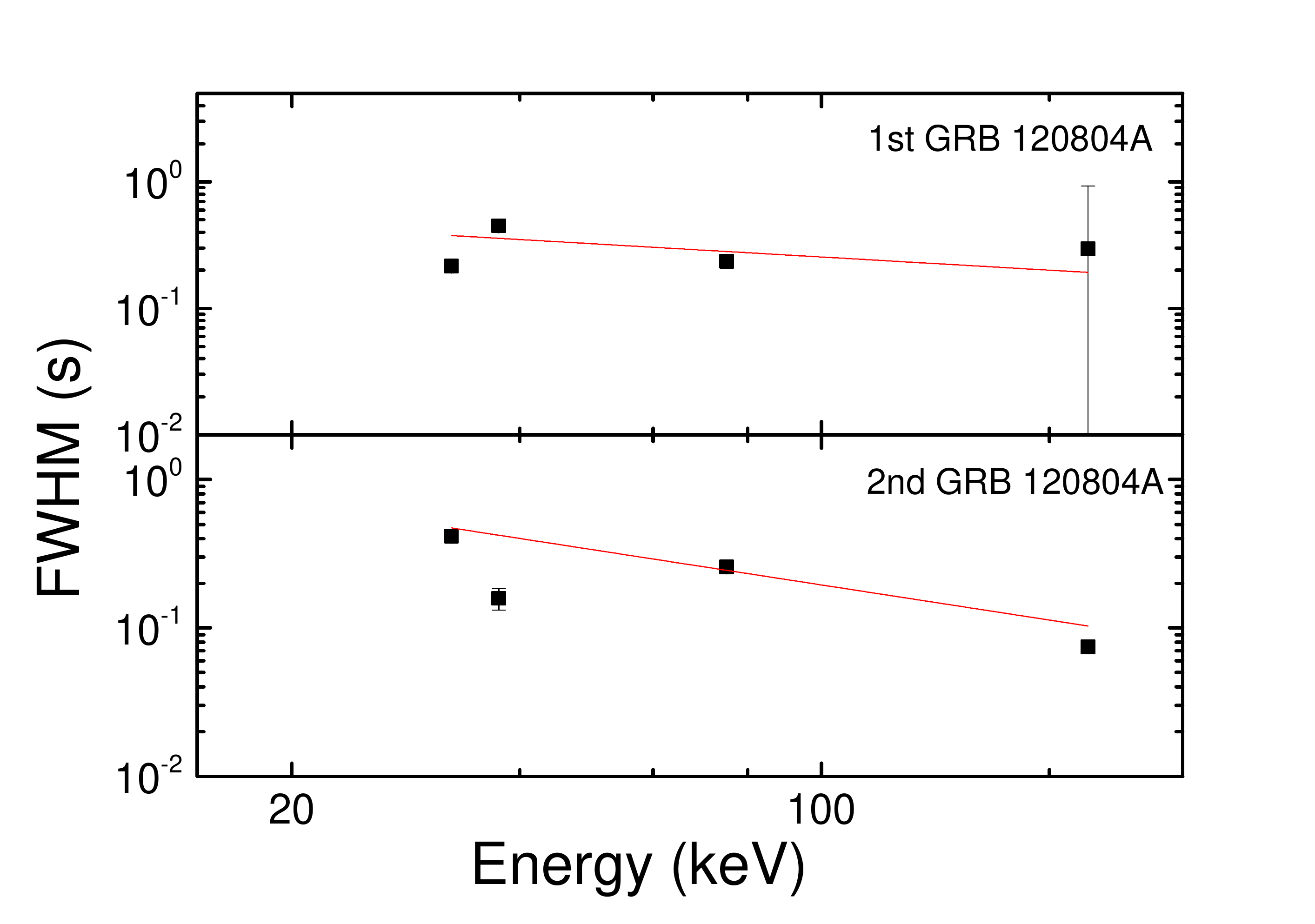}{0.3333\textwidth}{(b)}
\fig{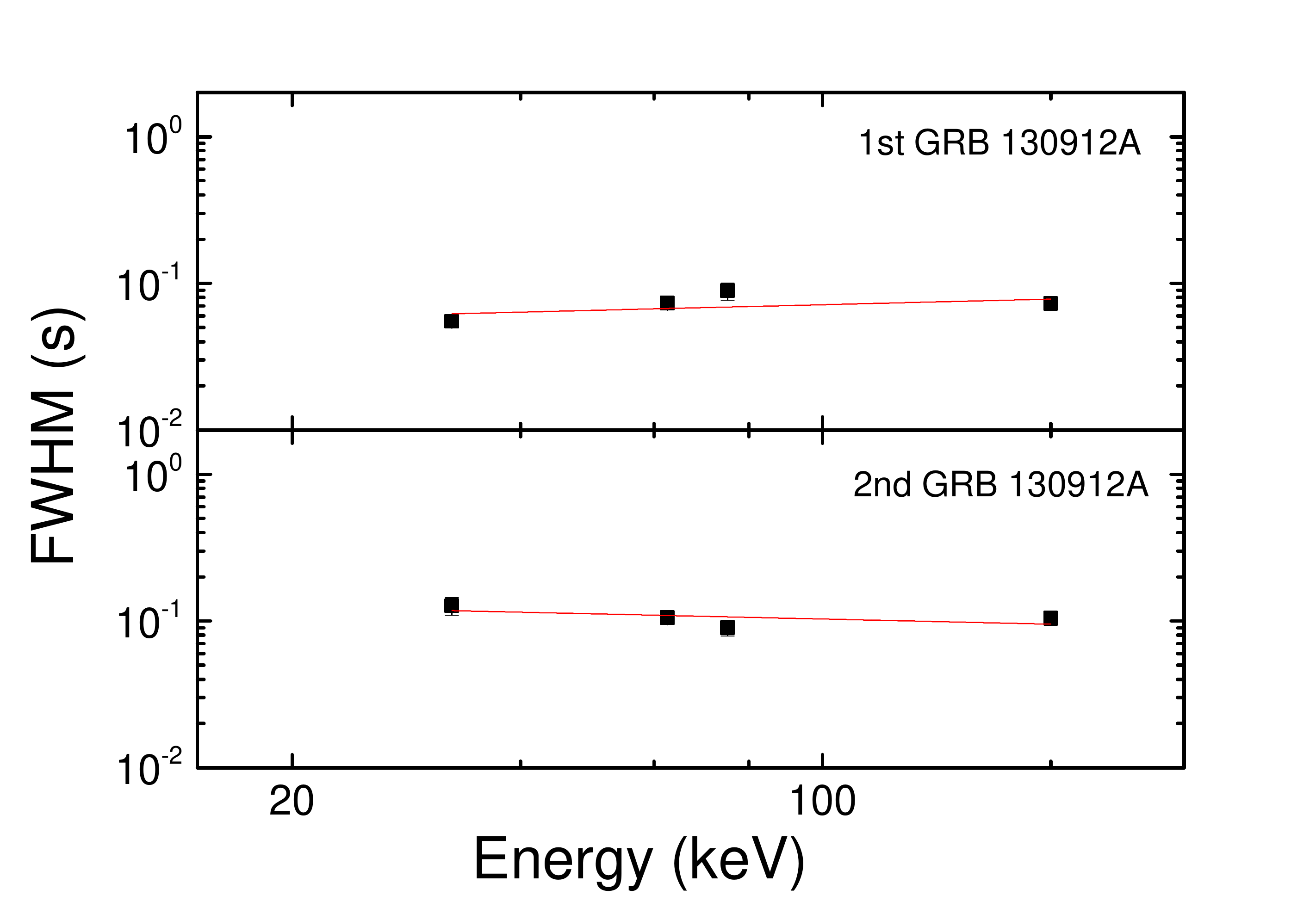}{0.3333\textwidth}{(c)}
          }
\caption{the FWHM vs. the average photon energy of the Mt-DPs. The examples are present for
(a) GRB 101219A, (b) GRB 120804A and (c) GRB 130912A. The solid line stands for the best power-law fit to the observations. \label{fig:mtFandE}}
\end{figure*}

\begin{figure*}[ht]
\centering
\gridline{
\fig{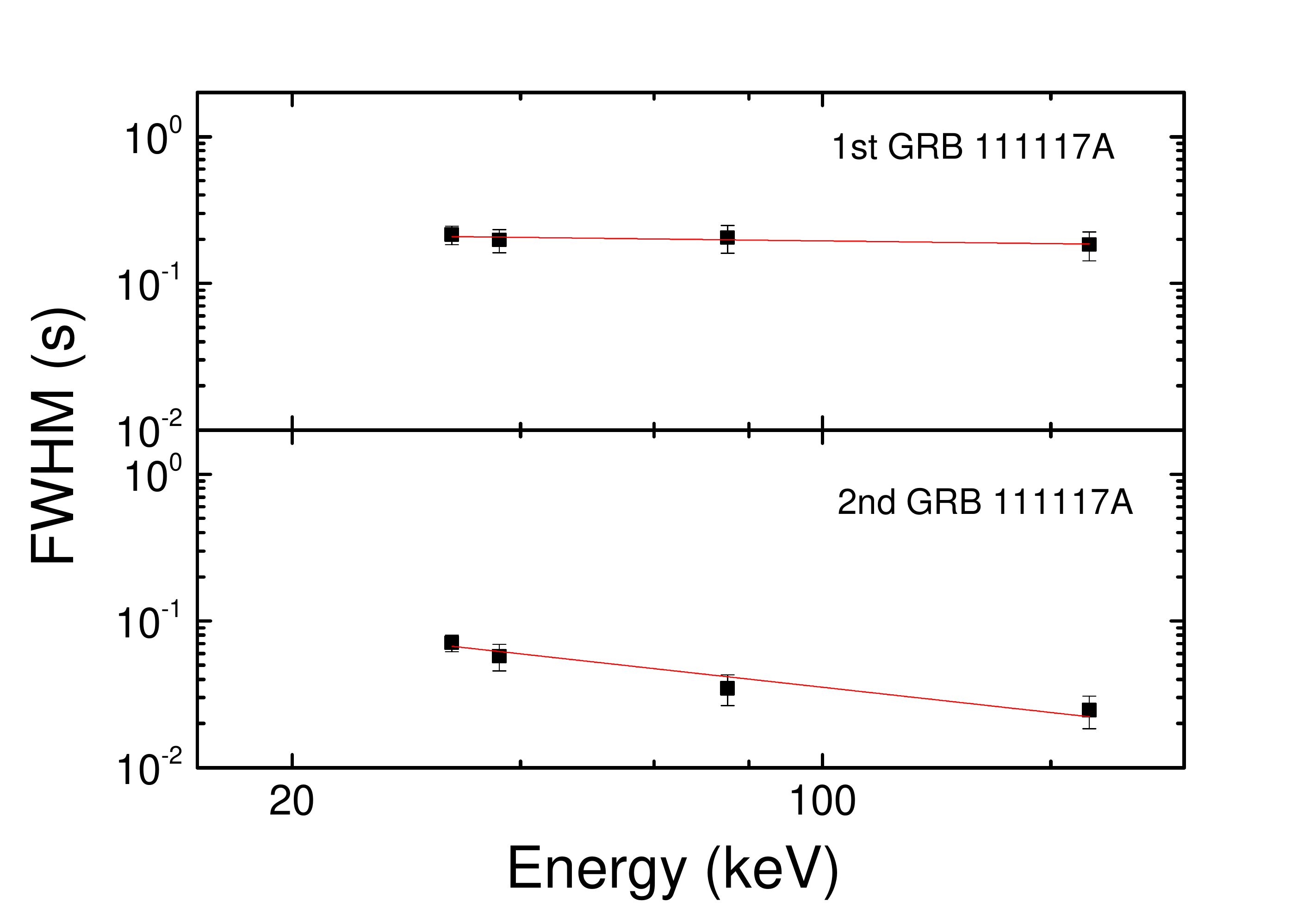}{0.3333\linewidth}{(a)}
\fig{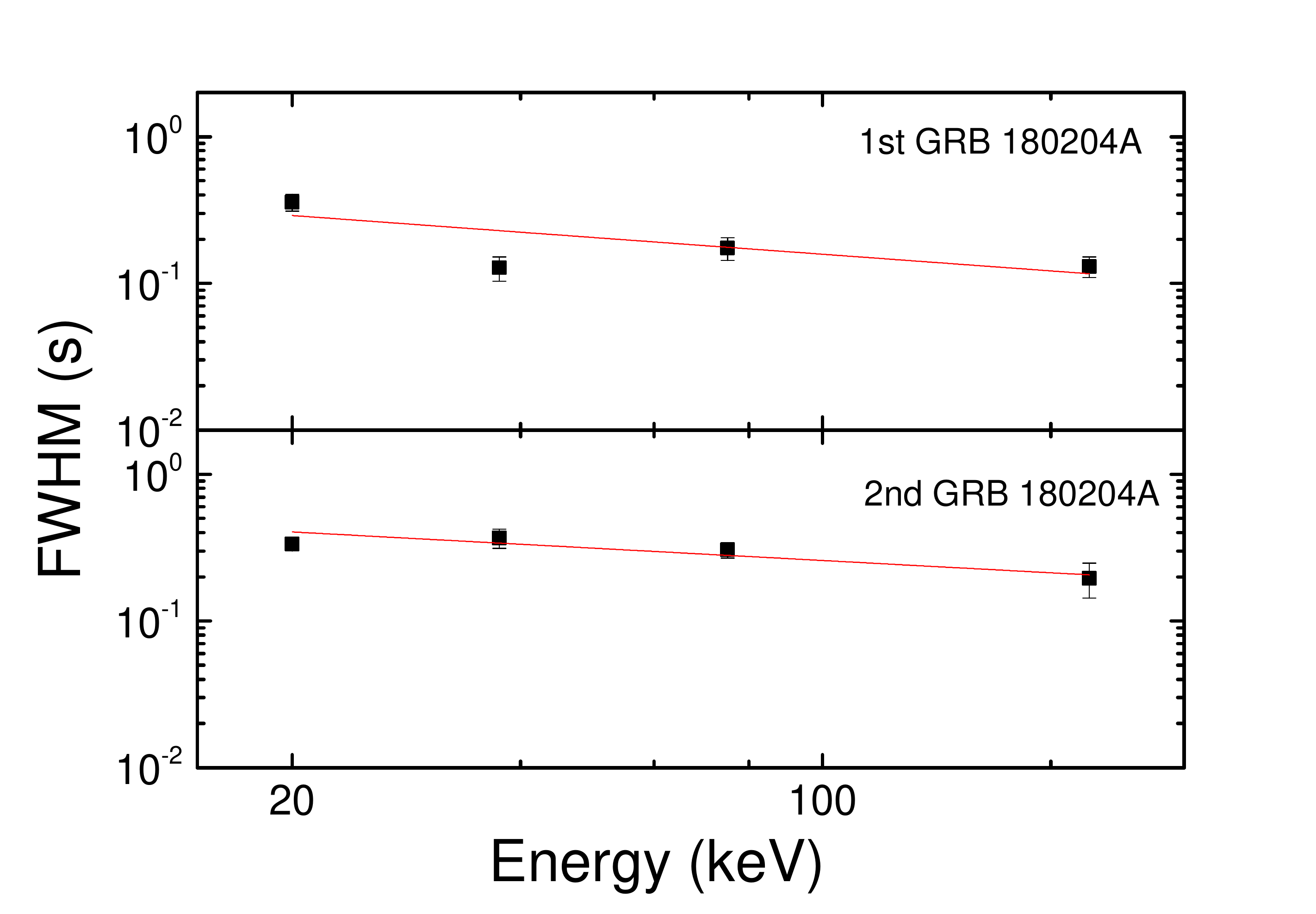}{0.3333\textwidth}{(b)}
          }
\caption{The FWHM vs. average photon energy of the Ml-DPs. The examples are shown for
(a) GRB 111117A and (b) GRB 180204A. The solid line stands for the best power-law fit to the observations. \label{fig:mlFandE}}
\end{figure*}

Figures \ref{fig:singleFandE} - Figure \ref{fig:mlFandE} show the relations of the width
(FWHM) dependence on the average photon energy
with a form as FWHM $\sim E^\alpha$ for the SPs and the
two kinds of the DPs. In Figure \ref{fig:singleFandE}, except the
GRB 070810B, the power-law indexes of the 16 SPs are negative and the mean value is $\alpha\simeq$ $-$0.32 $\pm$ 0.02
which is in close proximity to the results of $-$0.4 found by \cite{1995APJ...448...L101} and \cite{1996APJ...459...393} for the long
GRBs. Very excitingly, the mean value is almost equal
to the value of $-$0.32 for the BATSE SPs (see paper I). Figure \ref{fig:single-index} shows the distribution of these power-law indexes of the 16 SPs. The detailed
fitting results are summarized in Table \ref{tab:index}.

\begin{figure*}[ht]
\centering
\gridline{
\fig{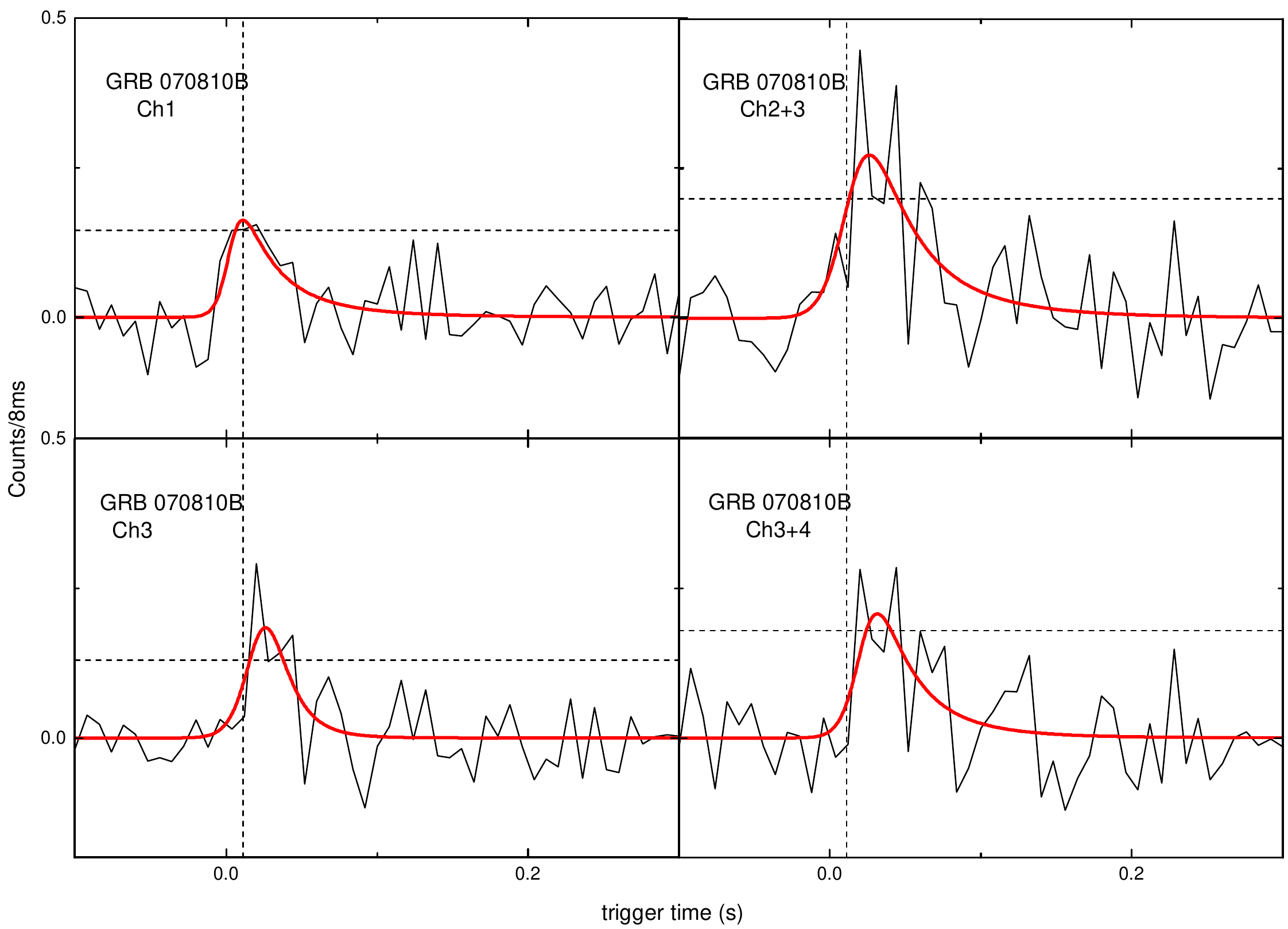}{0.495\linewidth}{(1) the pulse shape revolutions of GRB 070810B}
\fig{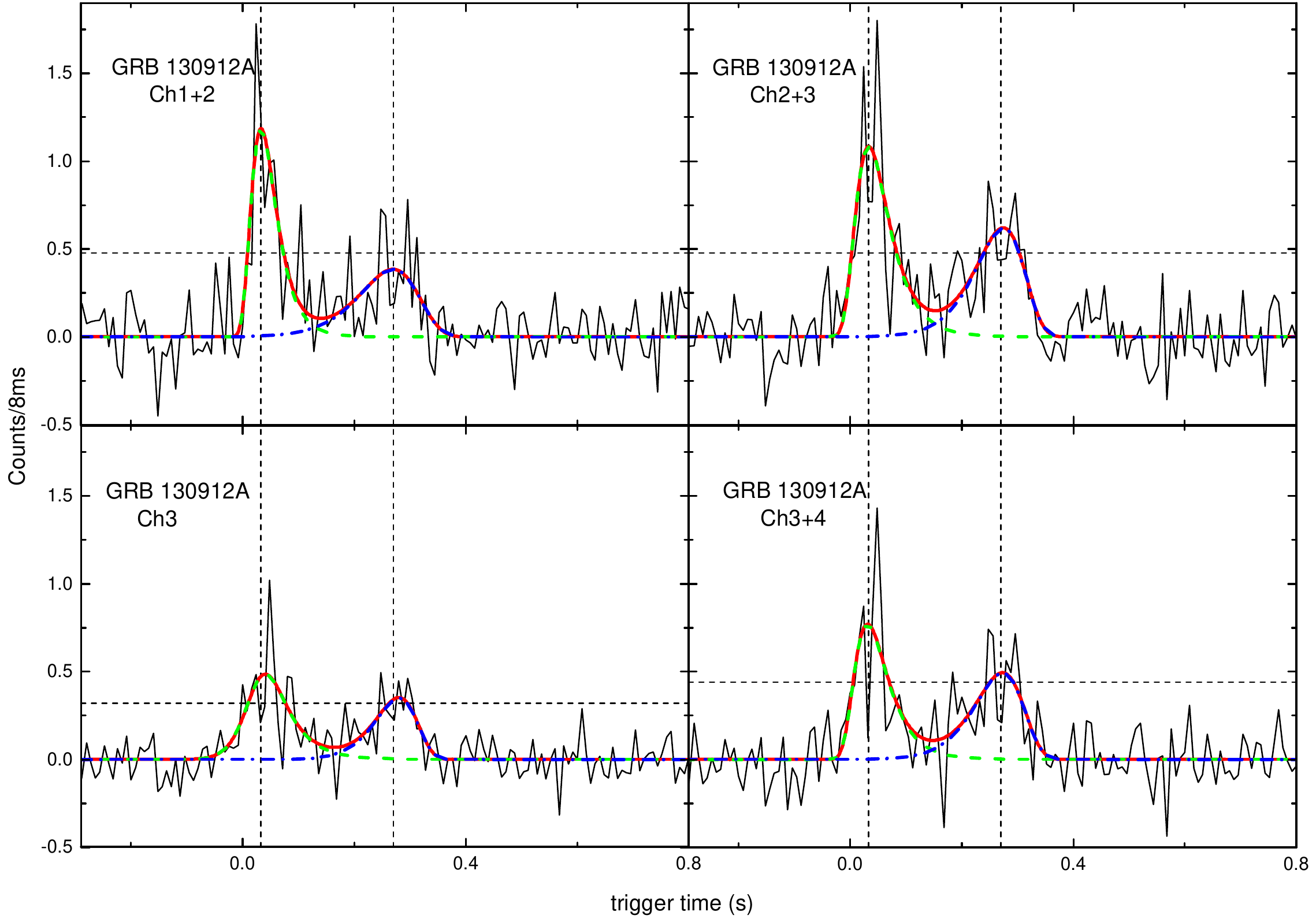}{0.5\textwidth}{(2) the pulse shape revolutions of GRB 130912A}}
\caption{The revolution from the lower to the higher energy channels. The vertical
black dash lines mark the peak time (t$_m$) of the main peaks in Ch1 (GRB 070810B) and Ch1+2 (GRB 130912A). The horizontal dotted black lines mark a 3$\sigma$ confidence level. \label{fig:pulse evo}}
\end{figure*}

For GRB 070810B, the value of the power-law index
is $\alpha\simeq$ $-$0.05 $\pm$ 0.16 and the pulse shape evolution from
low to high energy channel is shown in Figure \ref{fig:pulse evo} (1).
The values of t$_r/t_d$ from the low to the high energy channel are 0.49, 0.60, 0.82
and 0.60, respectively.

In Figures \ref{fig:mtFandE} and \ref{fig:mlFandE}, we find that most of the DPs possess the negative
power-law indexes except GRB 130912A whose
light curve shows two overlapping peaks \citep{2013GCN...15216...1}. The mean values of these
negative power-law indexes are $\alpha\simeq$ $-$0.38 $\pm$ 0.85 (1st)
and $\alpha\simeq$ $-$0.45 $\pm$ 0.33 (2nd) for the Mt-DPs and
$\alpha\simeq$ $-$0.22 $\pm$ 0.21 (1st) and $\alpha\simeq$ $-$0.42 $\pm$ 0.13 (2nd) for the Ml-DPs, respectively. For Mt-DP 130912A, the
values of the power-law indexes are $\alpha\simeq$ 0.13 $\pm$ 0.12
and $\alpha\simeq$ $-$0.12 $\pm$ 0.11 for the 1st pulse and the 2nd
pulse, respectively. The pulse shape evolution from low
to high energy channel is shown in Figure \ref{fig:pulse evo} (2). The
values of t$_r$/t$_d$ are 0.67, 0.65, 0.82 and 0.66 for the 1st
pulses and 1.39, 1.38, 1.44 and 1.37 for the 2nd pulses.
Obviously, we can see that there is no obvious shape
evolution for either the 1st or the 2nd pulses.

\subsection{Main Peaks versus Precursors or EEs}

\begin{figure*}
\centering
\gridline{\fig{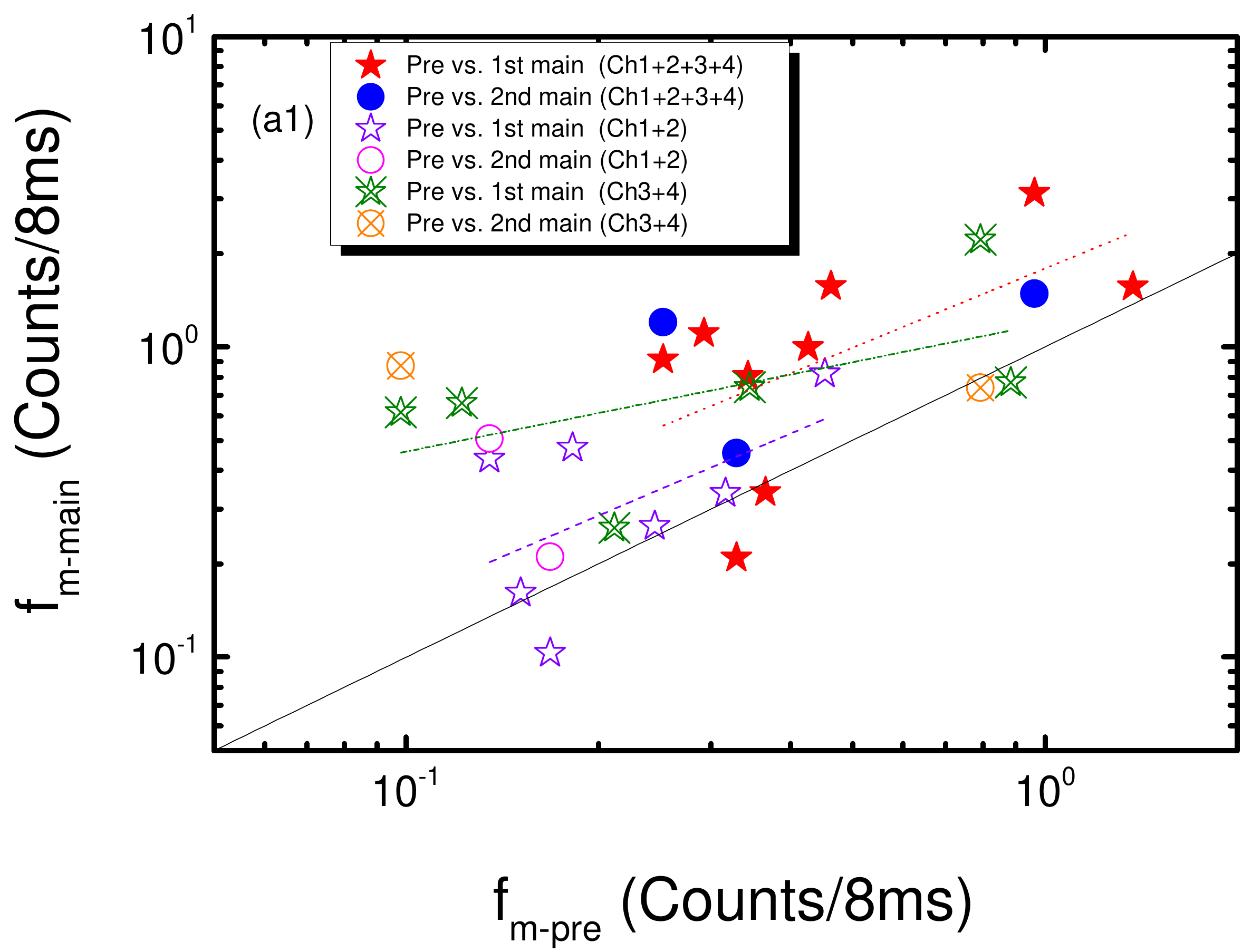}{0.33\textwidth}{(a1)}
          \fig{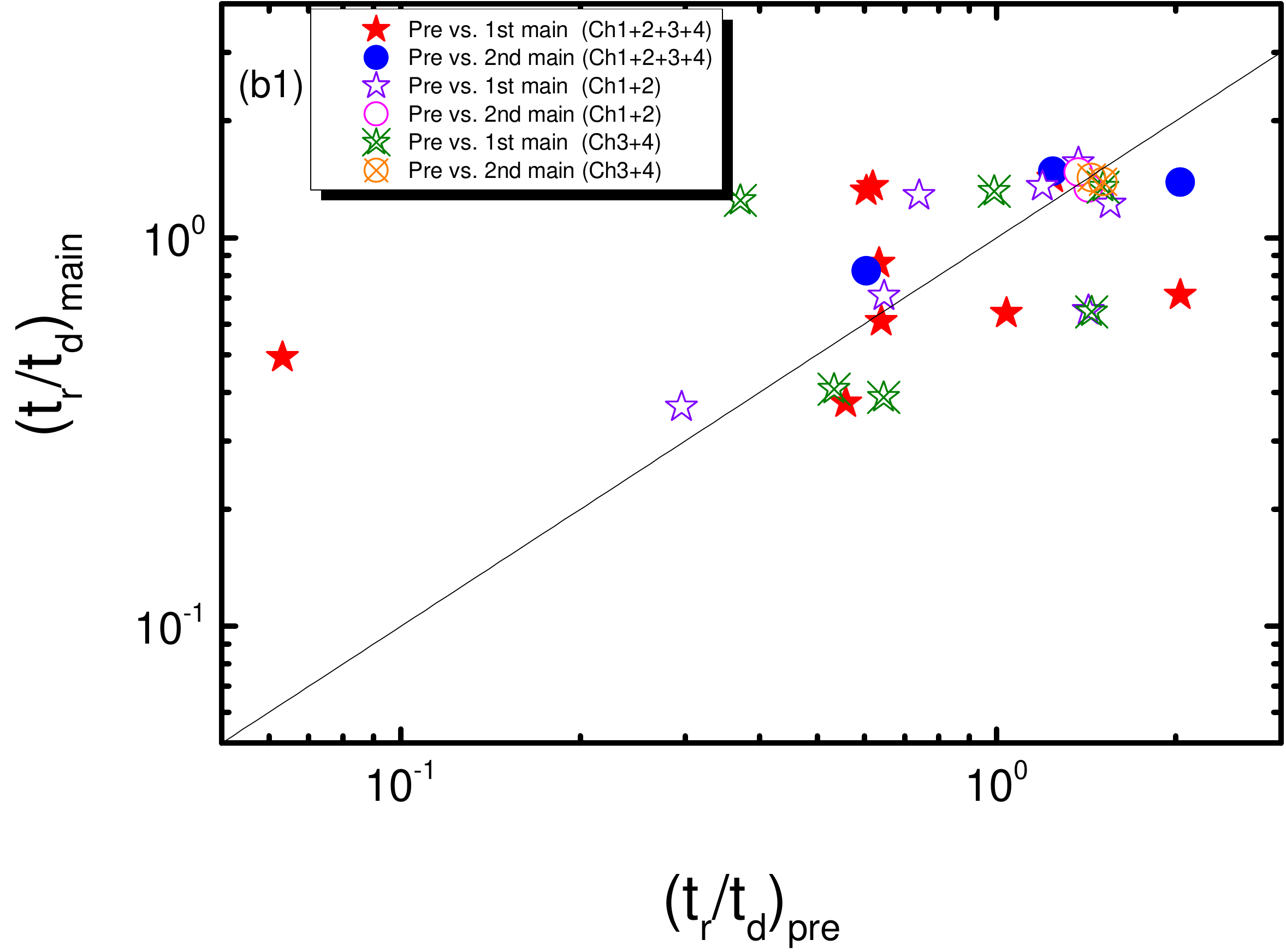}{0.33\textwidth}{(b1)}
          \fig{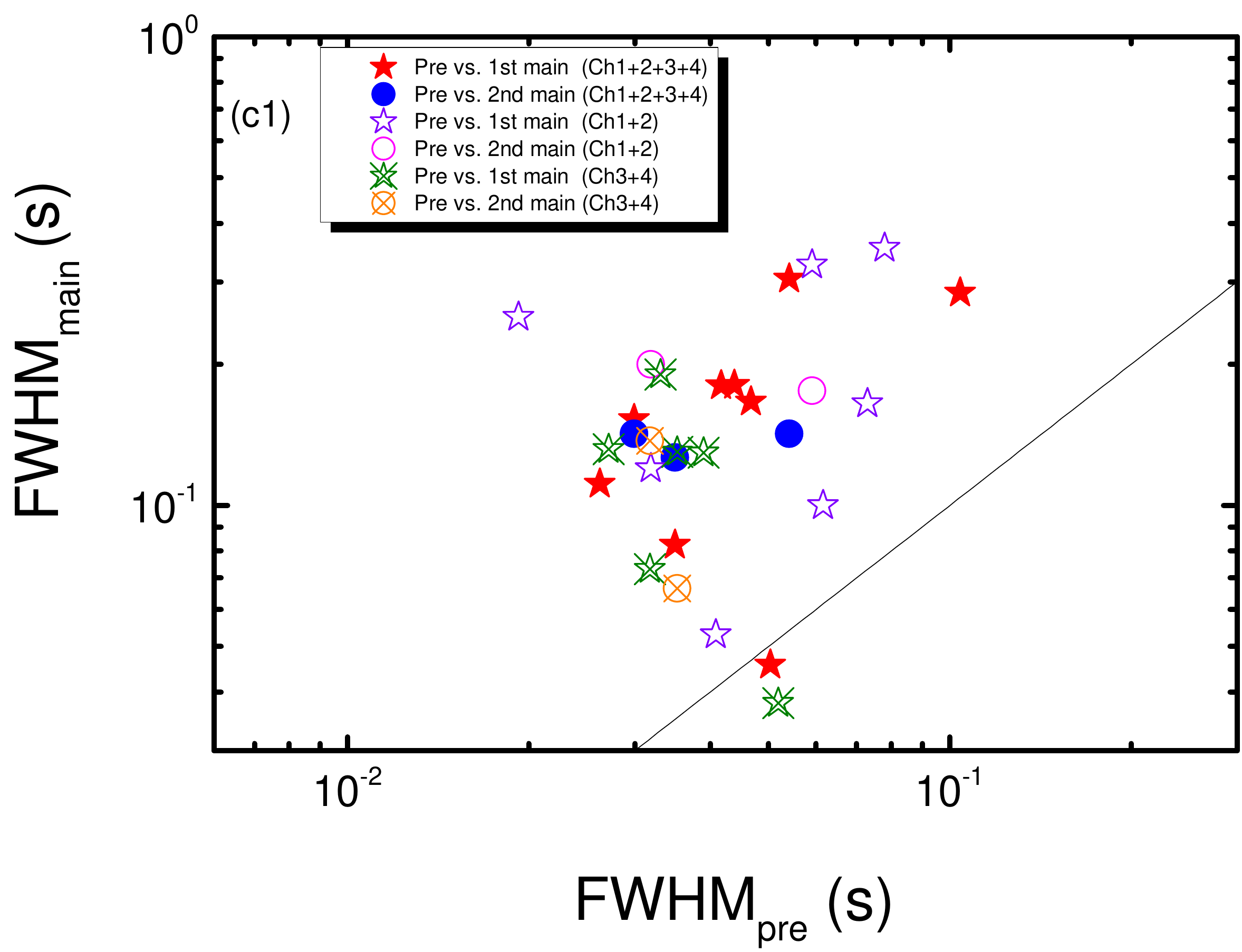}{0.33\textwidth}{(c1)}
          }
\gridline{\fig{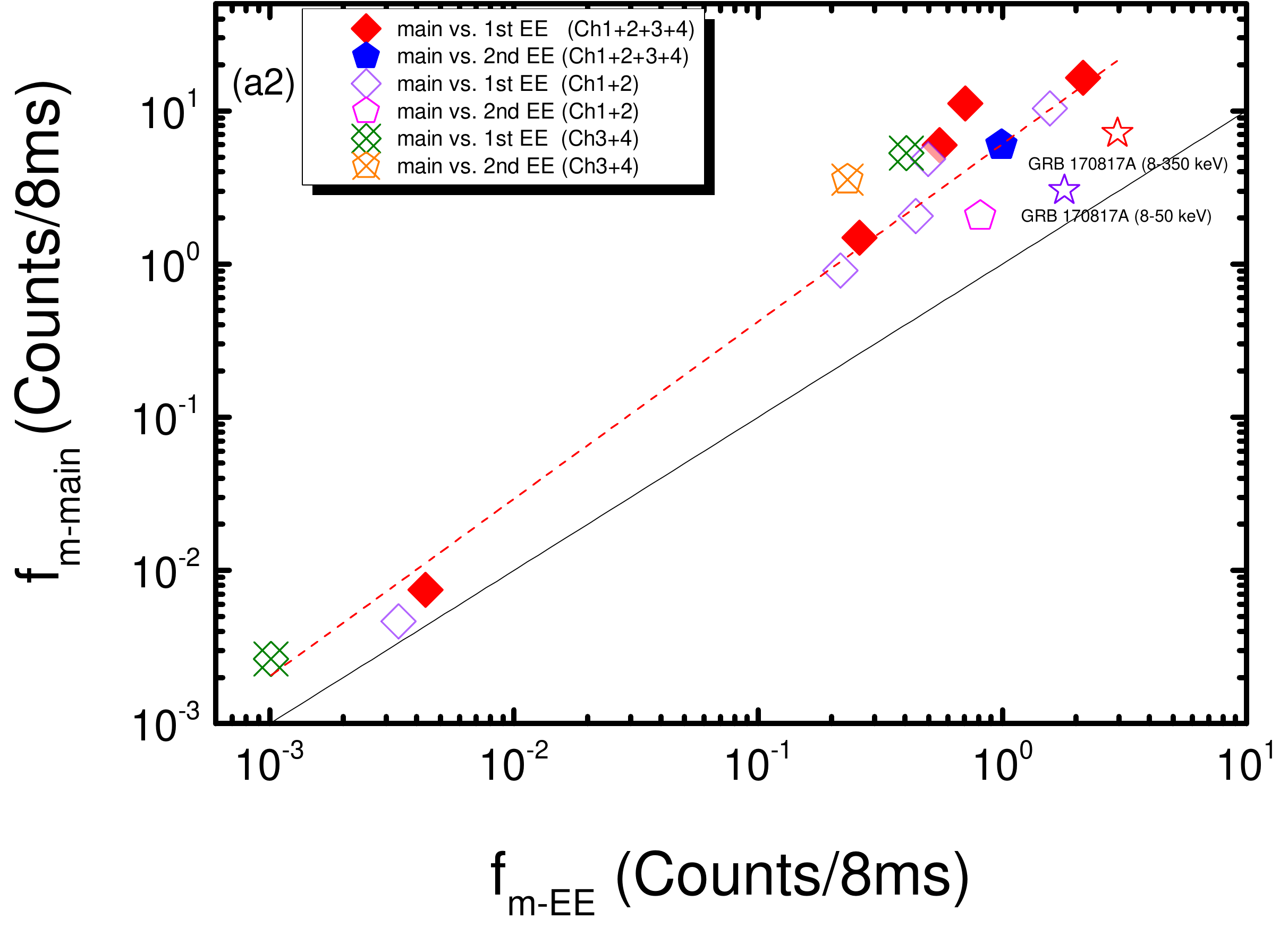}{0.33\textwidth}{(a2)}
          \fig{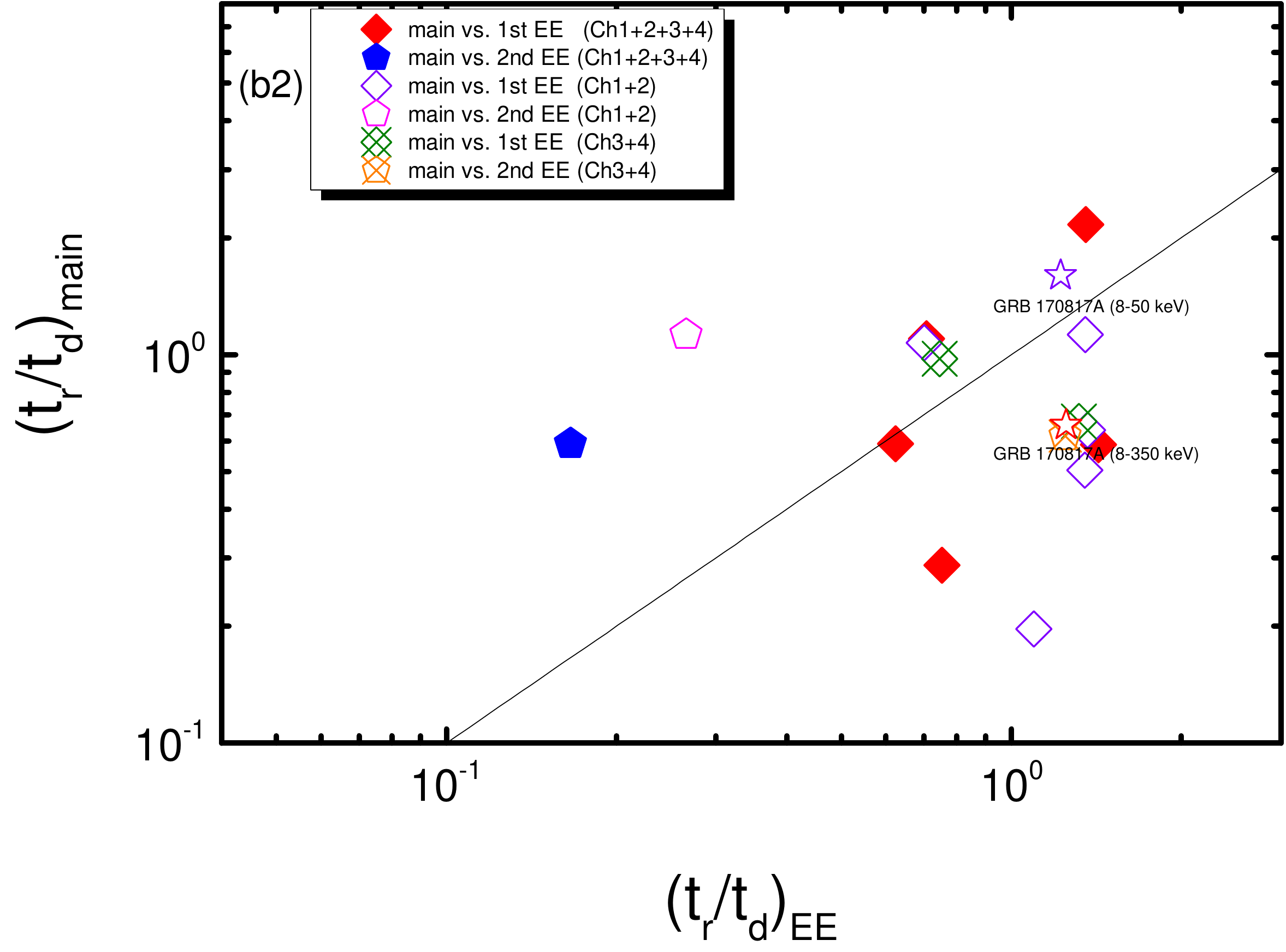}{0.33\textwidth}{(b2)}
          \fig{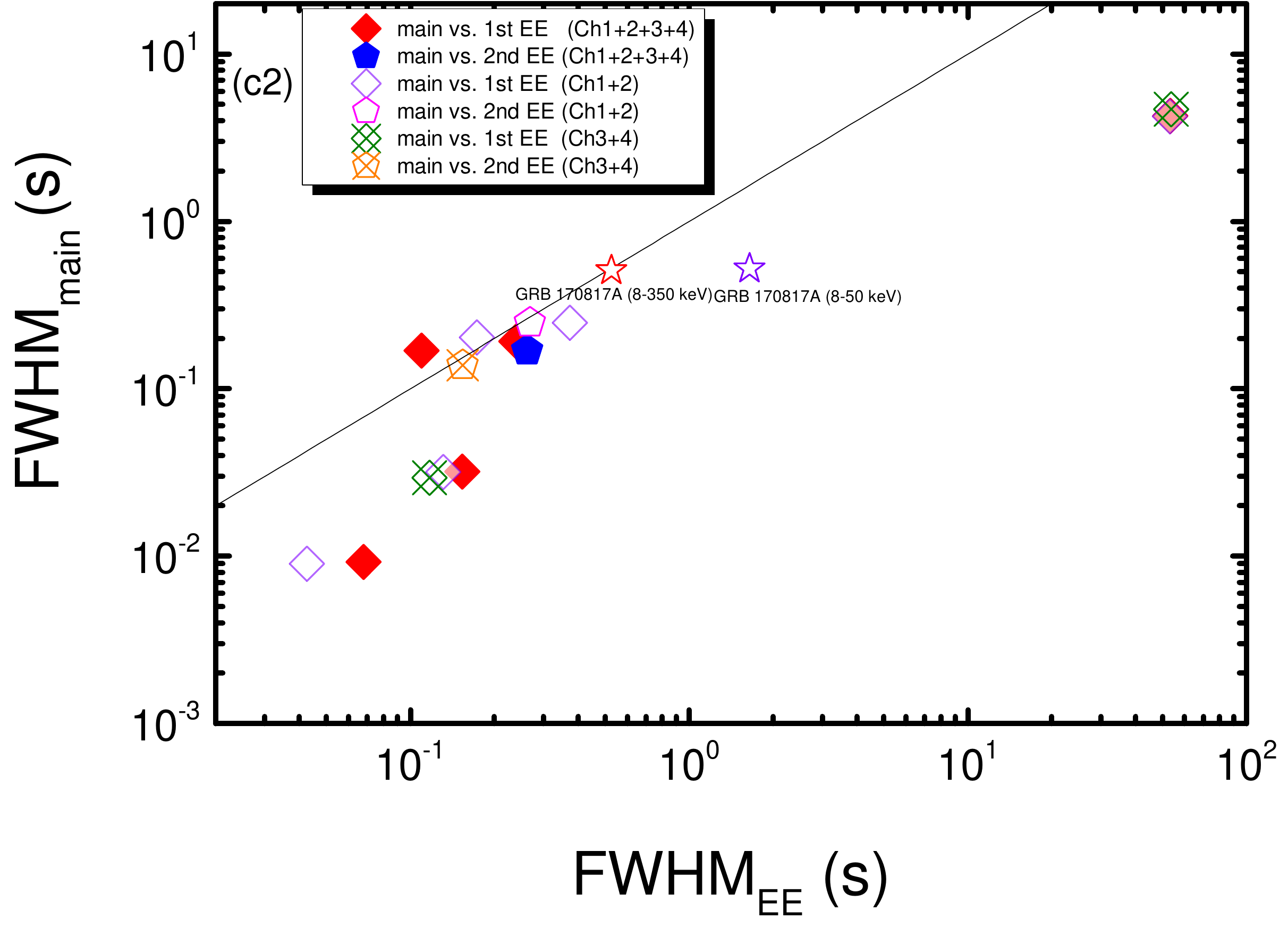}{0.33\textwidth}{(c2)}
          }
\gridline{\fig{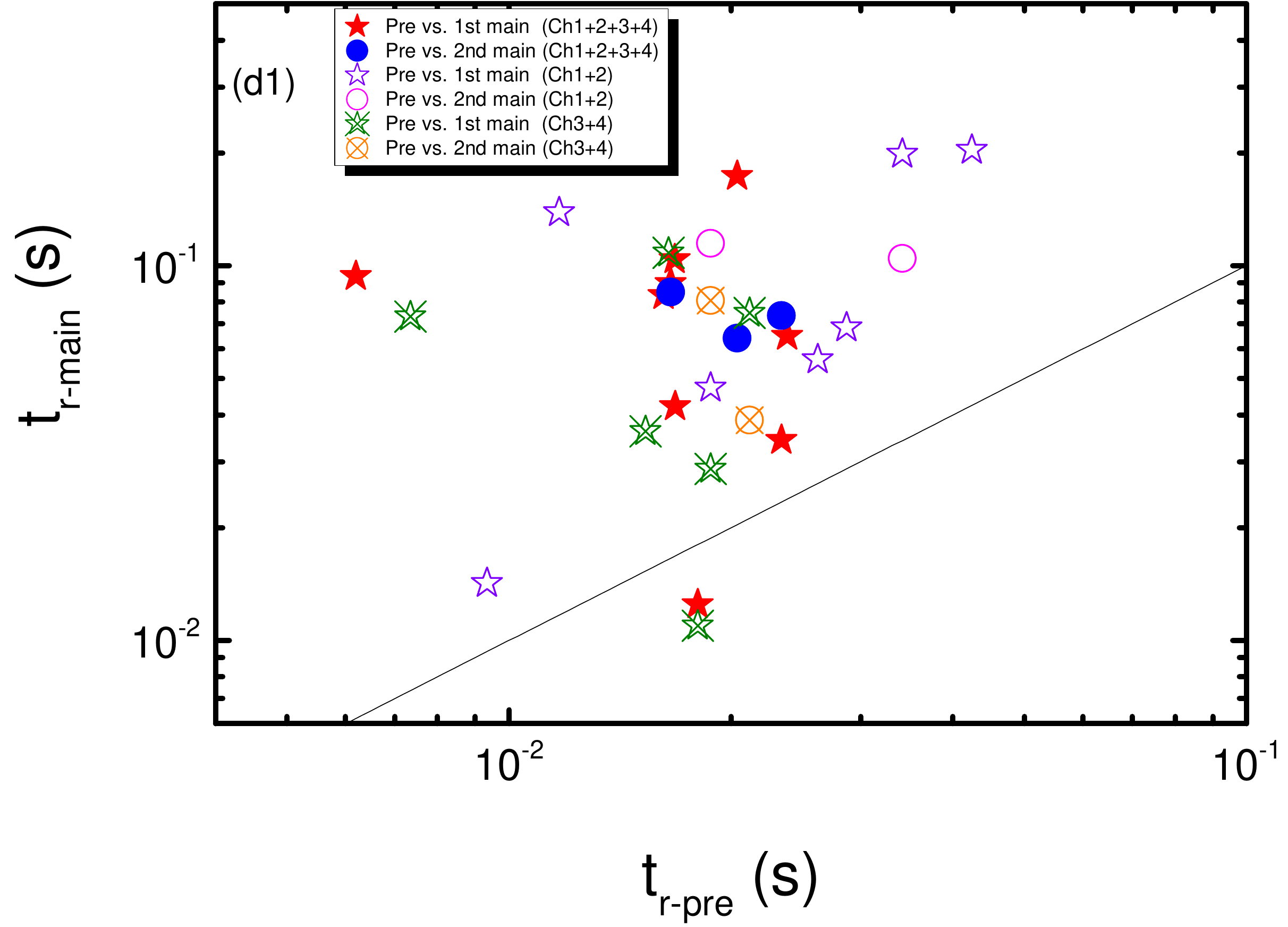}{0.33\textwidth}{(d1)}
          \fig{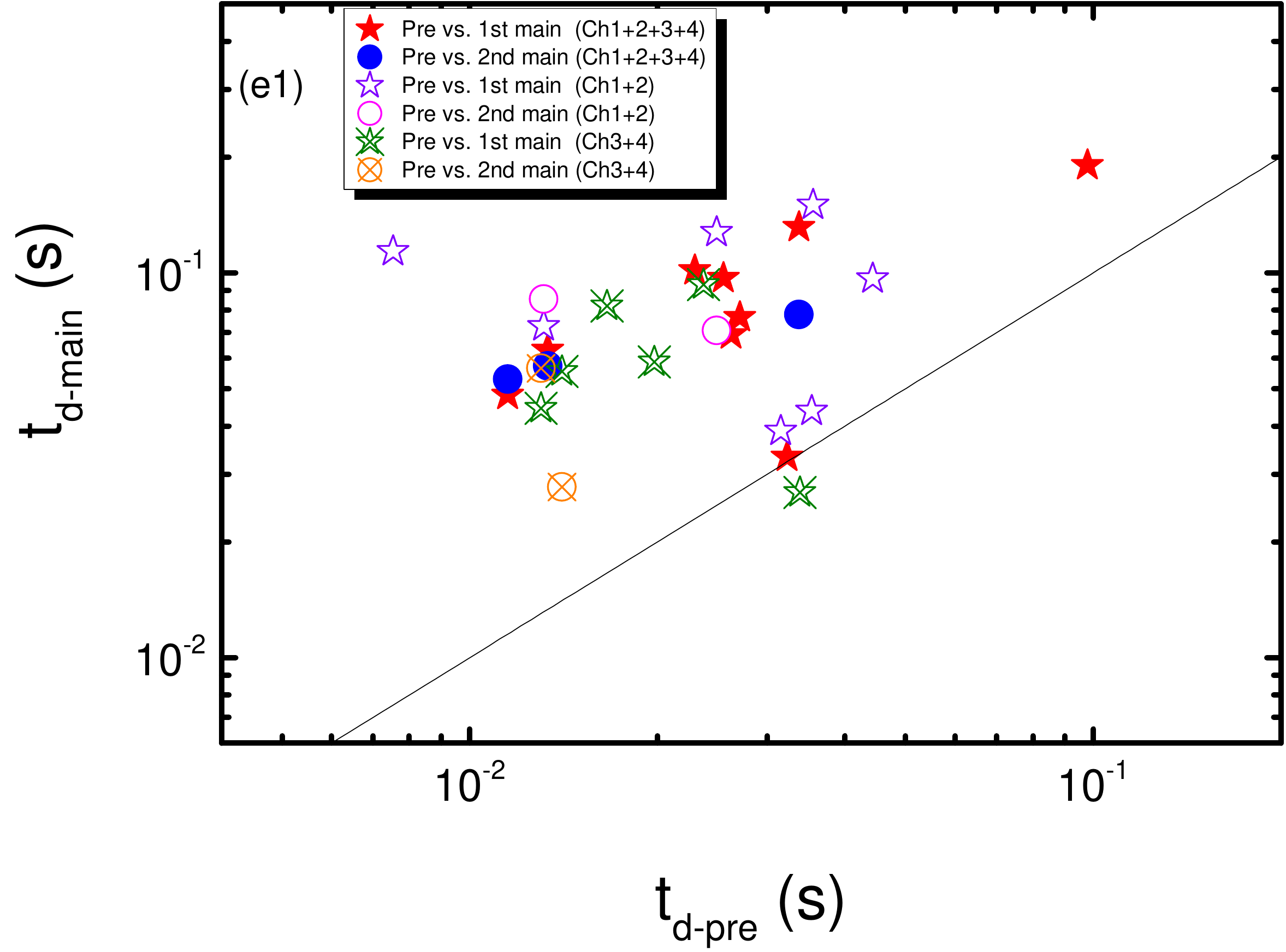}{0.33\textwidth}{(e1)}
          }
\gridline{\fig{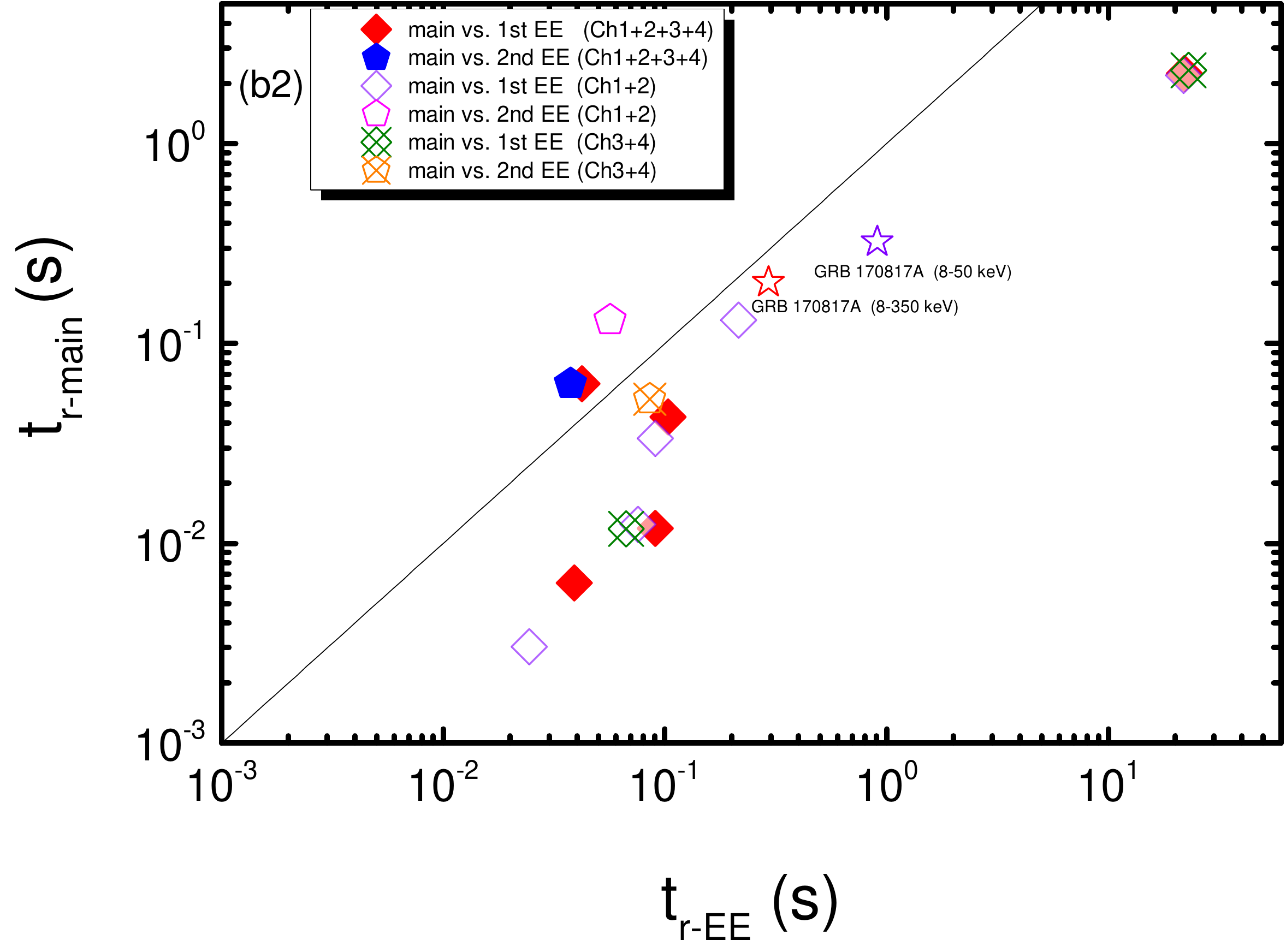}{0.33\textwidth}{(d2)}
          \fig{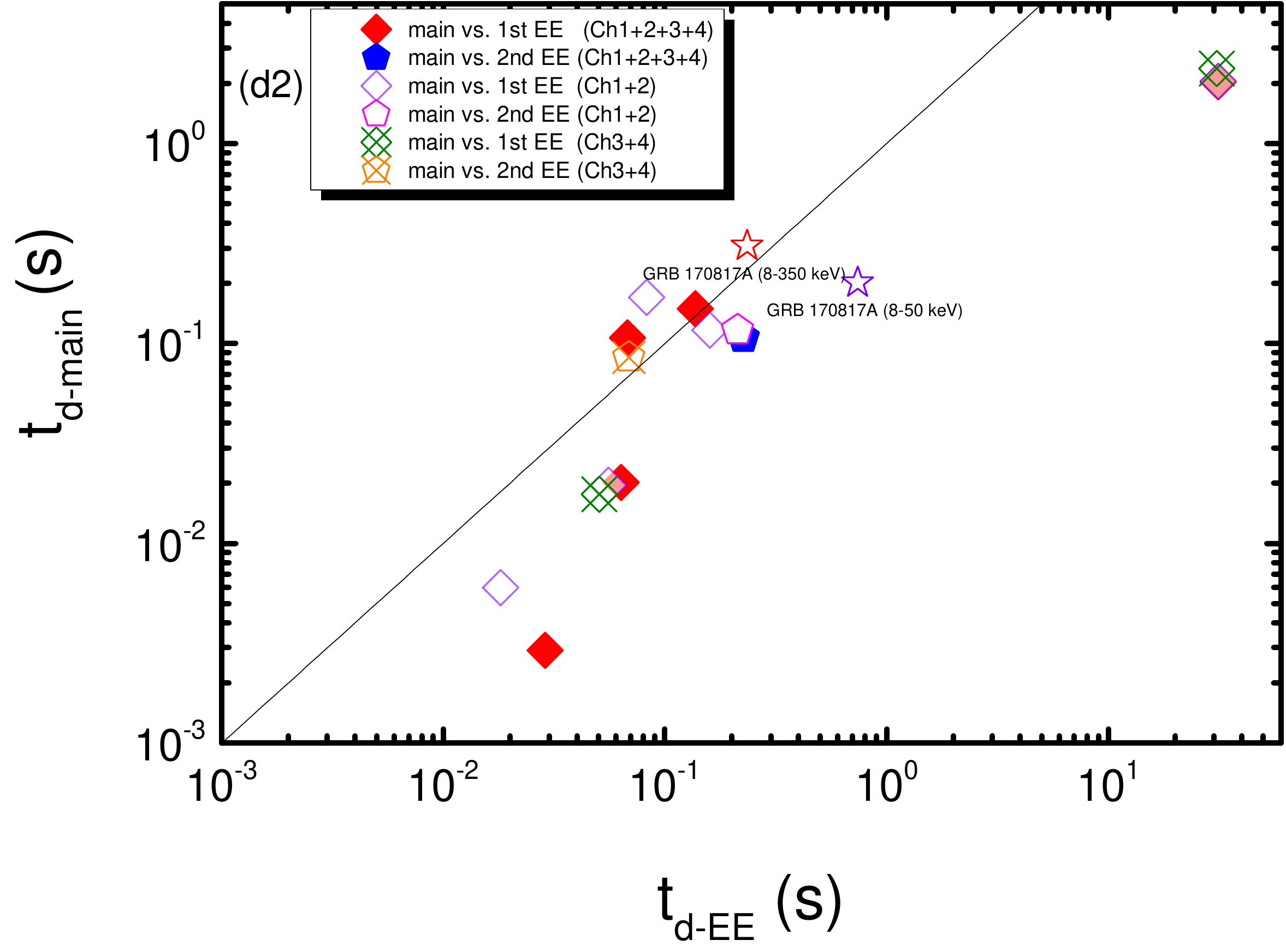}{0.33\textwidth}{(e2)}
          }
\caption{Comparisons of the pulse properties of the main peaks to those of the precursors (panels a(1) b(1) c(1) d(1) e(1)) the EE tails (panels a(2) b(2) c(2) d(2) e(2)). The solid black lines plotted
in both panels denote where the pulse properties are equal. In panel a(1), the lines stand for the best power-law fits to the data of the first main peak and its precursor, dotted line for Ch1+2+3+4, dashed line for Ch1+2, dash-dotted line for Ch3+4. In panel a(2), the dotted line stands for the best power-law fit to the data of the main peak and its first EE pulse.
\label{fig:preandee}}
\end{figure*}

In Figure \ref{fig:preandee} (a1), we find weak positive correlations
in the f$_m$ between the precursors and the main
peaks. For the three Pre+DPs, we just check the
first (1st) main peaks (main1) because the number
of the second (2nd) main peaks is relatively
limited. Power-law fits across different energy bands
give $logf_{m,main1}$=$(0.85\pm0.44)\times logf_{m,pre}+(0.25\pm0.18)$ with the correlation coefficient of Pearson's r=0.59
(15-350 keV), $logf_{m,main1}$=$(0.87\pm0.59)\times logf_{m,pre}+(0.07\pm0.41)$ with the correlation coefficient of Pearson's
r=0.55 (15-50 keV), $logf_{m,main1}$=$(0.41\pm0.34)\times logf_{m,pre}+(0.08\pm0.20)$  with the correlation coefficient of Pearson's r=0.56 (50-350 keV). The results are marginally in agreement with the recent conclusion
drawn by \cite{2019Zhong} for 18 sGRB candidates with precursor observed by Fermi/GBM and Swift/BAT.

In Figure \ref{fig:preandee} (b1)(c1)(d1)(e1), mild correlations are
found in the t$_r$/t$_d$, the FWHM, the t$_r$ and the t$_d$ for
the Pre+sGRBs, similar to the results of the ``Type I"
precursors reported by \cite{2015PHD...???...???}. Additionally, there
is general no event in the lower right region from the
solid line in Figure \ref{fig:preandee} (c1)(d1)(e1). The FWHM of the
precursors are found to be on average an order of magnitude
smaller than those of the main peaks. These
results all indicate that the widths of main peak pulses
tend to be wider than those of precursor pulses for
the Pre+sGRBs, suggesting that the main peaks tend
to last for longer time than the precursors, agreement
with result of the \cite{2019Zhong}.

Similarly, a positive correlation is found in the
f$_m$ between the EE and the main peaks. For the GRB
051221A with two EE pulses, we take into account the 1st
EE pulse. For Mt-DPs+EE 130603B, we also just consider
the first main pulse. The results of power-law fit gives
$logf_{m,main1}$=$(1.16\pm0.08)\times logf_{m,EE1}+(0.78\pm0.10)$
with the correlation coefficient of Pearson's r =0.97 (see
Figure \ref{fig:preandee} (a2)). The positive power-law index is larger than the result suggested by \cite{2019ApJ...876...89} for
the Fermi sGRBs with soft tail similar to GRB 170817A.
Here, note that the correlations in the f$_m$ between the
EE and the main peaks among different energy channels
behave no dependence on energy in Figure \ref{fig:preandee} (a2),
thus, all the first EE pulses across 15-350 keV are chosen
to increase the statistical reliability.

No distinct correlations are found in the t$_r$/t$_d$, the
FWHM, the t$_r$ and the t$_d$ not only for the Pre+sGRBs but also for
the sGRBs+EE (see Figure \ref{fig:preandee} (b2), (c2), (d2), (e2)).
However, there are less events in the upper left region
from the solid line in Figure \ref{fig:preandee} (c2)(d2)(e2). These
results indicate that the widths of main peak pulses
tend to be narrower than those of the EE pulses for the
sGRBs+EE, which is caused by both the rise and the decay times. Particularly, we compare the temporal properties with those of the GRB 170817A and no obvious differences are found.

Additionally,
Figure \ref{fig:preandee} (a1)(a2) illustrate that the f$_m$ values of the main
peaks are generally larger than those of the other two
components. On the other hand, the photon
fluxes of main
peaks and the other two components seem to be positively correlated, indicating that the luminosities of the main
peaks are linked to those of the precursors and the EE components, agreement
with the results of \cite{Zhangxiaolu2020} and \cite{2019ApJ...876...89}. Moreover, it in return hints that the three parts of prompt gamma ray emissions could be produced from the same progenitor. For example, recently, \cite{2018Nature...2...69} studied the properties
of a extraordinarily bright three-episode long GRB 160625B detected by Fermi. Although we have not
found three-episode sGRBs in our sample, the similarities may indicate that the two weak emission episodes
may be likely to exist, or, in other words, be intrinsic. The absences of the EEs or precursors might be related to sensitivity or energy coverage of the current GRB detectors.

\begin{figure*}[!h]
\centering
\gridline{
\fig{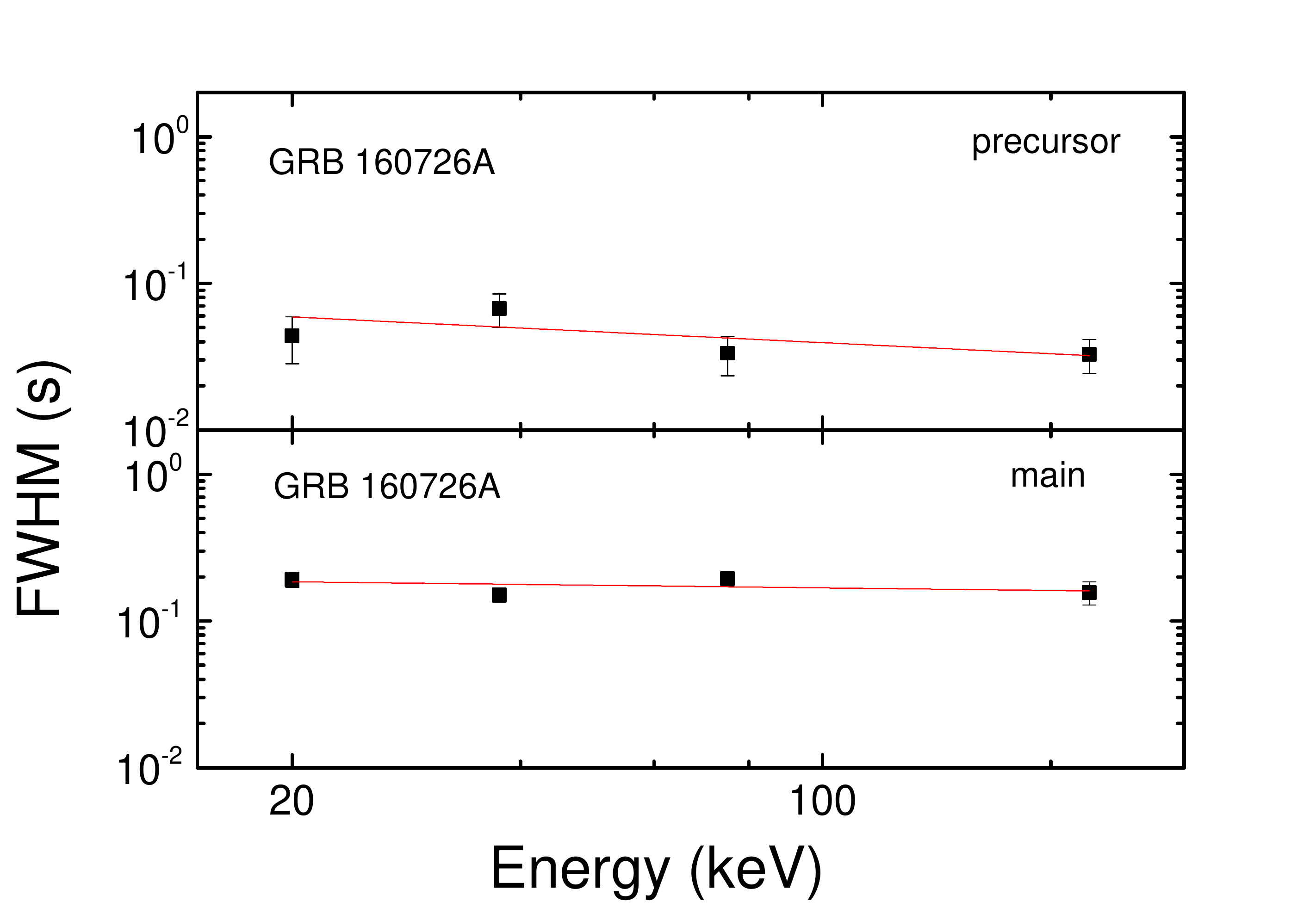}{0.5\textwidth}{(1)}
\fig{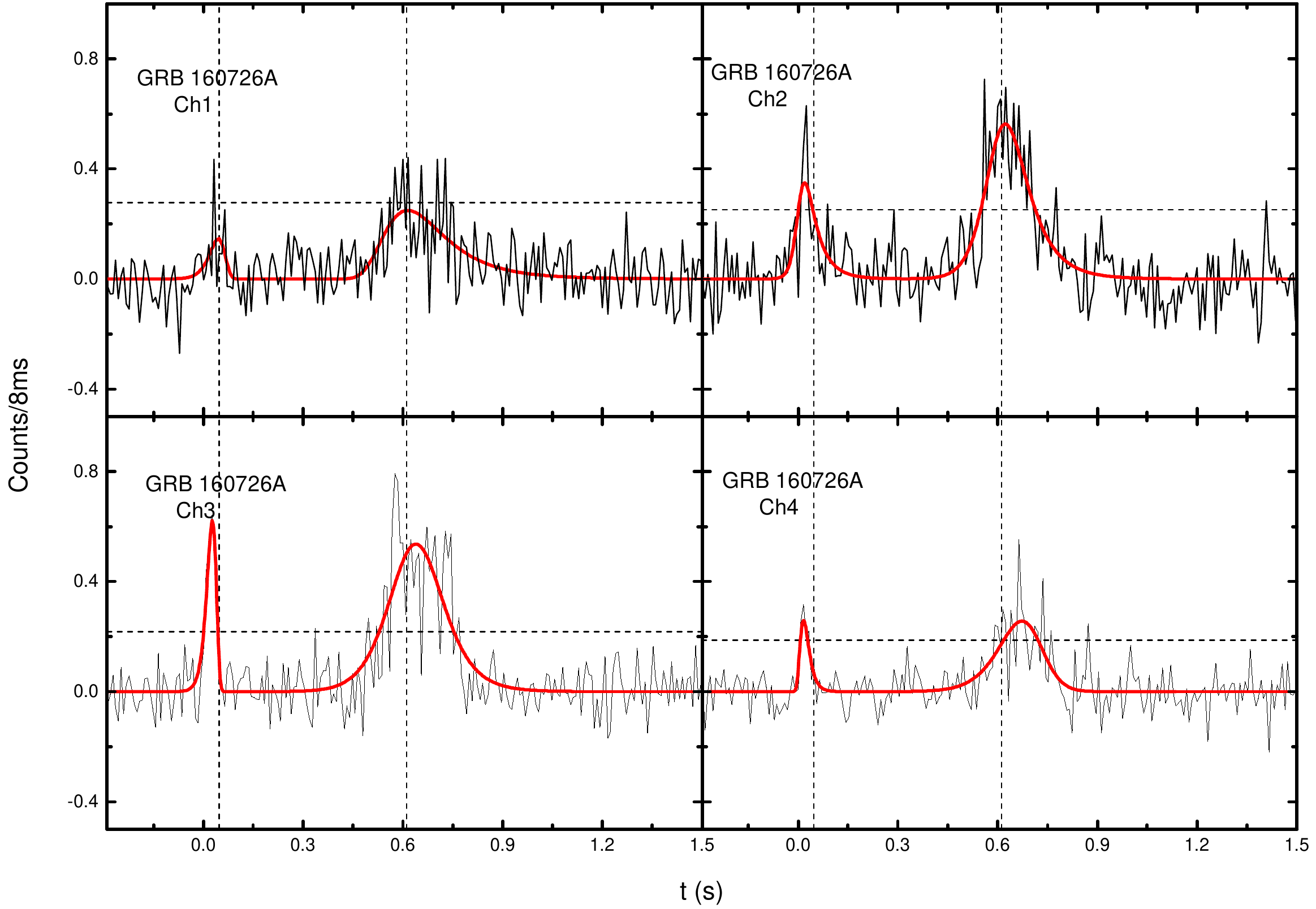}{0.5\textwidth}{(2)}
          }
\caption{The example of the Pre+SPs. (1) The FWHM vs. average photon energy for GRB 160726A. (2) The pulse shapes of GRB 160726A revolution from the lower to the higher energy channels. The vertical
black dash lines mark the peak time (t$_m$) of the main peaks in Ch1. The horizontal dotted black lines mark a 3$\sigma$ confidence level. \label{fig:pulse evo2}}
\end{figure*}

Especially, we check the relations of the width dependence
on the average photon energy and the pulse evolution
modes among different energy channels not only
in the precursor pulse but the main pulse of the Pre+SPs in Figure \ref{fig:pulse evo2}. The values of t$_r$/t$_d$ of the precursor
in GRB 160726A are 0.67, 0.64, 0.91, 0.69 while those
of the main pulse are 0.48, 0.84, 0.96, 1.21 corresponding
to channels from low to high individually. There is
almost no shape evolution across different energy bands
for precursor pulse. However, the shape evolution of the
main pulses is very different.

\section{DISCUSSION}
Except two regular evolving modes, ``MODE I" and
``MODE II" which are corresponding to the sGRBs with
positive and negative power-law index of the pulse width
with the photon energy (see paper I), we found that the
indexes of some sGRBs are marginally zero including
GRB 100206A, GRB 070810B, GRB 111117A (1st), GRB 101219A, GRB 130912A and GRB
160726A (main). Surprisingly, there are two possible cases for
these sGRBs. In one case, we find there are almost no
shape evolution across different energy bands, for example,
GRB 100206A, GRB 070810B and GRB 130912A.
In another case, pulse shape of the main peak for the GRB 160726A evolves from the low to the high channel
in the inverse ``MODE II" way. In case this evolution
mode is new for sGRBs. Therefore, it is worthy
to search for the same effect in events from the Fermi/
GBM, HXMT/HE catalogues to asses more robust
conclusions in the future.

On 2017 August 17, GRB 170817A was observed independently
by the Fermi GBM Monitor and the Anti-
Coincidence Shield for the Spectrometer for the International
Gamma-Ray Astrophysics Laboratory \citep{2017ApJL...848...L14,2017ApJL...848...L15,2017ApJL...848...L13}
$\sim$ 1.7 s posterior to the first binary neutron star (BNS)
merger event GW170817 observed by the Advanced
LIGO and Virgo detectors \citep{2017GCN21509}. The
joint detection of GW170817/GRB 170817A confirms
that at least some sGRBs are indeed originated from the
mergers of the compact binaries. So, which sGRBs show
similar characteristics to GRB 170817A becomes a hot topic \citep[e.g.][]{2018APJL...863...L34,2018ApJL...853...L10,2018NC...9...4089,2019ApJ...876...89,2019APJL...880...L63}. In our subsample with EE, the temporal
properties of GRB 150101B and GRB 050724, including
the apparent two-component signature and the EE
tails which are strongest below 50 keV energy range starting
approximately at the end of the main peak,
are very phenomenologically similar in shape to that of
GRB 170817A, strengthening the potential relation with
GRB 170817A (see \citealt{2018APJL...863...L34,2019ApJ...876...89} for similar conclusions).

\cite{2016MNRAS...461...3607} reported that a highly variable
light curve viewed on-axis will become smooth
and apparently single-pulsed (when viewed off-axis).
They suggested that low-luminosity GRBs are consistent
with being ordinary bursts seen off-axis. \cite{2019APJL...880...L63} investigated the outflow structure
of GRB 170817A and found 14 sGRBs share the
similar relativistic structured jets with GRB 170817A.
They modelled their afterglow light curves and generated
the on-axis light curve for GRB 170817A which is
consistent with those of common sGRBs, as suggested by \cite{2019AA...628...A18}. Therefore,
it is very interesting to compare the on-axis prompt
emission properties by further observations directly.

\emph{\section{CONCLUSIONS}}
We studied nine sGRBs with precursor in our sample. Simultaneously, five typical Swift sGRBs with EE and one Fermi GRB 170817A
have been analyzed. For the first time, we presented the joint temporal property analysis of the fitting pulses
among the main peak with the other two components in both one-component sGRBs and two-component sGRBs. Our major results are summarized as follows:

1. We confirm that the main peaks in either one-component or two-component sGRBs tend to have no significant difference and might be  generated from the similar physical mechanism.

2. We inspected the correlations among the temporal properties of the SPs and DPs and found that the results are essentially consistent with those in
CGRO/BATSE ones recently found in our paper I. For instance, the t$_r$ and the t$_d$ are in agreement with a power-law relation for the SPs and the DPs, except the 2nd pulses of the Mt-DPs. There are no evident correlation between the asymmetry and the FWHM, the t$_m$ as well as the f$_m$, marginally similar to the results of BATSE sGRBs. Particularly, the temporal properties of the SPs have been found to be quite close to these
of the 1st pulses of the sub-type DPs, especially the asymmetry.

3. The relations, FWHM $\sim E^\alpha$, of the width of the main peak dependence on the average photon energy have been compared with the results discovered in paper I. The negatively mean index of $a\sim-$0.32 for the SPs is consistent with the
results for the BATSE SPs. Relative to the negative and positive energy correlations, $\alpha$ is found to be marginally zero not only in the SPs but also in the DPs. We studied the pulse shape evolutions from low to high energy channel for the sGRBs with $\alpha$ which is in close proximity to zero. It is found that there is no obvious shape evolution in one case. In another case, pulse shape evolves from low to high channel in a new way, inverse ``MODE II'', demonstrating that there would have more evolution
modes of pulses across differently adjacent energy channels in view of the Swift/BAT observations.

4. Furthermore, we studied the correlations of the main peaks with either the precursors or the EEs. No distinct correlations have been found in the asymmetry, the FWHM, the t$_r$ and the t$_r$ not only for the Pre+sGRBs but also for the sGRBs+EE. We found that the widths of main peaks tend to be narrower than those of the EE pulses and wider than those of the precursors. In particular, we verified the power-law correlations in the f$_m$ of the three components, strongly suggesting that they are seem to origin from the similar central engine activities. Especially, we compared the
temporal properties of the GRB 170817A with the other sGRBs+EE and no obvious difference have been found.

On the basis of these studies, we hope that the most important role of our results could show new lights on the search of the possible connection among these three components. More observational data of the precursors and the EEs are extremely needed in the future to constrain the current physical models in order to interpret the complex GRB light curves in the new era of satellites.


Acknowledgements

We thank the referee for very helpful suggestion and comments.
This work makes use of the data supplied NASA's High Energy Astrophysics Science Archive Research Center (HEASARC). It was supported by the Youth Innovations and Talents Project of Shandong Provincial Colleges and Universities (Grant No. 201909118) and the Natural Science Foundations (ZR2018MA030, XKJJC201901, OP201511, 20165660 and 11104161).



\clearpage
\startlongtable
\begin{deluxetable*}{l| c|c|c|c|c|c|c}
\tablecaption{The sample of sGRBs with three emission components. \label{tableEE}}
\tablehead{
\emph{GRB}& T$_{90} (s)$ & $Redshift$ & $Energy$ $band$ & $Satellite$ & $Precursor$ & $Main peak$ & $EE$  }
\tabletypesize{\small}
\startdata
190326A&0.076&&15-350 keV&Swift&$\times$&$\surd$&$\times$\\
181123B&0.260&&15-350 keV&Swift&$\times$&$\surd$&$\times$\\
180727A&1.056&&15-350 keV&Swift&$\times$&$\surd$&$\times$\\
180718A&0.084&&15-350 keV&Swift&$\times$&$\surd$&$\times$\\
180402A&0.180&&15-350 keV&Swift&$\surd$&$\surd$&$\times$\\
180204A&1.164&&15-350 keV&Swift&$\times$&$\surd$&$\times$\\
170817A&2.048&0.009783&8-350 keV&Fermi&$\times$&$\surd$&$\surd$\\
170428A&0.200&0.454&15-350 keV&Swift&$\times$&$\surd$&$\times$\\
170325A&0.332&&15-350 keV&Swift&$\times$&$\surd$&$\times$\\
160726A&0.728&&15-350 keV&Swift&$\surd$&$\surd$&$\times$\\
160612A&0.248&&15-350 keV&Swift&$\times$&$\surd$&$\times$\\
160601A&0.120&&15-350 keV&Swift&$\times$&$\surd$&$\times$\\
160408A&0.320&&15-350 keV&Swift&$\surd$&$\surd$&$\times$\\
151229A&1.440&&15-350 keV&Swift&$\times$&$\surd$&$\times$\\
151228A&0.276&&15-350 keV&Swift&$\times$&$\surd$&$\times$\\
150831A&0.920&&15-350 keV&Swift&$\times$&$\surd$&$\times$\\
150710A&0.152&&15-350 keV&Swift&$\times$&$\surd$&$\times$\\
150423A$^\ddag$&0.216&1.394&15-350 keV&Swift&$\times$&$\surd$&$\surd$\\
150301A&0.484&&15-350 keV&Swift&$\times$&$\surd$&$\times$\\
150120A$^\ddag$&1.196&0.460&15-350 keV&Swift&$\times$&$\surd$&$\surd$\\
150101B&0.012&0.1343&15-350 keV&Swift&$\times$&$\surd$&$\surd$\\
141212A&0.288&0.596&15-350 keV&Swift&$\times$&$\surd$&$\times$\\
140930B&0.844&&15-350 keV&Swift&$\times$&$\surd$&$\times$\\
140903A&0.296&0.351&15-350 keV&Swift&$\times$&$\surd$&$\times$\\
131004A$^\ddag$&1.536&0.717&15-350 keV&Swift&$\times$&$\surd$&$\surd$\\
130912A&0.284&&15-350 keV&Swift&$\times$&$\surd$&$\times$\\
130626A&0.160&&15-350 keV&Swift&$\times$&$\surd$&$\times$\\
130603B&0.176&0.3565&15-350 keV&Swift&$\times$&$\surd$&$\surd$\\
130515A&0.296&&15-350 keV&Swift&$\times$&$\surd$&$\times$\\
120804A$^\ddag$&0.808&1.3&15-350 keV&Swift&$\times$&$\surd$&$\surd$\\
120630A&0.596&&15-350 keV&Swift&$\times$&$\surd$&$\times$\\
120521A&0.512&&15-350 keV&Swift&$\times$&$\surd$&$\times$\\
120305A&0.100&&15-350 keV&Swift&$\times$&$\surd$&$\times$\\
111117A$^\ddag$&0.464&2.211&15-350 keV&Swift&$\times$&$\surd$&$\surd$\\
111020A&0.384&&15-350 keV&Swift&$\times$&$\surd$&$\times$\\
110420B&0.084&&15-350 keV&Swift&$\times$&$\surd$&$\times$\\
101224A&0.244&&15-350 keV&Swift&$\times$&$\surd$&$\times$\\
101219A$^\ddag$&0.828&0.718&15-350 keV&Swift&$\times$&$\surd$&$\surd$\\
101129A&0.384&&15-350 keV&Swift&$\times$&$\surd$&$\times$\\
100724A$^\ddag$&1.388&1.288&15-350 keV&Swift&$\times$&$\surd$&$\surd$\\
100702A&0.512&&15-350 keV&Swift&$\surd$&$\surd$&$\times$\\
100625A$^\ddag$&0.332&0.452&15-350 keV&Swift&$\surd$&$\surd$&$\surd$\\
100206A&0.116&0.4068&15-350 keV&Swift&$\times$&$\surd$&$\times$\\
100117A$^\ddag$&0.292&0.915&15-350 keV&Swift&$\times$&$\surd$&$\surd$\\
091109B&0.272&&15-350 keV&Swift&$\times$&$\surd$&$\times$\\
090621B&0.140&&15-350 keV&Swift&$\times$&$\surd$&$\times$\\
090510$^\ddag$&5.664&0.903&15-350 keV&Swift&$\surd$&$\surd$&$\surd$\\
090426$^\ddag$&1.236&2.609&15-350 keV&Swift&$\times$&$\surd$&$\surd$\\
090417A&0.068&&15-350 keV&Swift&$\times$&$\surd$&$\times$\\
081226A&0.436&&15-350 keV&Swift&$\times$&$\surd$&$\times$\\
081101&0.180&&15-350 keV&Swift&$\times$&$\surd$&$\times$\\
081024A&1.824&&15-350 keV&Swift&$\surd$&$\surd$&$\times$\\
080426&1.732&&15-350 keV&Swift&$\times$&$\surd$&$\times$\\
071112B&0.304&&15-350 keV&Swift&$\surd$&$\surd$&$\times$\\
070923&0.040&&15-350 keV&Swift&$\times$&$\surd$&$\times$\\
070810B&0.072&&15-350 keV&Swift&$\times$&$\surd$&$\times$\\
070809&1.280&0.2187&15-350 keV&Swift&$\times$&$\surd$&$\times$\\
061217&0.224&0.827&15-350 keV&Swift&$\times$&$\surd$&$\times$\\
060614&109.104&0.1254 &15-350 keV&Swift&$\times$&$\surd$&$\surd$\\
060502B&0.144&0.287&15-350 keV&Swift&$\surd$&$\surd$&$\times$\\
051221A&1.392&0.5464&15-350 keV&Swift&$\times$&$\surd$&$\surd$\\
051105A&0.056&&15-350 keV&Swift&$\times$&$\surd$&$\times$\\
050925&0.092&&15-350 keV&Swift&$\times$&$\surd$&$\times$\\
050813&0.384&0.722&15-350 keV&Swift&$\times$&$\surd$&$\times$\\
050724$^\dagger$&98.684&0.257&15-350 keV&Swift&$\surd$&$\surd$&$\surd$\\
050509B&0.024&0.2249&15-350 keV&Swift&$\times$&$\surd$&$\times$\\
050202&0.112&&15-350 keV&Swift&$\times$&$\surd$&$\times$
\enddata
\tablecomments{ The precursor (or EE) components of these sGRBs are too weak to be fitted successfully and are marked with $\dag$ (or $\ddag$).}
\end{deluxetable*}

\startlongtable
\begin{deluxetable*}{clcccc}
\tabletypesize{\small}
\tablecaption{The best-fit parameters of the power-law correlation between the t$_r$ and the t$_d$.\label{tab:tabletrtd}}
\tablehead{
\emph{GRB}& $\emph{N}$&$\emph{Pearson's r}$ & \emph{$R^2$}& $\emph{$\beta$}$}
\startdata
\hline
SPs&157&0.82&0.67&0.86 $\pm$ 0.05 \\
\hline
Mt-DPs 1st&18&0.79&0.59&0.94 $\pm$ 0.18 \\
Mt-DPs 2nd&18&0.24&0&0.32 $\pm$ 0.31 \\
\hline
Ml-DPs 1st&	7&0.52&0.12&1.31 $\pm$ 0.96 \\
Ml-DPs 2nd&	7&0.93&0.83&1.07 $\pm$ 0.20 \\
\hline
\enddata
\end{deluxetable*}

\startlongtable
\begin{deluxetable*}{clcccc}
\tabletypesize{\small}
\tablecaption{The best-fit parameters of the correlation of the logt$_m$ with the logFWHM. \label{tab:tabletmFWHM}}
\tablehead{
\emph{GRB}& $\emph{N}$&$\emph{Pearson's r}$ & \emph{$R^2$}& $\emph{$\mu$}$}
\startdata
\hline
SPs&154*&0.78&0.60&0.93$\pm$ 0.06\\
\hline
Mt-DPs 1st	&15*&0.46&0.15&0.38 $\pm$ 0.21\\
Mt-DPs 2nd	&18*&0.43&0.14&0.67$\pm$ 0.35\\
\hline
Ml-DPs 1st	&6*&0.53&0.10&2.20 $\pm$ 1.76\\
Ml-DPs 2nd	&7*&0.95&0.89&0.33 $\pm$ 0.05\\
\hline
\enddata
\tablecomments{* The number of sGRB pulses whose t$_m$ is positive.}
\end{deluxetable*}

\startlongtable
\begin{deluxetable*}{cccccc}
\tabletypesize{\small}
\tablecaption{The best-fit parameters of the power-law correlation of the f$_m$ with the FWHM. \label{tab:tablefmFWHM}}
\tablehead{
\emph{GRB}&\emph{N}&\emph{Pearson's r} & \emph{$R^2$}& \emph{$\nu$}}
\startdata
SPs&157 &-0.51 &0.25 &-0.45 $\pm$ 0.06\\
\hline
Mt-DPs 1st&18  &-0.79 &0.61 &-1.08 $\pm$ 0.21\\
Mt-DPs 2nd&18  &-0.31&0.04&-0.65 $\pm$ 0.50\\
\hline
Ml-DPs 1st&7  &-0.67 &0.34 &-1.31 $\pm$ 0.66\\
Ml-DPs 2nd&7  &-0.30 &-0.09 &-0.13 $\pm$ 0.18 \\
\hline
\enddata
\end{deluxetable*}

\startlongtable
\begin{deluxetable*}{cccccc}
\tabletypesize{\small}
\tablecaption{Asymmetric properties of the main peaks for different kinds of sGRBs.\label{tab:tableasy}}
\tablehead{
\emph{GRB}&\emph{N}&\emph{mean}&\emph{Median} & \emph{Minimum}& \emph{Maximum}}
\startdata
SPs		&157&0.73&0.65&0.03&1.56\\
\hline
Mt-DPs	1st	&18&0.79&0.65&0.17&1.56\\
Mt-DPs	2nd	&18&1.12&1.28&0.07&1.48\\
\hline
Ml-DPs	1st	&7&0.87&0.80&0.33&1.30\\
Ml-DPs	2nd	&7&1.07&1.37&0.50&1.51\\
\hline
All&207&0.79&0.70&0.03&1.56\\
\enddata
\end{deluxetable*}

\startlongtable
\begin{deluxetable*}{clccccc}
\tabletypesize{\small}
\tablecaption{The best-fit parameters of the correlation between the FWHM and the average energy
of photons in each channel.\label{tab:index}}
\tablehead{
\emph{Type}&\emph{GRB} &\emph{Pearson's r}& \emph{$R^2$}& \emph{$\chi^2_\nu$}& \emph{$\alpha$}}
\startdata
SPs&190326A &-0.99 &0.97 &1.2 &-0.40 $\pm$ 0.04\\
SPs&180718A &-0.94 &0.82 &0.60&-0.47 $\pm$ 0.12\\
SPs&170428A  &-0.92 &0.77&5.15 &-0.34 $\pm$ 0.10\\
SPs&160601A   &-0.94 &0.82 &7.27&-0.66 $\pm$ 0.17\\
SPs&150710A  &-0.97 &0.92&0.29 &-0.19 $\pm$ 0.03\\
SPs& 150301A &-0.93 &0.78&6.32 &-0.21 $\pm$ 0.06\\
SPs&141212A &-0.90 &0.71&3.16 &-0.33 $\pm$ 0.11\\
SPs&140903A &-0.92 &0.76 &3.12&-0.26 $\pm$ 0.08\\
SPs& 131004A &-0.90 &0.70 &1.52&-0.21 $\pm$ 0.06\\
SPs& 120305A&-0.99 &0.98 &0.63&-0.26 $\pm$ 0.02\\
SPs& 110420B &-0.93 &0.79 &12.82&-0.80 $\pm$ 0.23\\
SPs& 100206A* &-0.90 &0.73 &0.08&-0.06 $\pm$ 0.02\\
SPs& 091109B &-0.95 &0.85 &0.91&-0.27 $\pm$ 0.06\\
SPs& 090621B &-0.98 &0.93 &0.63&-0.26 $\pm$ 0.04\\
SPs& 070923 &-0.99 &0.97&0.10 &-0.18 $\pm$ 0.02\\
SPs& 050925&-0.91 &0.75 &3.60&-0.29 $\pm$ 0.09\\
\hline
Pre+SPs& 160726A precursor&-0.82&0.50&0.89&-0.28$\pm$ 0.14\\
Pre+SPs& 160726A* main&-0.66&0.16&3.57&-0.11 $\pm$ 0.09\\
\hline
SPs&GRB 070810B* &0.23 &-0.42 &0.69&0.05 $\pm$ 0.16\\
\hline
\hline
Mt-DPs& 120804A 1st &-0.28 &-0.38 &33.88&-0.35 $\pm$ 0.83\\
Mt-DPs& 120804A 2nd &-0.86 &0.62 &64.79&-0.79 $\pm$ 0.32\\
Mt-DPs& 101219A 1st &-0.85 &0.59 &31.33&-0.42 $\pm$ 0.18\\
Mt-DPs&101219A* 2nd &-0.69 &0.21&0.66 &-0.11 $\pm$ 0.08\\
\hline
Mt-DPs& 130912A* 1st &0.61 &0.05 &4.23&0.13 $\pm$ 0.12\\
Mt-DPs& 130912A* 2nd &-0.60 &0.05 &1.76&-0.12 $\pm$ 0.11\\
\hline
Ml-DPs& 180204A 1st&-0.79&0.43&10.27&-0.38 $\pm$ 0.21\\
Ml-DPs& 180204A 2nd&-0.93&0.79&4.32&-0.28 $\pm$ 0.08\\
Ml-DPs& 111117A* 1st&-0.83&0.52&0.07&-0.06 $\pm$ 0.03\\
Ml-DPs& 111117A 2nd&-0.97&0.91&0.59&-0.57 $\pm$ 0.10\\
\enddata
\tablecomments{* The power-law indexes in $FWHM \sim E^\alpha$ of sGRBs are marginally zero.}
\end{deluxetable*}

\clearpage
\begin{longrotatetable}
\begin{deluxetable*}{lclcrccCccccccccccc}
\tabletypesize{\tiny}
\tablecaption{Analysis Results of Precursors\label{prefitting}}
\tablewidth{750pt}
\setlength{\tabcolsep}{0.15mm}
\tablehead{
\colhead{sGRBs}&\colhead{f$_{mp}$}&
\colhead{t$_{mp}$} &
\colhead{t$_{rp}$} & \colhead{t$_{dp}$} & \colhead{(t$_r$/t$_d$)$_p$} &
\colhead{FWHM$_{p}$}&
\colhead{f$_{mm1}$} &
\colhead{t$_{mm1}$} &
\colhead{t$_{rm1}$} & \colhead{t$_{dm1}$} & \colhead{(t$_r$/t$_d$)$_{m1}$} &
\colhead{FWHM$_{m1}$}&\colhead{f$_{mm2}$} &
\colhead{t$_{mm2}$} &
\colhead{t$_{rm2}$} & \colhead{t$_{dm2}$} & \colhead{(t$_r$/t$_d$)$_{m2}$} &
\colhead{FWHM$_{m2}$}\\
&\colhead{(Counts/bin)} &
\colhead{(s)} &
\colhead{(s)} & \colhead{(s)} & \colhead{-} &
\colhead{(s)}&
\colhead{(Counts/bin)} &
\colhead{(s)} &
\colhead{(s)} & \colhead{(s)} & \colhead{-} &
\colhead{(s)}&
\colhead{(Counts/bin)} &
\colhead{(s)} &
\colhead{(s)} & \colhead{(s)} & \colhead{-} &
\colhead{(s)}
}
\startdata
single-main-peaked Pre+sGRBs&&&&&&&&&&&&&&&&&&\\
\hline 060502B(Ch1+2+3+4)(8ms)&0.29254&-0.39565&0.01804&0.03231&0.55834&0.05035&1.10619&0.01680&0.01249&0.03320&0.37620&0.04569&\nodata&\nodata&\nodata&\nodata&\nodata&\nodata\\ 060502B(Ch1+2)&0.18210&-0.39826&0.00933&0.03153&0.29591&0.04086&0.47337&0.01878&0.01421&0.03878&0.36643&0.05299&\nodata&\nodata&\nodata&\nodata&\nodata&\nodata\\
 060502B(Ch3+4)&0.12216&-0.40005&0.01804&0.03385&0.53294&0.05189&0.65985&0.01582&0.01095&0.02689&0.40721&0.03784&\nodata&\nodata&\nodata&\nodata&\nodata&\nodata\\ 071112B(Ch1+2+3+4)(16ms)\tablenotemark{a}&0.36533&-0.57263&0.02383&0.02295&1.03834&0.04678&0.34022&0.05956&0.06489&0.10150&0.63931&0.16639&\nodata&\nodata&\nodata&\nodata&\nodata&\nodata\\ 071112B(Ch1+2)&0.15123&-0.58015&0.02623&0.03538&0.74138&0.06161&0.16127&0.06561&0.05627&0.04372&1.28705&0.09999&\nodata&\nodata&\nodata&\nodata&\nodata&\nodata\\
071112B(Ch3+4)&0.21181&-0.57754&0.01532&0.02371&0.64614&0.03903&0.26077&0.05354&0.03620&0.09331&0.38795&0.12951&\nodata&\nodata&\nodata&\nodata&\nodata&\nodata\\
100702A(Ch1+2+3+4)(8ms)&0.46258&-0.25755&0.01681&0.02625&0.64038&0.04306&1.57141&0.08286&0.04207&0.06910&0.60883&0.11117&\nodata&\nodata&\nodata&\nodata&\nodata&\nodata\\
100702A(Ch2)&0.18082&-0.25757&0.03834&0.02750&1.39418&0.06584&0.56600&0.07597&0.04315&0.07271&0.59345 &0.11586&\nodata&\nodata&\nodata&\nodata&\nodata&\nodata\\
100702A(Ch3)&0.09887&-0.26002&0.02406&0.01716&1.40210&0.04122&0.55149&0.07321&0.02712&0.06700&0.40478&0.09412&\nodata&\nodata&\nodata&\nodata&\nodata&\nodata\\
160408A(Ch1+2+3+4)(8ms)&0.34285&-0.92805&0.00620&0.09794&0.06330&0.10414&0.80350&0.12414&0.09374&0.19035&0.49246&0.28409&\nodata&\nodata&\nodata&\nodata&\nodata&\nodata\\
160408A(Ch1+2)&0.24496&-0.89193 &0.04242&0.03554&1.19358&0.07796&0.26421&0.23971&0.20388&0.15022&1.35721&0.35410&\nodata&\nodata&\nodata&\nodata&\nodata&\nodata\\
160408A(Ch3+4)&\nodata&\nodata&\nodata&\nodata&\nodata&\nodata&0.57607&0.07839&0.05323&0.17024&0.31268&0.22347&\nodata&\nodata&\nodata&\nodata&\nodata&\nodata\\	160726A(Ch1+2+3+4)(8ms)&1.37414&0.01983&0.01621&0.02554&0.63469&0.04175&1.56596&0.63190&0.08339&0.09692&0.86040&0.18031&\nodata&\nodata&\nodata&\nodata&\nodata&\nodata\\
160726A(Ch1+2)&0.45222&0.01986&0.02871&0.04433&0.64764&0.07304&0.81925&0.61585&0.06846&0.09671&0.70789&0.16517&\nodata&\nodata&\nodata&\nodata&\nodata&\nodata\\
160726A(Ch3+4)&0.88454&0.02220&0.01645&0.01661&0.99037&0.03306&0.76960&0.66611&0.10826&0.08215&1.31783&0.19041&\nodata&\nodata&\nodata&\nodata&\nodata&\nodata\\	180402A(Ch1+2+3+4)(8ms)&0.42598&-0.19677&0.01679&0.02710&0.61956&0.04389&0.99640&0.19356&0.10409&0.07669&1.35728&0.18078&\nodata&\nodata&\nodata&\nodata&\nodata&\nodata\\
180402A(Ch1+2)&0.31619&-0.19699&0.01170&0.00754&1.55172&0.01924&0.33803&0.17332&0.13891&0.11366&1.22215&0.25257&\nodata&\nodata&\nodata&\nodata&\nodata&\nodata\\
180402A(Ch3+4)&0.34479&-0.20039&0.00735&0.01979&0.37140&0.02714&0.74040&0.20217&0.07322&0.05872&1.24693&0.13194&\nodata&\nodata&\nodata&\nodata&\nodata&\nodata\\
\hline
double-main-peaked Pre+sGRBs &&&&&&&&&&&&&&&&&&\\
\hline
081024A(Ch1+2+3+4)(8ms)&0.32881&-1.62738 &0.02039&0.03372&0.60469&0.05411&0.20932&-0.26660&0.17324&0.1313&1.31932&0.30455&0.45532&0.07160 &0.06408&0.07796&0.82196&0.14204\\
081024A(Ch1+2)&0.16799&-1.61997&0.03414&0.02491&1.37053&0.05905&0.10299&-0.27799&0.19897&0.12758&1.55957&0.32655&0.21059&0.09102 &0.10461&0.07092&1.47504&0.17553\\
081024A(Ch3)&0.12958&-1.63504&0.01357&0.04425&0.30667&0.05782&0.06638&-0.24707&0.16134&0.11555&1.39628&0.27689&0.18884&0.09078&0.05191 &0.04091&1.26888&0.09282\\
090510(Ch1+2+3+4)(8ms)&0.96274&-0.52401&0.02343&0.01152&2.03385&0.03495&3.13425&0.03598&0.03423&0.04810&0.71164&0.08233&1.48482&0.29996&0.07362&0.05295&1.39037&0.12657\\
090510(Ch1+2)&\nodata&\nodata&\nodata&\nodata&\nodata&\nodata&0.98523&0.03116 &0.03742&0.05867&0.63780&0.09609&0.67289&0.28228&0.06988&0.08297&0.84223&0.15285\\
090510(Ch3+4)&0.79215&-0.52361&0.01878&0.01302&1.44240&0.03180&2.21937&0.03597&0.02867&0.04453&0.64384&0.07320&0.73833&0.29850&0.08080&0.05654&1.42908&0.13734\\
100625A(Ch1+2+3+4)(8ms)&0.25267&-0.37602&0.01657&0.01334&1.24213&0.02991&0.91095&0.04791&0.08971&0.06296&1.42487&0.15267&1.19950 &0.21595 &0.08502&0.05728&1.48429&0.14230\\
100625A(Ch1+2)&0.13500&-0.37508&0.01876&0.01315&1.42662&0.03191&0.43599&-0.02516&0.04723&0.07267&0.64992&0.11990&0.50705&0.21970 &0.11453&0.08541&1.34094&0.19994\\
100625A(Ch3+4)&0.09812&-0.37598&0.02120&0.01405&1.50890&0.03525&0.61602&0.07495&0.07481&0.05539&1.35052&0.13020&0.87084&0.21245&0.03871 &0.02780&1.39245&0.06651\\
\enddata
\tablecomments{ We give the peak amplitude f$_{m}$, the peak time t$_{m}$, the rise time t$_{r}$, the decay time t$_{d}$, the asymmetry t$_r$/t$_d$ and the FWHM of each component. The subscript m and p identify the main peaks and precursor components.
}
\tablenotetext{a}{For this GRB, we adopt the average amplitude to estimate the peak amplitudes of the 8 ms light curve data. }
\end{deluxetable*}
\end{longrotatetable}

\clearpage
\begin{longrotatetable}
\begin{deluxetable*}{lccccccCccccccllrcc}
\tabletypesize{\tiny}
\tablecaption{Analysis Results of Extended Emissions\label{EEfitting1}}
\tablewidth{500pt}
\setlength{\tabcolsep}{0.25mm}
\tablehead{
\colhead{sGRBs} &
\colhead{f$_{mm}$} &
\colhead{t$_{mm}$} &
\colhead{t$_{rm}$} & \colhead{t$_{dm}$} & \colhead{(t$_r$/t$_d$)$_m$} &
\colhead{FWHM$_{m}$}&
\colhead{f$_{mE1}$} &
\colhead{t$_{mE1}$} &
\colhead{t$_{rE1}$} & \colhead{t$_{dE1}$} & \colhead{(t$_r$/t$_d$)$_{E1}$} &
\colhead{FWHM$_{E1}$}&\colhead{Cp$_{E2}$} &
\colhead{Tp$_{E2}$} &
\colhead{t$_{rE2}$} & \colhead{t$_{dE2}$} & \colhead{(t$_r$/t$_d$)$_{E2}$} &
\colhead{FWHM$_{E2}$}\\
&\colhead{(Counts/bin)} &
\colhead{(s)} &
\colhead{(s)} & \colhead{(s)} & \colhead{-} &
\colhead{(s)}&
\colhead{(Counts/bin)} &
\colhead{(s)} &
\colhead{(s)} & \colhead{(s)} & \colhead{-} &
\colhead{(s)}&
\colhead{(Counts/bin)} &
\colhead{(s)} &
\colhead{(s)} & \colhead{(s)} & \colhead{-} &
\colhead{(s)}
}
\startdata
sGRBs with single EE&&&&&&&&&&&&&&&&&&\\
\hline
050724(Ch1+2+3+4)(8ms)&1.48424&0.03066&0.04292&0.14942&0.28724&0.19234&0.25967&1.06785&0.10299&0.13674&0.75318&0.23973&\nodata&\nodata&\nodata&\nodata&\nodata&\nodata\\					050724(Ch1+2)&0.90669&0.01477&0.03347&0.17023&0.19662&0.20370&0.21770&1.08060&0.09051&0.08264&1.09523&0.17315&\nodata&\nodata&\nodata&\nodata&\nodata&\nodata\\ 050724(Ch3+4)&0.63496&0.06784&0.05864&0.10271&0.57093&0.16135&\nodata&\nodata&\nodata&\nodata&\nodata&\nodata&\nodata&\nodata&\nodata&\nodata&\nodata&\nodata\\
060614(Ch1+2+3+4)(1s)\tablenotemark{a}&0.93613&0.73466&2.24341&2.04035&1.09952&4.28376&0.54390&37.72733&22.07336&31.22141&0.70699&53.29477&\nodata&\nodata&\nodata&\nodata&\nodata&\nodata\\					060614(Ch1+2)&0.58041&0.73500&2.20074&2.04868&1.07422&4.24942&0.42033&38.73400&21.86396&31.21840&0.70035&53.08236&\nodata&\nodata&\nodata&\nodata&\nodata&\nodata\\					060614(Ch3+4)&0.33087&0.74033&2.32277&2.37680&0.97727&4.69957&0.12627&33.75601&22.95666&30.75025&0.74655&53.70691&\nodata&\nodata&\nodata&\nodata&\nodata&\nodata\\					150101B(Ch1+2+3+4)(2ms)\tablenotemark{a}&4.10741&0.00899&0.00632&0.00292&2.16438&0.00924&0.53550&0.03902&0.03900&0.02880&1.35411&0.06780&\nodata&\nodata&\nodata&\nodata&\nodata&\nodata\\					150101B(Ch1+2)&2.59968&0.00701&0.00303&0.00601&0.50416&0.00904&0.39149&0.05099&0.02439&0.01809&1.34826&0.04248&\nodata&\nodata&\nodata&\nodata&\nodata&\nodata\\					150101B(Ch3+4)&2.14224&0.00678&0.00634&0.00439&1.44465&0.01073&\nodata&\nodata&\nodata&\nodata&\nodata&\nodata\\
170817A(8-350 KeV)(64ms)\tablenotemark{a}{$^,$}\tablenotemark{b}&57.07469&-0.06509&0.20199&0.30698&0.65799&0.50897&23.73346&1.66293&0.29296&0.23442&1.24970&0.52738&\nodata&\nodata&\nodata&\nodata&\nodata&\nodata\\					170817A(8-50 KeV)&24.19345&0.12665&0.32209&0.20113&1.60140&0.52322&14.38594&1.53474&0.90567&0.74143&1.22152&1.64710&\nodata&\nodata&\nodata&\nodata&\nodata&\nodata\\					170817A(50-350 KeV)&52.45493&-0.16450&0.05621&0.15697&0.35809&0.21318&\nodata&\nodata&\nodata&\nodata&\nodata&\nodata\\											\hline
sGRBs with double EE&&&&&&&&&&&&&&&&&&\\
\hline
051221A(Ch1+2+3+4)(8ms)&5.96541&0.11585&0.06280&0.10660&0.58912&0.16940&0.55322&0.62791&0.04217&0.06762&0.62363&0.10979&0.99519&0.89204&0.03746&0.22567&0.16599&0.26313\\
051221A(Ch1+2)&2.06378&0.16418&0.13124&0.11657&1.12585&0.24781&0.44105&0.58814&0.21491&0.15913&1.35053&0.37404&0.81210&0.90792&0.05647&0.21265&0.26555&0.26912\\
051221A(Ch3+4)&3.54495&0.11599&0.05270&0.08532&0.61767&0.13802&\nodata&\nodata&\nodata&\nodata&\nodata&\nodata&0.2260&0.96211&0.08544&0.06866&1.24439&0.15410\\
\enddata
\tablecomments{We give the peak amplitude f$_{m}$, the peak time t$_{m}$, the rise time t$_{r}$, the decay time t$_{d}$, the asymmetry t$_r$/t$_d$ and the FWHM of each component. The subscript m and E identify the main peaks and EE components.
}
\tablenotetext{a}{For these GRBs, we adopt the average amplitude to estimate the peak amplitudes of the 8 ms light curve data. }
\tablenotetext{b} {For GRB
170817A, the summed GBM lightcurves for sodium iodide
(NaI) detectors 1, 2, and 5 with 64 ms resolution
between 8 and 350 keV have been used. We estimate the backgrounds
using the model \emph{$f_0$(t) = at+b} from 20 s prior to the trigger time.}
\end{deluxetable*}
\end{longrotatetable}

\begin{longrotatetable}
\setlength{\tabcolsep}{6pt}
\begin{deluxetable*}{lccccccCccccccllrcc}
\tabletypesize{\tiny}
\tablecaption{Analysis Results of Extended Emissions (Continued) \label{EEfitting2}}
\tablewidth{500pt}
\setlength{\tabcolsep}{0.25mm}
\tablehead{
\colhead{sGRBs} &
\colhead{f$_{mm1}$} &
\colhead{t$_{mm1}$} &
\colhead{t$_{rm1}$} & \colhead{t$_{dm1}$} & \colhead{(t$_r$/t$_d$)$_{m1}$} &
\colhead{FWHM$_{m1}$}&
\colhead{f$_{mm2}$} &
\colhead{t$_{mm2}$} &
\colhead{t$_{rm2}$} & \colhead{t$_{dm2}$} & \colhead{(t$_r$/t$_d$)$_{m2}$} &
\colhead{FWHM$_{m2}$}&\colhead{f$_{mE}$} &
\colhead{t$_{mE}$} &
\colhead{t$_{rE}$} & \colhead{t$_{dE}$} & \colhead{(t$_r$/t$_d$)$_{E}$} &
\colhead{FWHM$_{E}$}\\
&\colhead{(Counts/bin)} &
\colhead{(s)} &
\colhead{(s)} & \colhead{(s)} & \colhead{-} &
\colhead{(s)}&
\colhead{(Counts/bin)} &
\colhead{(s)} &
\colhead{(s)} & \colhead{(s)} & \colhead{-} &
\colhead{(s)}&
\colhead{(Counts/bin)} &
\colhead{(s)} &
\colhead{(s)} & \colhead{(s)} & \colhead{-} &
\colhead{(s)}
}
\startdata
sGRBs with single EE&&&&&&&&&&&&&&&&&&\\
130603B(Ch1+2+3+4)(8ms)&11.14306&0.02243&0.01184&0.02019&0.58643&0.03203&5.42614&0.07061&0.02032&0.02692&0.75483&0.04724&0.70613&0.19549&0.09033&0.06332&1.42656&0.15365\\
130603B(Ch1+2)&4.86977&0.02241&0.01246&0.01949&0.63930&0.03195&2.13590&0.07964&0.02673&0.03681&0.72616&0.06354&0.49596&0.21647&0.07554&0.05542&1.36305&0.13096\\
130603B(Ch3+4)&5.29256&0.02239&0.01183&0.01758&0.67292&0.02941&3.77207&0.06909&0.02843&0.02086&1.36290&0.04929&0.40396&0.11880&0.06665&0.05060&1.31719&0.11725\\
\enddata
\tablecomments{We give the peak amplitude f$_{m}$, the peak time t$_{m}$, the rise time t$_{r}$, the decay time t$_{d}$, the asymmetry t$_r$/t$_d$ and the FWHM of each component. The subscript m and E identify the main peak and EE components.
}
\end{deluxetable*}
\end{longrotatetable}

\end{document}